\documentclass[12pt]{article}
\usepackage{amsmath,amssymb,amsfonts}
\usepackage{graphicx}
\usepackage{hyperref}
\hypersetup{colorlinks,linkcolor={blue},citecolor={blue},urlcolor={red}} 
\usepackage{cite}
\usepackage[english]{babel}
\usepackage{bm}
\usepackage[utf8]{inputenc}
\usepackage{listings}
\usepackage[T1]{fontenc}
\usepackage{lmodern}
\usepackage{caption}
\usepackage{floatrow}
\usepackage{subcaption}
\usepackage{mathrsfs}
\usepackage{enumerate}
\usepackage[normalem]{ulem}
\usepackage{color}
\usepackage{array}
\def\beq{\begin{equation}}
\def\eeq{\end{equation}}
\def\be{\begin{equation}}
\def\ee{\end{equation}}
\def\bea{\begin{eqnarray}}
\def\eea{\end{eqnarray}}

\begin{document}

\title{\hfill\mbox{\small}\\[-1mm]
\hfill~\\[0mm]
       \textbf{Predictions for the Dirac CP-Violating Phase from Sum Rules}        }
\date{}
\author{\\[1mm]Luis A.~Delgadillo$^{1\,}$\footnote{E-mail: {\tt ldelgadillo4@ucol.mx}}~,~Lisa L.~Everett$^{3\,}$\footnote{E-mail: {\tt
leverett@wisc.edu}}~,~Raymundo~Ramos$^{4\,}$\footnote{E-mail: {\tt raramos@gate.sinica.edu.tw}}~,
\\and Alexander J.~Stuart$^{1,2\,}$\footnote{E-mail: {\tt astuart@ucol.mx}}\\
\\[1mm] \textit{\small $^1$Facultad de Ciencias-CUICBAS, Universidad de Colima,}\\
  \textit{\small C.P.~28045, Colima, Mexico}\\[3mm]
  \textit{\small $^2$ Dual CP Institute of High Energy Physics,}\\
  \textit{\small C.P.~28045, Colima, Mexico}\\[3mm]
  \textit{\small $^3$Department of Physics, University of Wisconsin,}\\
  \textit{\small Madison, WI 53706, USA}\\[3mm]
  \textit{\small $^4$Institute of Physics, Academia Sinica,}\\
  \textit{\small Nangang, Taipei 11529, Taiwan}\\[3mm]
 }
 
\maketitle

\vspace{-0.5cm}

\begin{abstract}
\noindent
We explore the implications of recent results relating the Dirac CP-violating
phase to predicted and measured leptonic mixing angles within a standard set of theoretical scenarios in which charged lepton corrections are responsible for generating a non-zero value of the reactor mixing angle.  We employ a full set of leptonic sum rules as required by the unitarity of the lepton mixing matrix, which can be reduced to predictions for the observable mixing angles and the Dirac CP-violating phase in terms of model parameters.  These sum rules are investigated within a given set of theoretical scenarios for the neutrino sector diagonalization matrix for several known classes of charged lepton corrections.  The results provide explicit maps of the allowed model parameter space within each given scenario and assumed form of charged lepton perturbations.
\end{abstract}


\thispagestyle{empty}
\vfill
\newpage
\setcounter{page}{1}

\newpage


\section{Introduction}
\label{sec:intro}

The decisive measurement of a non-zero reactor mixing angle\cite{dayabay,reno,doublechooz} has marked the beginning of an era of precision lepton measurements.   The current neutrino oscillation data have determined
the neutrino mass squared differences, $\Delta m_{21}^{2}$ and $\Delta m_{32}^{2}$, and the mixing parameters $\sin^2(\theta_{ij})$ (with $ij=\{12,13,23\}$) of the Maki-Nakagawa-Sakata-Pontecorvo (MNSP) lepton mixing matrix, $U_\mathsf{MNSP}$ \cite{Pontecorvo:1957cp,Maki:1962mu}. 
While  there are tantalizing recent hints of a nonvanishing value of the Dirac CP-violating phase $\delta \sim \pm \pi/2$ released by the T2K \cite{c5} and NO$\nu$A \cite{c6} collaborations,
current global fits allow the full [$0, 2 \pi$) range for $\delta$ at 3$\sigma$\cite{Gonzalez-Garcia:2014bfa,c1,c11}. 

With this wealth of current and anticipated lepton data, it is an ideal time to explore the many theoretical issues that remain unresolved in neutrino physics. A dominant question is the  origin and dynamics of the neutrino masses and the lepton mixing parameters. One popular
approach to addressing this question is to invoke a discrete flavor symmetry to explain the
observed patterns in the lepton data. In this framework, a given mixing pattern is related to
certain residual symmetries of the leptonic mass matrices.  These residual
symmetries may arise from the breaking of the flavor symmetry group.

Flavor models based on discrete flavor symmetries can be classified according to their leading order predictions for the solar mixing angle.  Usually these models are constructed in the basis where the charged leptons are diagonal. Indeed, one class of models that have been considered, and that in fact were quite standard prior to the reactor angle measurement, are situations in which the atmospheric neutrino mixing angle is maximal at leading order, and in which the reactor angle $\theta_{13}$ is zero to leading order.  Such scenarios include tribimaximal (TBM) mixing\cite{tbm1, tbm2, tbm3, tbm4,
tbm5}, bimaximal (BM) mixing\cite{bm1, bm2, bm3, bm4, bm5},  two golden ratio mixing schemes (GR1 \cite{gr1} and  GR2 \cite{gr2}), and hexagonal (HEX) \cite{hex} mixing. As the presence of a non-zero reactor angle $\theta_{13}$ in $U_\mathsf{MNSP}=U_e^\dagger U_\nu$ is incompatible with the predictions of these scenarios at leading order, these scenarios are not viable unless there are sufficient corrections to the mixing angle predictions at higher order. Such corrections can be attributed to perturbations in the charged lepton mixing matrix $U_{e}$ from a diagonal form at leading order, among other possibilities. $U_e$ can then consist of one rotation or two or more non-commuting rotations; in all viable cases considered, all of the measured mixing angles shift from their theoretical starting points.   For single rotations, which can arise, for example, within certain grand unified theories (GUTs) or specific discrete family symmetry models, it is well known that achieving a non-zero value of $\theta_{13}$ requires a rotation either in the $1-2$ or $1-3$ planes, as a single $2-3$ rotation {\it only} shifts $\theta_{23}$. Double rotations can be in (i) the $1-2$ and $2-3$ planes, (ii) the $1-3$ and $2-3$ planes, and (iii) the $1-2$ and $1-3$ planes. The results obtained from these different sets of possible perturbations can then be compared to data to constrain the parameters of the underlying theory. 

One method of characterizing the mixing angle predictions within this class of theoretical models has been to consider two 
specific types of sum rules that are referred to collectively as atmospheric and solar sum rules.   Atmospheric sum rules \cite{atm2,atm1} arise from a
variety of scenarios, including what are sometimes called semi-direct models (see e.g., \cite{atm3,atm4,atm5,atm6,atm7,atm8,atm9}), while solar sum rules \cite{sol1,sol2,sol3,sol4} are characteristic of models in which the leading order $U_{\nu}$ matrix is corrected by charged lepton contributions. The idea of correcting the leading order neutrino sector mixing angles by such charged lepton effects has been developed in
\cite{sum1,sum2a,sum2}.  Recent literature on such sum rules also includes \cite{sum3,sum4,sum5,sum6,sum7,sum8,sum9,sum10}, as well as the related work of  \cite{geref1,geref2,geref3,geref4,geref5}.

In this work, we describe an approach for obtaining a comprehensive set of sum rules that must all be satisfied to ensure the unitarity of $U_\mathsf{MNSP}$ within the assumption of three active neutrino species and no sterile neutrinos. The procedure is that for a given theoretical scenario, we first calculate all the possible ratios between the entries of the mixing matrix in question and equate them to the corresponding entries in the standard (PDG)
parametrization of $U_\mathsf{MNSP}$ \cite{Patrignani:2016xqp}. This results in several different expressions for
$\cos\delta$. Requiring that all such expressions are satisfied reduces these constraints to a single relation for $\cos\delta$, together with specific predicted relations for the lepton mixing angles (and for correlations among the mixing angles) that must be satisfied in any specific theory under consideration.  

Here we will consider theoretical scenarios for which at leading order, $U_{\nu}$ consists of the mixing angles $\theta_{23}^{\nu}$ and $\theta_{12}^{\nu}$, and consider sets of either one or two nontrivial rotations in $U_e$.  From these starting points, we will obtain general sets of sum rules for each case, and show how imposing these rules simultaneously results in nontrivial information about the allowed parameter space.
Using these results, we will then 
determine the favorable parameter space and numerical values for the MNSP phase $\delta$ for each combination of charged lepton mixing matrices and neutrino
mixing patterns, using the data from neutrino oscillation experiments as reported in the most recent global fit \cite{c1} and summarized in Table~\ref{tab:1}.
\begin{table}[H]
\centering
\begin{tabular}{|c|c|c|}
\hline 
 &  $3\sigma$ range \textbf{NO} &  $3 \sigma$ range \textbf{IO}\\
\hline 
$\sin^{2}\theta_{12}$ & 0.273 $-$ 0.379 &0.273 $    -$ 0.379\\
$\sin^{2}\theta_{23}$ & 0.384 $-$ 0.635 & 0.388 $   -$ 0.638\\
$\sin^{2}\theta_{13}$ & 0.0189 $-$ 0.0239 & 0.0189 $-$ 0.0239\\
\hline \hline
\end{tabular}
\caption{The current status of the lepton mixing angles for the case of normal ordering (NO) and inverted ordering (IO), as taken from the global fit of \cite{c1}.}
\label{tab:1}
\end{table} 
This paper is organized as follows. In Section~\ref{sec:frame}, we review the derivation of a well-known sum rule and present, for the case of a single rotation in $U_e$, our method for determining a broad set of sum rules that encompasses and extends this specific sum rule.  We then turn to the case of two rotations for the charged leptons in 
Section~\ref{sec:harderrot}.  In Section~\ref{sec:NA}, we present a numerical analysis of the predictions for $\cos\delta$ and $\sin\delta$ that correspond to regions in the parameter space of
$\theta_{ij}^{e}$ and $\delta_{ij}^{e}$ that satisfy the global fit values of the lepton mixing angles as given 
in Table~\ref{tab:1}.    
Conclusions are drawn in Section~\ref{sec:con}.


\section{Framework and one single rotation from the charged lepton sector}
\label{sec:frame}


We will begin this section by reviewing a well-known solar sum rule \cite{sum3} (see also \cite{sum2a,sum2,sum4,sum5} for related literature).  This sum rule is as follows:
\begin{equation}
\label{eq:sumruleorig}
\cos\delta=\frac{t_{23}s_{12}^2+s_{13}^2c_{12}^2/t_{23}-(s_{12}^{\nu})^
2(t_{23}+s_{13}^2/t_{23})}{s^\prime_{12}s_{13}},
\end{equation}
in which $c_{ij}=\cos\theta_{ij}$,
$s_{ij}=\sin\theta_{ij}$, $t_{ij} = \tan\theta_{ij}$, and we have used primed letters to represent the corresponding trigonometric functions of twice the argument, e.g., $s'_{ij}=\sin(2\theta_{ij})$.  

The sum rule of Eq.~(\ref{eq:sumruleorig}) holds for a particular class of theoretical scenarios in which the starting point is the assumption that
$U_{\nu}=R_{23}(\theta_{23}^{\nu})R_{12}(\theta_{12}^{\nu})$, in which the $R_{ij}$ are given by
\begin{equation}
\label{eq:neumix}
\begin{aligned}
R_{23}^{\nu}=\left(
\begin{array}{ccc}
 1 & 0 & 0 \\
 0 & c^{\nu}_{23} &  s^{\nu}_{23} \\
 0 & - s^{\nu}_{23} & c^{\nu}_{23} \\
\end{array}
\right), \;\;\; R_{12}^{\nu}=\left(
\begin{array}{ccc}
  c^{\nu}_{12} &  s^{\nu}_{12}&0 \\
  - s^{\nu}_{12} & c^{\nu}_{12}&0 \\
 0 & 0&1
\end{array}
\right).
\end{aligned}
\end{equation} 
It is well known that the measurement of a sizable $\theta_{13}$, as well as the possibility of nontrivial CP violation of the Dirac type, requires the presence of perturbations to this leading order structure. In particular, for the symmetric forms BM, TBM, HEX, GR1, and GR2, $\theta^{\nu}_{23}=\pi/4$, and $\theta_{12}^{\nu}$ satisfies
$\sin^{2} \theta_{12}^{\nu}=1/2$ for BM, $\sin^{2} \theta_{12}^{\nu}=1/3$ for
TBM, $\sin^{2} \theta_{12}^{\nu}=1/4$ for HEX, $\sin^{2}
\theta_{12}^{\nu}=(2+\phi)^{-1}$ for GR1 and $\sin^{2}
\theta_{12}^{\nu}=(3-\phi)/4$ for GR2, in which $\phi=(1+\sqrt{5})/2$ is the golden ratio. Hence, the leading order predictions of these scenarios must be modified for consistency with experimental data.  The needed modifications can take place, for example, if there are either one or two nontrivial rotation angles in the charged lepton mixing matrix $U_e$.  The solar sum rule of Eq.~(\ref{eq:sumruleorig}) holds in the case in which there is a single rotation in $U_e$ of the form $U_{e}=U^e_{12}$, or two successive rotations of the form $U_{e}=U^e_{23}U^e_{12}$ \cite{sum3,sum4}, in which the   $U^e_{ij}$ are defined as
\begin{equation}\label{rotations}
\begin{aligned}
	U_{23}^e & =\left(
\begin{array}{ccc}
 1 & 0        &                                0 \\
 0 & c^e_{23} &  s^e_{23}e^{-i \delta _{23}^{e}} \\
 0 & - s^e_{23}e^{i \delta _{23}^{e}} & c^e_{23} \\
\end{array}
\right), \;\;\; U_{12}^e=\left(
\begin{array}{ccc}
 c^e_{12} &  s^e_{12}e^{-i \delta _{12}^{e}}&0 \\
 - s^e_{12}e^{i \delta _{12}^{e}} & c^e_{12}&0 \\
 0 & 0&1
\end{array}
\right),\\
& \;\;\;\;\;\;\;\;\;\;\;\;\;\;\;\;\;\;\;\;\;\;\;
U_{13}^e  =\left(
 \begin{array}{ccc}
   c^e_{13} & 0& s^e_{13}e^{-i \delta _{13}^{e}}\\
   0 & 1&0\\
   - s^e_{13}e^{i \delta _{13}^{e}}&0 & c^e_{13} 
 \end{array}
 \right),
\end{aligned}
\end{equation}
in which $s^e_{ij} = \sin\theta^e_{ij}$ and $c^e_{ij} =
\cos\theta^e_{ij}$.\footnote{We note that these definitions  have an intrinsic degeneracy, i.e., 
$\delta_{ij}^{\prime e}\rightarrow \delta_{ij}^{e}-\pi$ and
$\theta_{ij}^{\prime e}\rightarrow \theta_{ij}^{e}-\pi /2$ yield the same
rotation matrix.  This degeneracy will be commented on later and taken into
account in our numerical analysis of the sum rules in Section~\ref{sec:NA}.}

We can also consider alternative structures for $U_{e}$ which lead to different sum rules than Eq.~(\ref{eq:sumruleorig}).  For example, it is well known that in the cases in which $U_{e}=U^e_{13}$ or $U_{e}=U^e_{23}U^e_{13}$, the analogous sum rule takes the very similar form
 \begin{equation}
 \label{eq:sumruleorig13}
\cos\delta=\frac{(1/t_{23}+s_{13}^2 t_{23}) (s_{12}^{\nu})^2 -( s_{12}^2/t_{23}+c_{12}^2 s_{13}^2 t_{23})}{s^\prime_{12}s_{13}}.
 \end{equation}
In contrast, in the case that $U_e$ is given by $U_{e}=U^e_{13}U^e_{12}$, it is not possible to write a sum rule that does not explicitly depend on the phase parameters $\delta^{e}_{ij}$ (as we will discuss later). 

In what follows, we will begin by discussing the origin of Eq.~(\ref{eq:sumruleorig}) for the scenario in which $U_{e}$ is given by a single rotation in the $1-2$ plane. We will then discuss how in this context, we can generate a broader set of sum rules that must all hold at once to guarantee unitarity and the proper predictions for the lepton mixing angles.  As a result, we see immediately the ways in which Eq.~(\ref{eq:sumruleorig}) can be naturally expressed in a form that takes into account specific mixing angle correlations that result from the underlying theory.  Using this approach, we will then turn to the other forms of $U_{e}$ just mentioned, considering first the second case of a single rotation in the $1-3$ plane, then turning to double rotations.


\subsection{The case of one rotation in the 1--2 sector}
\label{sec:case12}


In the case that $U_{e}=U^e_{12}(\theta_{12}^{e},\delta_{12}^{e})$, the MNSP matrix is given by  
\begin{equation}
U_\textsf{MNSP}\equiv U = U_e^\dagger U_\nu =U_{12}^{e\dagger}R_{23}^{\nu}R_{12}^{\nu}, 
\end{equation}
which yields the result that
\begin{equation}\label{eq:Ue12R23nuR12nu}
\begin{aligned}
	U_{e1} &=c^e_{12} c^{\nu }_{12}+c^{\nu }_{23} e^{-i \delta _{12}^e} s^e_{12}
	s^{\nu }_{12}, &
	U_{e2} &=c^e_{12} s^{\nu }_{12}-c^{\nu }_{12} c^{\nu }_{23} e^{-i \delta _{12}^e} s^e_{12},\\
	U_{e3} &=-e^{-i \delta _{12}^e} s^e_{12} s^{\nu }_{23},&
	U_{\mu1} &=-c^e_{12} c^{\nu }_{23} s^{\nu }_{12}+c^{\nu }_{12} e^{i \delta _{12}^e} s^e_{12},\\
		  U_{\mu2} &=c^e_{12} c^{\nu }_{12} c^{\nu }_{23}+e^{i \delta _{12}^e}
		  s^e_{12} s^{\nu }_{12},&
		  U_{\mu3} &=c^e_{12} s^{\nu }_{23},\\
		  U_{\tau1} &=s^{\nu }_{12} s^{\nu }_{23},&
U_{\tau2} &=-c^{\nu }_{12} s^{\nu }_{23},\\
U_{\tau3} &=c^{\nu }_{23}.
\end{aligned}
\end{equation}
With these results, it was noticed in \cite{sum4} that taking the ratio of the absolute values of $U_{\tau1}$ and $U_{\tau2}$ yields the simple relation
\begin{equation}
	\label{eq:ratiotau1tau2}
	\frac{|U_{\tau1}|}{|U_{\tau2}|}=t_{12}^{\nu}.
\end{equation}
Given the PDG form of the MNSP matrix, which is given by
\begin{equation}
	\label{eq:mnsppdg}
	U^{\mathsf{PDG}}=\left(
	\begin{array}{ccc}
	 c_{12} c_{13} & c_{13} s_{12} & e^{-i \delta } s_{13} \\
	 -c_{23} s_{12}-c_{12} e^{i \delta } s_{13} s_{23} & c_{12} c_{23}-e^{i \delta } s_{12} s_{13} s_{23} & c_{13} s_{23} \\
	 s_{12} s_{23}-c_{12} c_{23} e^{i \delta } s_{13} & -c_{12} s_{23}-c_{23}e^{i \delta }s_{12} s_{13} & c_{13} c_{23} \\
	\end{array}
\right)P_{\mathsf{Maj}},
\end{equation}
in which $P_{\mathsf{Maj}}$ is the diagonal Majorana phase matrix,
 the corresponding ratio takes the form
\begin{equation}\label{eq:ratiotau1tau2pdg}
\frac{|U^{PDG}_{\tau1}|}{|U^{PDG}_{\tau2}|}=\frac{|s_{12} s_{23}-c_{12} c_{23}  s_{13}e^{i \delta }|}{|c_{12} s_{23}+c_{23}  s_{12} s_{13}e^{i \delta }|}.
\end{equation}
Eqs.~\eqref{eq:ratiotau1tau2} and \eqref{eq:ratiotau1tau2pdg} can then be equated to solve for $\cos\delta$, which immediately yields Eq.~\eqref{eq:sumruleorig}.

This result was obtained with only one ratio of the entries of the MNSP matrix.  However, in principle there exist
$9!/7!2!=36$ possible ratios that must all hold simultaneously in order to preserve the unitarity of this matrix.  Let us now explore these additional ratios for the case at hand, for which the MNSP entries are given in Eq.~(\ref{eq:Ue12R23nuR12nu}). We will begin by considering the ratios of the remaining third row
elements of Eq.~\eqref{eq:Ue12R23nuR12nu}, which are
\begin{equation}\label{eq:thirdrowratios}
\frac{|U_{\tau1}|}{|U_{\tau2}|}=t^{\nu}_{12},~\frac{|U_{\tau1}|}{|U_{\tau3}|}=s^{\nu}_{12}t^{\nu}_{23},\frac{|U_{\tau2}|}{|U_{\tau3}|}=c_{12}^{\nu}t_{23}^{\nu},
\end{equation}
and here we have included $|U_{\tau1}|/|U_{\tau2}|$ once again for completeness.  Clearly,
we see now the appearance of two additional sum rules:
\begin{equation}\label{eq:extrasumrules3rdrow}
\begin{aligned}
s_{12}^{\nu}t_{23}^{\nu}=\frac{|s_{12} s_{23}-c_{12} c_{23}  s_{13}e^{i \delta
}|}{|c_{13} c_{23}|}, \;\;
c_{12}^{\nu}t_{23}^{\nu}=\frac{|c_{12} s_{23}+c_{23}  s_{12} s_{13}e^{i \delta }|}{|c_{13} c_{23}|}.
\end{aligned}
\end{equation}
Hence, we can extract relations for $\cos\delta$ from these sum rules in a similar way as in the first case.  More precisely, the 
$|U_{\tau1}|/|U_{\tau3}|$ sum rule for $\cos\delta$ takes the form
\begin{equation}\label{eq:Ut1/Ut3}
\cos\delta=\frac{1}{t_{23}s_{13}\sin(2\theta_{12})}(s_{12}^2t_{23}^2+c_{12}^2s_{13}^2-s_{12}^{\nu2}t_{23}^{\nu2}c_{13}^2),
\end{equation}
while the $|U_{\tau2}|/|U_{\tau3}|$ sum rule is given by
\begin{equation}\label{eq:Ut2/Ut3}
\cos\delta=\frac{1}{t_{23}s_{13}\sin(2\theta_{12})}(c_{12}^{\nu2}t_{23}^{\nu2}c_{13}^2-c_{12}^2t_{23}^2-s_{12}^2s_{13}^2).
\end{equation}
Since these two distinct predictions for $\cos\delta$ result from the same
unitary matrix, it is necessary to demand that Eqs.~\eqref{eq:Ut1/Ut3} and
\eqref{eq:Ut2/Ut3} predict the same result for $\cos\delta$, which yields
\begin{equation}\label{eq:angrel1}
t_{23}^{\nu2}=\frac{t_{23}^2+s_{13}^2}{c_{13}^2}.
\end{equation}
This relation constrains the possible values that the
measured atmospheric and reactor mixing angles can take when $\theta^{\nu}_{23}$ is fixed by theory, for example when
$\theta^{\nu}_{23}=\pi/4$.

As a second simple example of novel sum rules, let us consider $U_{\mu3}$ and $U_{\tau3}$.  Following an identical procedure, these terms can be shown to lead to the sum rule
\begin{equation}\label{eq:U23/U33}
\frac{|U_{\mu3}|}{|U_{\tau3}|}=|c_{12}^e t^{\nu}_{23}|=|t_{23}|,
\end{equation}
constraining $\theta_{12}^e$ with the experimentally measured $\theta_{23}$,
when $\theta_{23}^{\nu}$ is given by a specific flavor model. This line of reasoning thus leads
to the conclusion that all mixing matrix entry ratios must
be simultaneously satisfied while using the sum rule or else problematic
predictions may result. Let us thus analyze the $9!/7!2!=36$ possible sum rules that result
 from equating the theoretical predicted MNSP matrix $U=U_{12}^{e\dagger}R_{23}^{\nu}R_{12}^{\nu}$ to the PDG parametrization of the same matrix, $U^{\mathsf{PDG}}$. By comparing
Eqs.~\eqref{eq:Ue12R23nuR12nu}--\eqref{eq:mnsppdg}, it is seen that there are
four possible categories of sum rules, which are as follows: the sum rule
involves only either (i) $\cos\delta$ or (ii) $\cos(\delta^e_{12})$, (iii) it involves both
phases, or (iv) it involves none of them. 
Let us begin with case (iv).  It can readily be seen that there are three sum rules of this type:
\begin{equation}\label{eq:cd0list}
\begin{aligned}
|U_{e3}|/|U_{\tau 3}|&:&(s^e_{12})^2 (t^{\nu }_{23})^2&=sc_{23}^2 t_{13}^2,\\
|U_{\mu 3}|/|U_{\tau 3}|&:&(c^e_{12})^2 (t^{\nu }_{23})^2&=t_{23}^2,\\
|U_{e3}|/|U_{\mu 3}|&:&(t^e_{12})^2&=cs_{23}^2 t_{13}^2,
\end{aligned}
\end{equation}
in which $cs_{ij}=\csc(\theta_{ij})$, $sc_{ij}=\sec(\theta_{ij})$, and $ct_{ij}=\cot(\theta_{ij})$ (we will use this notation later).
Case (i) consists of sum rules that only contain $\cos\delta$; i.e., they do not
explicitly involve $\delta_{12}^e$.  There are seven of these sum rules:
\begin{equation}\label{eq:cdlist}
\begin{aligned}
|U_{\tau1}|/|U_{\tau2}|&:&2\cos\delta&=(c^{\nu }_{12})^2 \left[{cs}_{13} {ct}_{12}
	t_{23} (t_{12}^2-(t^{\nu }_{12})^2)+{ct}_{23} s_{13} t_{12}
	({ct}_{12}^2-(t^{\nu}_{12})^2)\right],\\
|U_{\tau1}|/|U_{\tau3}|&:&2\cos\delta&={ct}_{23} \left[{ct}_{12}
	s_{13}-c_{13} {cs}_{12} {ct}_{13} {sc}_{12} (s^{\nu }_{12})^2
	(t^{\nu}_{23})^2\right]+{cs}_{13} t_{12} t_{23},\\
|U_{\tau2}|/|U_{\tau 3}|&:&2\cos\delta&={ct}_{23} \left[c_{13} {cs}_{12}
	{ct}_{13} {sc}_{12} (c^{\nu }_{12})^2 (t^{\nu }_{23})^2-s_{13}
	t_{12}\right]-{cs}_{13} {ct}_{12} t_{23},\\
|U_{e3}|/|U_{\tau 1}|&:&2\cos\delta&=-{cs}_{12} {cs}_{23} s_{13} {sc}_{12} {sc}_{23} ({cs}^e_{12})^2 (s^{\nu }_{12})^2+{cs}_{13} t_{12} t_{23}+{ct}_{12} {ct}_{23} s_{13},\\
|U_{e3}|/|U_{\tau2}|&:&2\cos\delta&={cs}_{12} {cs}_{23} s_{13} {sc}_{12} {sc}_{23} (c^{\nu }_{12})^2 ({cs}^e_{12})^2-{cs}_{13} {ct}_{12} t_{23}-{ct}_{23} s_{13} t_{12},\\
|U_{\mu3}|/|U_{\tau1}|&:&2\cos\delta&=t_{23} \left[{cs}_{13} t_{12}-c_{13}
{cs}_{12} {ct}_{13} {sc}_{12} ({sc}^e_{12})^2 (s^{\nu }_{12})^2\right]+{ct}_{12} {ct}_{23} s_{13},\\
|U_{\mu3}|/|U_{\tau2}|&:&2\cos\delta&=c_{13} {cs}_{12} {ct}_{13} {sc}_{12} t_{23} (c^{\nu }_{12})^2 ({sc}^e_{12})^2-{cs}_{13} {ct}_{12} t_{23}-{ct}_{23} s_{13} t_{12}.
\end{aligned}
\end{equation}
Here we note that the sum rule corresponding to $|U_{\tau1}|/|U_{\tau2}|$ in Eq.~\eqref{eq:cdlist} is
the same sum rule as in Eq.~\eqref{eq:sumruleorig}, just written in a slightly more compact form. In case (ii), we have  the sum rules that contain $\cos(\delta^e_{12})$, and are independent of $\delta$. Here, again, there are seven sum rules:
\begin{equation}\label{eq:cdelist}
\begin{aligned}
&|U_{e1}|/|U_{e3}|: 2\cos(\delta^e_{12})=t^e_{12} \left(c_{12}^2 {ct}_{13}^2 {cs}^{\nu }_{12} s^{\nu }_{23} {sc}^{\nu }_{12} t^{\nu }_{23}-c^{\nu }_{23} t^{\nu}_{12}\right)-{ct}^e_{12} {ct}^{\nu}_{12} {sc}^{\nu }_{23},\\
& |U_{e2}|/|U_{e3}|:2\cos(\delta^e_{12})=c^{\nu }_{23} {ct}^{\nu }_{12} t^e_{12}-{ct}_{13}^2 s_{12}^2{cs}^{\nu }_{12} t^e_{12} s^{\nu }_{23} {sc}^{\nu }_{12} t^{\nu}_{23}+{ct}^e_{12} {sc}^{\nu }_{23} t^{\nu }_{12}, \\
&|U_{e2}|/|U_{e1}|:2\cos(\delta^e_{12})=c_{12}^2 \left[c^{\nu }_{23} t^e_{12} t^{\nu }_{12} \left(({ct}^{\nu }_{12})^2-t_{12}^2\right)+{ct}^e_{12} \left({sc}^{\nu }_{23} t^{\nu }_{12}-t_{12}^2{ct}^{\nu }_{12} {sc}^{\nu }_{23}\right)\right], \\
&|U_{e1}|/|U_{\mu3}|:2\cos(\delta^e_{12})={ct}^e_{12} \left(c_{12}^2 {cs}_{23}^2 {cs}^{\nu }_{12} s^{\nu }_{23} {sc}^{\nu }_{12} t^{\nu }_{23}-{ct}^{\nu }_{12} {sc}^{\nu }_{23}\right)-c^{\nu}_{23} t^e_{12} t^{\nu }_{12}, \\
&|U_{e1}|/|U_{\tau3}|:2\cos(\delta^e_{12})=c^{\nu }_{23} \left(c_{12}^2 {sc}_{23}^2 {cs}^e_{12} {cs}^{\nu }_{12} {sc}^e_{12} {sc}^{\nu }_{12}-t^e_{12} t^{\nu }_{12}\right)-{ct}^e_{12}{ct}^{\nu }_{12} {sc}^{\nu }_{23}, \\
&|U_{e2}|/|U_{\mu3}|:2\cos(\delta^e_{12})=c^{\nu }_{23} {ct}^{\nu }_{12} t^e_{12}+{ct}^e_{12} \left({sc}^{\nu }_{23} t^{\nu }_{12}-{cs}_{23}^2 s_{12}^2 {cs}^{\nu }_{12} s^{\nu }_{23} {sc}^{\nu}_{12} t^{\nu }_{23}\right), \\
&|U_{e2}|/|U_{\tau3}|:2\cos(\delta^e_{12})=c^{\nu }_{23} \left({ct}^{\nu }_{12} t^e_{12}-s_{12}^2 {sc}_{23}^2 {cs}^e_{12} {cs}^{\nu }_{12} {sc}^e_{12} {sc}^{\nu }_{12}\right)+{ct}^e_{12}{sc}^{\nu }_{23} t^{\nu }_{12}. \\
\end{aligned}
\end{equation}
In case (iii), there are 19 additional sum rules that depend on $\delta$ and $\delta^{e}_{12}$.  These are not presented here for simplicity, though they will be used in the analysis below.

We now consider the implications of considering the full set of 36 sum rules.  We will see that many of these relations lead to redundant information, which is expected given the smaller number of parameters than constraints.  Solving the third equation of Eq.~\eqref{eq:cd0list} yields a relationship between $\theta^e_{12}$, $\theta_{23}$, and $\theta_{13}$, which takes the form
\begin{equation}\label{eq:cd0listsol1}
(t^{e}_{12})^2=\frac{t_{13}^2}{s_{23}^2}.
\end{equation}
Incorporating this result in the other two relations in Eq.~\eqref{eq:cd0list} yields two conditions that are proportional to each other, as follows:
\begin{equation}\label{eq:consprop}
(s^e_{12})^2 (t^{\nu }_{23})^2-\text{sc}_{23}^2 t_{13}^2=\xi\left[(c^e_{12})^2 (t^{\nu }_{23})^2-t_{23}^2\right]=0 \text{~with~} \xi=\frac{t_{13}^2}{s_{23}^2}.
\end{equation}
Eq.~(\ref{eq:consprop}) thus clearly provides a single constraint on the mixing angles.\footnote{$\xi=0$ could be another constraint, but this leaves a single
remaining constraint which also must vanish.}  Solving this remaining
constraint gives the relation
\begin{equation}\label{eq:cd0listsol2}
t_{13}^2=c^2_{23}\left[(t^{\nu}_{23})^2-t_{23}^2\right].
\end{equation}
It is straightforward to see that Eq.~(\ref{eq:cd0listsol2}) is equivalent to Eq.~\eqref{eq:angrel1}, as expected. 

Imposing these constraints reduces all of the sum rules of Eq.~\eqref{eq:cdlist} to a single relation that depends on $\theta^\nu_{23}$, $\theta^\nu_{12}$, $\theta_{12}$, and either $\theta_{23}$ or $\theta_{13}$ (which are related by Eq.~(\ref{eq:cd0listsol2})).   The result that depends on $\theta_{23}$ takes the form
\begin{equation}
\label{eq:cdlistred}
\cos{\delta}=\frac{\text{sign}({sc}^{\nu }_{23}) \left[s^{\nu }_{23} t^{\nu }_{23} \left({cs}_{12} {cs}_{23} {sc}_{12} {sc}_{23} (c^{\nu }_{12})^2-{ct}_{23} t_{12}\right)+t_{12} t_{23} c^{\nu }_{23}-{ct}_{12} t_{23} {sc}^{\nu }_{23}\right]}{2 \sqrt{(t^{\nu }_{23})^2-t_{23}^2}},
\end{equation}
in which $\text{sign}({sc}^{\nu }_{23})=-1,0,1$ depending on whether $({sc}^{\nu}_{23})$ is negative, zero, or positive, respectively. 
This result can be expressed more compactly as follows:
\begin{equation}\label{eq:cdlistredcompact}
\cos{\delta}=\frac{\vert c_{23} \vert }{s^\prime_{12}s_{23}c_{23}\sqrt{(s^\nu_{23})^2-s_{23}^2}}\left (c_{12}^2 (s^\nu_{23})^2-c^\prime_{12}s_{23}^2-(s^\nu_{12})^2 (s^\nu_{23})^2 \right ).
\end{equation}
In the case that we eliminate $\theta_{23}$ in favor of $\theta_{13}$, we obtain
\begin{equation}
\label{eq:cdlistred13}
\begin{aligned}
\cos\delta &= -\frac{\sqrt{(c_{13} sc^\nu_{23})^2-1}}{2s_{13} t_{12}}-\frac{1}{2\sqrt{(c_{13} sc^\nu_{23})^2-1}}\left [s_{13}t_{12}-\frac{c_{13}(c^\nu_{12} t^\nu_{23})^2}{c_{12}s_{12}t_{13}} \right ] \\
&= \frac{1}{s^\prime_{12}s_{13} \vert c^\nu_{23}\vert\sqrt{(s^\nu_{23})^2-s_{13}^2}}\left( ((s^\nu_{23})^2-s_{13}^2)s_{12}^2+s^2_{13}c^2_{12}(c^\nu_{23})^2-(s^\nu_{12})^2 (s^\nu_{23})^2c^2_{13} \right ).
\end{aligned}
\end{equation}
A similar analysis can be carried out starting from the sum rules that involve only $\cos(\delta^e_{12})$.  More precisely, 
after applying Eqs.~\eqref{eq:cd0listsol1}--\eqref{eq:cd0listsol2} and, for example, writing the result in terms of $\theta_{23}$ rather than $\theta_{13}$ as before, Eq.~\eqref{eq:cdelist} once again can be reduced to the single sum rule:\footnote{An alternate expression for $\cos(\delta^e_{12})$ in which $\theta_{23}$ has been eliminated in favor of $\theta_{13}$ can be obtained in a straightforward manner; we neglect to present it here for simplicity.}
\begin{equation}
\label{eq:cdelistred}
\cos(\delta^e_{12})=\frac{\text{sign}({sc}_{23}){cs}_{23} {sc}_{23} {cs}^{\nu }_{12} {sc}^{\nu }_{12}  \left[{sc}^{\nu }_{23} {c'}_{12}^{\nu}(c'_{23}-(c^{\nu }_{23})^2)+s^{\nu }_{23} t^{\nu }_{23} c'_{12}\right]}{4 \sqrt{(t^{\nu }_{23})^2-t_{23}^2}}.
\end{equation}
Applying Eqs.~\eqref{eq:cd0listsol1}--\eqref{eq:cd0listsol2} to 
Eqs.~\eqref{eq:cdlist} and \eqref{eq:cdelist}  collapses each set down
to one remaining sum rule for $\cos\delta$, as given either in Eq.~(\ref{eq:cdlistredcompact}) or Eq.~(\ref{eq:cdlistred13}), and $\cos (\delta^e_{12})$, as seen, for example, in  Eq.~(\ref{eq:cdelistred}).  
These results can also be obtained directly from the 19 sum rules that include both $\delta$ and $\delta_{12}^e$. 
We can also use these constraints to replace $\theta_{12}$ in favor of the model parameters in any of the expressions we have obtained for $\cos\delta$ (Eq.~(\ref{eq:sumruleorig}), Eq.~(\ref{eq:cdlistred}), Eq.~(\ref{eq:cdlistredcompact}), or Eq.~(\ref{eq:cdlistred13})); we refrain from doing so until later when we focus on scenarios with specific model parameters.

In summary, for $U=U^{e\dagger}_{12}U^{\nu}_{23}U^{\nu}_{12}$, we have shown that we can reduce the 36 possible ratios of the entries of $U_{\mathsf{MNSP}}$  to two relations among the observable mixing angles,
Eqs.~\eqref{eq:cd0listsol1} and \eqref{eq:cd0listsol2}, one relation between $\cos(\delta^e_{12})$ and the
mixing angles as discussed, and one sum rule for $\cos\delta$, as expected. Furthermore, it is straightforward to see that this machinery reproduces the well-known relations among the mixing angles and the elements of $U_{\mathsf{MNSP}}$:
\begin{equation}\label{eq:alttoratios2}
	\sin^2(\theta_{13})=|U_{e3}|^2,\;\;\;
	\sin^2(\theta_{23})=\frac{|U_{\mu3}|^2}{1-|U_{e3}|^2},\;\;\;
	\sin^2(\theta_{12})=\frac{|U_{e2}|^2}{1-|U_{e3}|^2}.
\end{equation}
The relevant angle relations are encoded in Eqs.~(\ref{eq:cd0listsol1}), (\ref{eq:cd0listsol2}), and (\ref{eq:cdelistred}).  These relations reproduce the well-known results that also follow from Eq.~(\ref{eq:alttoratios2}):
\begin{equation}\label{eq:sij2}
\begin{aligned}
	\sin^2(\theta_{13}) &=(s^e_{12})^2 (s^{\nu }_{23})^2,\;\;\;\;\; \sin^2(\theta_{23}) =\frac{(c^e_{12})^2 (s^{\nu }_{23})^2}{1-(s^e_{12})^2
		(s^{\nu }_{23})^2},\\
		\sin^2(\theta_{12}) &=\frac{(c^{\nu }_{12})^2 (c^{\nu }_{23})^2 (s^e_{12})^2+(c^e_{12})^2 (s^{\nu }_{12})^2-2 c^e_{12} c^{\nu }_{12} c^{\nu }_{23} \cos
   (\delta^e_{12}) s^e_{12} s^{\nu }_{12}}{1-(s^e_{12})^2 (s^{\nu }_{23})^2}.
\end{aligned}
\end{equation}
It is also straightforward to see that our results for $\cos\delta$ as given in Eq.~(\ref{eq:cdlistredcompact}) and Eq.~(\ref{eq:cdlistred13}) also follow from imposing the mixing angle relations between $\theta_{13}$ and $\theta_{23}$ as encoded in Eq.~\eqref{eq:sij2} in the original sum rule of Eq.~\eqref{eq:sumruleorig}.

As a cross-check, we can derive a constraint on $\sin\delta$ from relating the Jarlskog determinants in a similar manner.  Carrying out this process and once again eliminating $\theta_{13}$ yields 
\begin{equation}\label{eq:jarlskog}
\begin{aligned}
&\sin\delta=
-cs'_{12} {s'}^{\nu}_{12} \sqrt{1-\frac{cs_{23}^2 (cs^{\nu }_{12})^2 (sc^{\nu }_{12})^2 \left[{c'}^{\nu}_{12} (-2c'_{23} +{c'}^{\nu}_{23}+1)-2 (s^{\nu }_{23})^2 c'_{12}\right]^2}{32 (c'_{23}-{c'}^{\nu}_{23})}}.
\end{aligned}
\end{equation}
Upon squaring the above relationship and adding it to any of results for the square of $\cos\delta$ in Eq.~\eqref{eq:cdlistredcompact} the result trivially yields the identity, as expected.

\subsection{The case of one rotation in the 1--3 sector} \label{sec:case13}

We now turn to the situation in which the single rotation in $U_e$ is in the $1-3$ plane, such that $U_{e}=U^e_{13}(\theta_{13}^{e},\delta_{13}^{e})$.  Implementing the same methodology as in the previous section, we can determine the $\cos\delta$ sum rule as well as the relations among the MNSP mixing angles in this scenario.  The sum rules that are independent of $\cos\delta$ and $\cos(\delta^e_{13})$ are given by:
\begin{equation}\label{eq:cd0list13}
\begin{aligned}
|U_{e3}|/|U_{\tau 3}|&:&(s^e_{13})^2 (t^{\nu }_{23})^2&=cs_{23}^2 t_{13}^2,\\
|U_{\mu 3}|/|U_{\tau 3}|&:&(sc^e_{13})^2 (t^{\nu }_{23})^2&=t_{23}^2,\\
|U_{e3}|/|U_{\mu 3}|&:&(t^e_{13})^2&=sc_{23}^2 t_{13}^2.
\end{aligned}
\end{equation}
The full set of $\cos\delta$-dependent sum rules is as follows:
\begin{equation}\label{eq:cd1list13}
\begin{aligned}
&\vert U_{e 3} \vert / \vert U_{\mu 1} \vert: 2 \cos \delta= cs_{12}(cs_{13}^{e})^{2}cs_{23}sc_{12}sc_{23}(s_{12}^{\nu})^{2}s_{13}-ct_{23}cs_{13}t_{12}-ct_{12}s_{13}t_{23}, \\
&\vert U_{e 3} \vert / \vert U_{\mu 2} \vert:  2 \cos \delta=ct_{12}ct_{23}cs_{13}+s_{13}\big(t_{12}t_{23}-(c_{12}^{\nu})^{2}cs_{12}(cs_{13}^{e})^{2}cs_{23}sc_{12}sc_{23} \big), \\
&\vert U_{\mu 1} \vert / \vert U_{\mu 2} \vert:  2 \cos \delta= (c_{12}^{\nu})^{2}\Big[ct_{23}cs_{13}\big(ct_{12}(t_{12}^{\nu})^{2}-t_{12} \big)+t_{23}ct_{12}s_{13}\big(t_{12}^{2}(t_{12}^{\nu})^{2}-1 \big) \Big], \\
&\vert U_{\mu 1} \vert / \vert U_{\mu 3} \vert:  2 \cos \delta= -ct_{23}cs_{13}t_{12}+t_{23}\big(c_{13}ct_{13}(ct_{23}^{\nu})^{2}cs_{12}sc_{12}(s_{12}^{\nu})^{2}-ct_{12}s_{13} \big), \\
&\vert U_{\mu 2} \vert / \vert U_{\mu 3} \vert:  2 \cos \delta= ct_{12}ct_{23}cs_{13}+t_{23}\big(s_{13}t_{12}-(c_{12}^{\nu})^{2}c_{13}ct_{13}(ct_{23}^{\nu})^{2}cs_{12}sc_{12} \big), \\
&\vert U_{\mu 1} \vert / \vert U_{\tau 3} \vert:  2 \cos \delta=c_{13}ct_{13}ct_{23}cs_{12}sc_{12}(sc_{13}^{e})^{2}(s_{12}^{\nu})^{2}-ct_{23}cs_{13}t_{12}-ct_{12}s_{13}t_{23}, \\ 
&\vert U_{\mu 2} \vert / \vert U_{\tau 3} \vert:  2 \cos \delta=ct_{12}ct_{23}cs_{13}-(c_{12}^{\nu})^{2}c_{13}ct_{13}ct_{23}cs_{12}sc_{12}(sc_{13}^{e})^{2}+s_{13}t_{12}t_{23}. \\
\end{aligned}
\end{equation}
The sum rules that contain $\cos(\delta^e_{13})$, but are independent of $\delta$, are:
\begin{equation}\label{eq:cdelist13}
\begin{aligned}
&|U_{e1}|/|U_{e3}|: 2\cos(\delta^e_{13})=t^e_{13} \left(s_{23}^{\nu}t_{12}^{\nu}-c_{12}^{2}c_{23}^{\nu}ct_{13}^{2}ct_{23}^{\nu}cs_{12}^{\nu}sc_{12}^{\nu}\right)+ct_{12}^{\nu}ct_{13}^{e}cs_{23}^{\nu},\\
& |U_{e2}|/|U_{e3}|:2\cos(\delta^e_{13})=t_{13}^{e} \left(c_{23}^{\nu}ct_{13}^{2}ct_{23}^{\nu}cs_{12}^{\nu}sc_{12}^{\nu}s_{12}^{2}-ct_{12}^{\nu}s_{23}^{\nu} \right)-ct_{13}^{e}cs_{23}^{\nu}t_{12}^{\nu}, \\
&|U_{e2}|/|U_{e1}|:2\cos(\delta^e_{13})=c_{12}^2 \left[s_{23}^{\nu}t_{12}^{2}t_{12}^{\nu}t_{13}^{e}-ct_{13}^{e}cs_{23}^{\nu}t_{12}^{\nu}+{ct}^\nu_{12} \left(ct^e_{13}cs^\nu_{23}t^2_{12}-s^\nu_{23}t^e_{13} \right)\right], \\
&|U_{e1}|/|U_{\mu3}|:2\cos(\delta^e_{13})=ct_{12}^{\nu}ct_{13}^{e}cs_{23}^{\nu}+s_{23}^{\nu} \left(t^\nu_{12}t^e_{13}-c_{12}^{2}cs_{12}^{\nu}cs_{13}^{e}cs_{23}^{2}sc^\nu_{12}sc^e_{13} \right), \\
&|U_{e1}|/|U_{\tau3}|:2\cos(\delta^e_{13})=ct_{12}^{\nu}ct_{13}^{e}cs_{23}^{\nu}+s_{23}^{\nu}t^\nu_{12}t^e_{13}-c^2_{12}c_{23}^{\nu}ct_{13}^{e}ct_{23}^{\nu}cs_{12}^{\nu}sc_{12}^{\nu}sc_{23}^{2}, \\
&|U_{e2}|/|U_{\mu3}|:2\cos(\delta^e_{13})=cs^\nu_{12}cs^e_{13}cs^2_{23}sc^\nu_{12}sc^{e}_{13}s^2_{12}s^\nu_{23}-ct^{e}_{13}cs^{\nu}_{23}t^\nu_{12}-ct^\nu_{12}s^\nu_{23}t^e_{13}, \\
&|U_{e2}|/|U_{\tau3}|:2\cos(\delta^e_{13})=c_{23}^{\nu}ct_{13}^{e}ct^\nu_{23}cs^\nu_{12}sc^\nu_{12}sc_{23}^{2}s^2_{12}-ct^e_{13}cs^\nu_{23}t^\nu_{12}-ct^\nu_{12}s^\nu_{23}t^e_{13}. \\
\end{aligned}
\end{equation}
As before, we do not present the remaining sum rules that depend on both $\cos\delta$ and $\cos(\delta^e_{13})$.  
The mixing angle relations that result from this full set of sum rules are given by
\begin{equation}\label{eq:sij213}
\begin{aligned}
\sin^{2}(\theta_{13}) &=(c_{23}^{\nu})^{2}(s_{13}^{e})^{2}\\
\sin^{2}(\theta_{23}) &=\frac{(s_{23}^{\nu})^{2}}{1-(c_{23}^{\nu})^{2}(s_{13}^{e})^{2}}=\frac{(s_{23}^{\nu})^{2}}{1-\sin^{2}\theta_{13}}\\
\sin^{2}(\theta_{12}) &=\frac{(c_{13}^{e})^{2}(s_{12}^{\nu})^{2}+2\cos \delta_{13}^ec_{12}^{\nu}c_{13}^{e}s_{12}^{\nu}s_{13}^{e}s_{23}^{\nu}+(c_{12}^{\nu})^{2}(s_{13}^{e})^{2}(s_{23}^{\nu})^{2}}{1-\sin^{2}\theta_{13}},
\end{aligned}
\end{equation}
 as is well known in the literature.  Here we note that as in the single $12$ rotation case, the relation between $\sin^2\theta_{13}$ and $\sin^2\theta_{23}$ follows from the phase-independent sum rules of Eq.~(\ref{eq:cd0list13}), while the expression for $\sin^2\theta_{12}$ requires further input from Eq.~(\ref{eq:cdelist13}).  The sum rule for $\cos\delta$, after using the mixing angle constraints to solve for $\theta_{13}$ in terms of $\theta_{23}$, takes the form
\begin{equation}\label{eq:cosdelta13v1}
\cos\delta = \frac{\vert s_{23} \vert}{s^\prime_{12}c_{23}s_{23} \sqrt{s_{23}^2 -(s^\nu_{23})^2}} \left (
(s^\nu_{12})^2(c^\nu_{23})^2-(s^2_{12}c^2_{23}+c^2_{12}s^2_{23}-c^2_{12}(s^\nu_{23})^2) \right ).
\end{equation}
As before, we can also express this relation as a function of $\theta_{12}$ and $\theta_{13}$, as follows:
\begin{equation}\label{eq:cosdelta13v2}
\cos\delta = \frac{1}{s^\prime_{12}s_{13} \vert s^\nu_{23} \vert \sqrt{c_{13}^2 -(s^\nu_{23})^2}} \left ((s^\nu_{12})^2c_{13}^2 (c^\nu_{23})^2-s_{12}^2c_{13}^2+(s_{12}^2-s_{13}^2c_{12}^2)(s^\nu_{23})^2 \right ).
\end{equation}
We note that as expected, Eqs.~(\ref{eq:cosdelta13v1}) and (\ref{eq:cosdelta13v2}) both follow directly from Eq.~(\ref{eq:sumruleorig13}) upon using the mixing angle relations of Eq.~(\ref{eq:sij213}) to eliminate either $\theta_{13}$ or $\theta_{23}$, respectively.


\section{Two unitary rotations from the charged lepton sector}
\label{sec:harderrot}

We now consider the case of two successive rotations in $U_e$.  As previously
mentioned, in light of the non-zero value for the reactor angle and the current
experimental hints for a possibly non-maximal atmospheric mixing angle, we
will consider the most economic forms of $U_{e}$ that achieve these goals as
previously discussed in \cite{sum5}, though we will use a slightly different
parametrization.  With this set of starting points, we will carry out the
analogous machinery as given in the previous section, generalized slightly to account for the additional parameters that result in the case of double rotations.  The results, as we will see, will again be a set of four relations, three for the
mixing angles, and one sum rule for $\cos\delta$.


\subsection{The case of two rotations in the 1--2 and 2--3 sectors}
\label{sec:1223}

We first consider the case in which $U_{e}=U_{23}^{e}(\theta_{23}^{e},\delta_{23}^{e})
U_{12}^{e}(\theta_{12}^{e},\delta_{12}^{e})$. We will present the sum rules that depend only on
$\cos\delta$ or $\cos (\delta^e_{23})$, or on no phase parameters.  For simplicity, we will not present the sum rules that depend on more than one phase parameter, though they will be incorporated into the analysis.  

We will start with the one sum
rule that does not depend on any of the phases: 
\begin{equation}
	\vert U_{e3}\vert / \vert U_{\mu 3}\vert: t^e_{12} = t_{13} cs_{23}.
\end{equation}
The $\cos\delta$ sum rules are given by
\begin{equation}\label{eq:cosdel1223}
\begin{aligned}
& \vert U_{\tau 1} \vert / \vert U_{\tau 2} \vert: 2 \cos \delta=(c_{12}^{\nu})^{2}\left[ -ct_{23}s_{13}t_{12}(t_{12}^{\nu})^{2}+cs_{13}t_{12}t_{23}+ct_{12}\big(ct_{23}s_{13}-cs_{13}(t_{12}^{\nu})^{2}t_{23} \big)\right], \\
&\vert U_{e 3} \vert/ \vert U_{\tau 2} \vert: 2 \cos \delta = s_{13} sc_{12}
	cs_{23}[sc_{23} cs_{12}  (c^\nu_{12} cs^e_{12})^{2} - c_{23} s_{12} ]- c_{12} s_{23}sc_{23} cs_{12} cs_{13}, \\
&\vert U_{\mu 3} \vert/ \vert U_{\tau 1} \vert :  2 \cos
	\delta=s_{13}ct_{12} ct_{23} + t_{23}cs_{13} \left[t_{12} -
	(s^\nu_{12} c_{13} sc^e_{12})^{2} cs_{12}  sc_{12}\right],\\
&\vert U_{e3} \vert/ \vert U_{\tau 1} \vert: 2 \cos \delta=
	s_{13} \left[ ct_{12} ct_{23} - (s^\nu_{12} cs^e_{12})^{2} sc_{12} sc_{23}
	cs_{12} cs_{23}\right] + t_{12} t_{23}cs_{13}, \\
&\vert U_{\mu 3} \vert / \vert U_{\tau 2} \vert: 2 \cos \delta =
	t_{23}\left[(c_{12}^{\nu}sc_{12}^{e})^{2}c_{13}ct_{13}cs_{12}sc_{12}-ct_{12}cs_{13}
	\right]-ct_{23}s_{13}t_{12}.
\end{aligned}
\end{equation}
Here we note that the first expression of Eq.~(\ref{eq:cosdel1223}) is equivalent to Eq.~\eqref{eq:sumruleorig} as previously noted in the literature. The $\cos(\delta^e_{23})$ sum rules are given by
\begin{equation}
\begin{aligned}
\vert U_{e3}\vert / \vert U_{\tau 3}\vert:
2\cos(\delta_{23}^{e})= {}&\frac{4}{s^{\prime e}_{23} s^{\prime\nu}_{23}
	\left[s^2_{12} + (s_{23} s^e_{12})^2\right]}
	\big\{(c_{13} c_{23} s^e_{12})^2 \big[( c^e_{23} s^\nu_{23})^{2}+(c^\nu_{23} s^e_{23})^{2}\big] \\
			& -  s^2_{13}\big[(c^e_{23}
	c^\nu_{23})^{2} + ( s^e_{23} s^\nu_{23})^{2}\big]\big\},\\
\vert U_{\mu 3}\vert / \vert U_{\tau 3}\vert:
2\cos(\delta_{23}^{e})= {}& \frac{4}{s^{\prime e}_{23} s^{\prime\nu}_{23}
	\left[1-(c_{23} s^e_{12})^2\right]} \big\{(c_{23} c^e_{12})^2
	\big[(c^e_{23} s^\nu_{23})^{2} + ( c^\nu_{23} s^e_{23})^{2}\big] \\
	&-
	s^2_{23}\big[(c^e_{23} c^\nu_{23})^{2} + (s^e_{23}
s^\nu_{23})^{2}\big]\big\}.\\
\end{aligned}
\end{equation} 
Once these sum rules and the remaining sum rules that depend on $\cos(\delta^e_{12})$ and on more than one phase factor are implemented simultaneously, the $\cos\delta$-dependent sum rules can be expressed as a function of $\theta_{12}$ and $\theta_{23}$ (as well as model parameters), as follows:
\begin{equation}\label{eq:cddouble1223form1}
\cos\delta = \frac{\vert c_{23} \vert}{s^\prime_{12}s_{23}c_{23}\sqrt{\sin^2(\tilde{\theta}_{23})-s_{23}^2}}\left (\sin^2(\tilde{\theta}_{23})c^2_{12}-s_{23}^2c^\prime_{12}
-(s_{12}^\nu)^2 \sin^2(\tilde{\theta}_{23})\right ),
\end{equation}
or equivalently in terms of $\theta_{12}$ and $\theta_{13}$, as
\begin{equation}\label{eq:cddouble1223form2}
\cos\delta= \frac{(\sin^2(\tilde{\theta}_{23})-s_{13}^2)s_{12}^2+s_{13}^2c_{12}^2(1-\sin^2(\tilde{\theta}_{23}))-(s^\nu_{12})^2\sin^2(\tilde{\theta}_{23})c_{13}^2}{s^\prime_{12}s_{13}\sqrt{1-\sin^2(\tilde{\theta}_{23})}\sqrt{(\sin^2(\tilde{\theta}_{23})-s_{13}^2)}},
\end{equation}
in which the angle $\tilde{\theta}_{23}$ has been defined analogously to Eq.~(A.1) of
\cite{sum2}, as follows:
\begin{equation}\label{eq:sintilde}
\sin^{2}(\tilde{\theta}_{23})=(c_{23}^{\nu})^2 (s_{23}^{e})^2-\frac{1}{2}\cos(\delta_{23}^{e})s_{23}^{\prime e}s_{23}^{\prime \nu}+(c_{23}^{e})^{2}(s_{23}^{\nu})^{2}.
\end{equation} 
Here we note that Eqs.~(\ref{eq:cddouble1223form1}) and (\ref{eq:cddouble1223form2}) are similar to the analogous expressions for the $1-2$ mixing scheme as given in Eqs.~(\ref{eq:cdlistredcompact}) and (\ref{eq:cdlistred13}) with the replacement of $\sin^2(\theta^\nu_{23})$ with $\sin^2(\tilde{\theta}_{23})$.

This procedure also results in the expected  relationships between the MNSP mixing angles and the theoretical parameters in this scheme, which are given by
\begin{align}
	\label{eq:sin131223}
	\sin^{2}(\theta_{13}) &= (s_{12}^{e})^2 \sin^{2}\tilde{\theta}_{23}\\
	\label{eq:sin231223}
	\sin^{2}(\theta_{23}) &=
		\frac{ \sin^{2}\tilde{\theta}_{23}(c_{12}^{e})^{2}}
			{1-(s_{12}^{e})^{2}\sin^{2}\tilde{\theta}_{23}}=
		\frac{\sin^{2}\tilde{\theta}_{23}(c_{12}^{e})^{2}}
			{1-\sin^{2}(\theta_{13})} ,\\
	\label{eq:sin121223}
	\sin^{2}(\theta_{12}) &= \frac{\Omega_{12}}{1-\sin^{2}(\theta_{13})},
\end{align}
in which $\Omega_{12}$ has been defined to be
\begin{equation}
\begin{split}
\Omega_{12}={}&
(c_{12}^es_{12}^\nu)^2+(c_{12}^\nu s_{12}^e)^2(
	(c_{23}^e c_{23}^\nu)^2+(s_{23}^e c_{23}^\nu)^2+\cos(\delta^e_{23})s^{\prime e}_{23}s_{23}^\nu c_{23}^\nu)\\
	& - s^{\prime e}_{12} c_{12}^\nu s_{12}^\nu (c_{23}^e c_{23}^\nu \cos(\delta^e_{12})+s_{23}^e s_{23}^\nu \cos(\delta^e_{12}+\delta^e_{23})).
\end{split}	
\end{equation}
Here we note that as before, Eqs.~(\ref{eq:cddouble1223form1}) and (\ref{eq:cddouble1223form2}) follow from Eq.~\eqref{eq:sumruleorig}, once the angle relations of Eqs.~(\ref{eq:sin131223}) and (\ref{eq:sin231223}) are incorporated. 


\subsection{The case of two rotations in the 1--3 and 2--3 sectors}
\label{case1323}

For this case, $U_{e}=U_{23}^{e}(\theta_{23}^{e},\delta_{23}^{e})
U_{13}^{e}(\theta_{13}^{e},\delta_{13}^{e})$. As before, we begin with the
sum rule that has no phase dependence:
\begin{equation}
	\vert U_{e3}\vert / \vert U_{\tau 3}\vert: t^e_{13} = t_{13} sc_{23}.
\end{equation}
We next present both the $\cos\delta$- and $\cos (\delta^e_{23})$-dependent sum rules, which are as follows:
\begin{equation}\label{eq:cosd1323}
\begin{aligned}
& \vert U_{e 3} \vert / \vert U_{\mu 1} \vert: 2 \cos \delta =
- {s_{13} t_{23}}{ct_{12}} - {t_{12}}{cs_{13} ct_{23}} +
{s_{13} (s^\nu_{12} cs^e_{13})^{2}}{sc_{12} sc_{23} cs_{12} cs_{23}},\\
& \vert U_{e 3} \vert / \vert U_{\mu 2} \vert: 2 \cos \delta = s_{13}
t_{12}t_{23} + cs_{13} ct_{12} ct_{23} - (c^\nu_{12} cs^e_{13})^{2}
s_{13} sc_{12} sc_{23} cs_{12} cs_{23},\\
& \vert U_{\mu 1} \vert / \vert U_{\mu 2} \vert: 2 \cos \delta= (c_{12}^{\nu})^{2}\Big[ct_{23}cs_{13}\big(ct_{12}(t_{12}^{\nu})^{2}-t_{12}\big)+ct_{12}s_{13}t_{23}\big(t_{12}^{2}(t_{12}^{\nu})^{2}-1\big) \Big], \\
&\vert U_{\mu 1} \vert/ \vert U_{\tau 3} \vert: 2 \cos \delta=- {s_{13}
t_{23}}{ct_{12}} - {t_{12}}{cs_{13} ct_{23}} + (c_{13}
s^\nu_{12}sc^e_{13})^{2}sc_{12}  cs_{12} cs_{13} ct_{23},\\
&\vert U_{\mu 2} \vert / \vert U_{\tau 3} \vert: 2 \cos \delta =s_{13} t_{12}
	t_{23} + s_{13} ct_{12} ct_{23} - (c_{13} c^\nu_{12} sc^e_{13})^{2} sc_{12}
	cs_{12} cs_{13} ct_{23},
\end{aligned}
\end{equation}
and
\begin{equation}
\begin{aligned}
\vert U_{23}\vert / \vert U_{33}\vert: 2 \cos(\delta_{23}^{e}) =
	{}&\frac{ct_{23}^{\nu}(sc_{13}^{e})^{2}t_{23}^{e}
		-t_{23}^{2}t_{23}^{e}t_{23}^{\nu}
		+ct_{23}^{e} \big((sc_{13}^{e})^{2}t_{23}^{\nu}
			-ct_{23}^{\nu}t_{23}^{2}\big)}
		{(sc_{13}^{e})^{2}+t_{23}^{2}} ,\\
\vert U_{13}\vert / \vert U_{23}\vert: 2 \cos(\delta_{23}^{e}) = {}&\frac{
	s^2_{13}sc^e_{23} sc^\nu_{23} cs^e_{23} cs^\nu_{23}}{\left[s_{13}^{2} +
		(c_{13} s_{23} s^e_{13})^{2}\right]}-{\left[
		\frac{c^e_{23} c^\nu_{23}}{s^e_{23} s^\nu_{23}}
		+ \frac{s^e_{23}s^\nu_{23}}{c^e_{23} c^\nu_{23}}\right]}.
\end{aligned}
\end{equation} 
Note that the first relation in Eq.~(\ref{eq:cosd1323}) is equivalent to Eq.~(\ref{eq:sumruleorig13}), as previously found in \cite{sum6}.  Incorporating all constraints, we thus obtain further equivalent relations for $\cos\delta$ as functions of two of the three observed mixing angles, as follows:
\begin{equation}\label{eq:cddouble1323form1}
\cos\delta=\frac{\vert s_{23}\vert }{s_{23}c_{23}s^\prime_{12}\sqrt{s_{23}^2-\sin^2(\tilde{\theta}_{23})}}\left [(s^\nu_{12})^2(1-\sin^2(\tilde{\theta}_{23}))-(s_{12}^2c_{23}^2+c_{12}^2s_{23}^2-c_{12}^2\sin^2(\tilde{\theta}_{23}))\right ],
\end{equation}
and
\begin{equation}\label{eq:cddouble1323form2}
\cos\delta = \frac{(s^\nu_{12})^2c_{13}^2 (1-\sin^2(\tilde{\theta}_{23}))-s_{12}^2c_{13}^2+(s_{12}^2-s_{13}^2c_{12}^2)\sin^2(\tilde{\theta}_{23}))}{s^\prime_{12}s_{13} \sqrt{\sin^2(\tilde{\theta}_{23})} \sqrt{c_{13}^2 -\sin^2(\tilde{\theta}_{23})}},  
\end{equation}
in which $\sin^2(\tilde{\theta}_{23})$ is given in Eq.~(\ref{eq:sintilde}).  Here again we see the similarities between Eqs.~(\ref{eq:cddouble1323form1})--(\ref{eq:cddouble1323form2}) and their counterparts for $1-3$ perturbations in Eqs.~(\ref{eq:cosdelta13v1})--(\ref{eq:cosdelta13v2}).

The relations for the mixing angles that also follow from the full set of sum rules are
\begin{align}
         \label{eq:sin131323}
	\sin^{2}(\theta_{13}) &=(s_{13}^{e})^{2}\cos^{2}\tilde{\theta}_{23},\\
	\label{eq:sin231323}
	\sin^{2}(\theta_{23})
	&=\frac{\sin^{2}\tilde{\theta}_{23}}{1-\sin^{2}(\theta_{13})},\\
	\label{eq:sin121323}
	\sin^{2}(\theta_{12}) &=\frac{\Theta_{12}}{1-\sin^{2}(\theta_{13})},
\end{align}
in which
\begin{equation}
\begin{split}
	\Theta_{12} = {}&
	(c_{13}^{e})^{2}(s_{12}^{\nu})^{2}+
	(c_{12}^{\nu})^{2}(s_{13}^{e})^{2} \left[
		(c_{23}^{\nu})^{2}(s_{23}^{e})^{2}
		-\frac{1}{2}\cos \delta_{23}^{e} s_{23}^{\prime e}s_{23}^{\prime \nu}
		+(c_{23}^{e})^{2}(s_{23}^{\nu})^{2}
	\right]\\
	& -2c_{12}^{\nu}c_{13}^{e}s_{12}^{\nu}s_{13}^{e}\left[
		c_{23}^{\nu}\sin \delta_{13}^{e}\sin \delta_{23}^{e}s_{23}^{e}+
		\cos\delta_{13}^{e} (
			\cos \delta_{23}^{e}c_{23}^{\nu}s_{23}^{e}
			-c_{23}^{e}s_{23}^{\nu}) 
	\right].
\end{split}
\end{equation}
It can be clearly seen that Eqs.~(\ref{eq:cddouble1323form1}) and (\ref{eq:cddouble1323form2}) follow from  Eq.~(\ref{eq:sumruleorig13}) together with Eqs.~(\ref{eq:sin131323}) and (\ref{eq:sin231323}), as expected.


\subsection{The case of two rotations in the 1--2 and 1--3 sectors} \label{case1213}

The case of $U_{e} = U_{13}^{e}(\theta_{13}^{e}, \delta_{13}^{e})
U_{12}^{e}(\theta_{12}^{e}, \delta_{12}^{e})$ is different than the previous cases in that there are no sum rules that depend only on $\cos\delta$ and no other phase parameters. Instead, the sum rules fall into the following categories: those that involve $\cos\delta$ and $\cos(\delta^e_{13})$, those that involve $\cos(\delta^e_{12}-\delta^e_{13})$, those that individually constrain $\cos(\delta^e_{12})$ and $\cos(\delta^e_{13})$,  and those that involve all three phases. 
Two of the three sum rules that depend on $\delta$ and $\delta^e_{13}$ are as follows:
\begin{equation}
\begin{aligned}
\vert U_{\tau 1} \vert / \vert U_{\tau 3} \vert: 2 \cos \delta = {}&
\frac{cs_{12} cs_{13} cs_{23}}{c_{12} c_{23} (c^e_{13} c^\nu_{23})^{2}}
\big\{- c_{13}^{2} c_{23}^{2} \big[(c^e_{13} s^\nu_{12} s^\nu_{23})^{2} +
	2 c^e_{13} c^\nu_{12} s^e_{13} s^\nu_{12} s^\nu_{23}\cos(\delta^e_{13}) \\
& + (c^\nu_{12} s^e_{13})^{2}\big] + (c^e_{13} c^\nu_{23})^{2}
	\left(c_{12}^{2} c_{23}^{2} s_{13}^{2} + s_{12}^{2}
s_{23}^{2}\right)\big\},\\
\vert U_{\tau 2} \vert / \vert U_{\tau 3} \vert: 2 \cos \delta = {}&
\frac{cs_{12} cs_{13} cs_{23}}{c_{12} c_{23} (c^e_{13} c^\nu_{23})^{2}}
\big\{c_{13}^{2} c_{23}^{2} \big[(c^e_{13} c^\nu_{12} s^\nu_{23})^{2} - 2
c^e_{13} c^\nu_{12} s^e_{13} s^\nu_{12} s^\nu_{23}\cos(\delta^e_{13})\\
&+ (s^e_{13} s^\nu_{12})^{2}\big] - (c^e_{13} c^\nu_{23})^{2} \left(c_{12}^{2}
s_{23}^{2} + c_{23}^{2} s_{12}^{2} s_{13}^{2}\right)\big\}.
\end{aligned}
\end{equation}
We do not present the (somewhat cumbersome) third sum rule of this type, which can be obtained from $\vert U_{\tau 1}/U_{\tau 2}\vert$. The set of sum rules that involve just $\cos(\delta^e_{12}-\delta^e_{13})$ is as follows:
\begin{equation}
\begin{aligned}
\vert U_{e 3} \vert / \vert U_{\mu 3} \vert: 2 \cos(\delta^e_{12}-\delta^e_{13}) = {}&
	\frac{sc^e_{12} cs^e_{12} cs^e_{13}}{c^\nu_{23} s^\nu_{23} \left(c_{13}^{2}
			s_{23}^{2} + s_{13}^{2}\right)} \big\{- c_{13}^{2}
			s_{23}^{2}\big[\left(c^e_{12} c^\nu_{23}
			s^e_{13}\right)^{2}+ \left(s^e_{12} s^\nu_{23}\right)^{2}\big]\\
 & + s_{13}^{2}\big[\left(c^\nu_{23} s^e_{12} s^e_{13}\right)^{2}+\left(c^e_{12}
s^\nu_{23}\right)^{2}\big]\big\},\\
\vert U_{e 3} \vert / \vert U_{\tau 3} \vert: 2 \cos(\delta^e_{12}-\delta^e_{13}) = {}&
	{ct^e_{13}c^e_{13} ct^\nu_{23} t_{13}^{2}sc_{23}^{2} sc^e_{12} cs^e_{12}}
	- {ct^e_{12} ct^\nu_{23} s^e_{13}} - {t^e_{12} t^\nu_{23} cs^e_{13}},\\
\vert U_{\mu 3} \vert / \vert U_{\tau 3} \vert: 2 \cos(\delta^e_{12}-\delta^e_{13}) = {}&
	{ct^e_{12} t^\nu_{23}cs^e_{13}} + {ct^\nu_{23} t^e_{12} s^e_{13}} - { ct^e_{13}
	c^e_{13} ct^\nu_{23} t_{23}^{2}sc^e_{12} cs^e_{12}}.\\
\end{aligned}
\end{equation}
The remaining sum rules, which either relate the individual phases $\delta^e_{12}$ and $\delta^e_{13}$, or relate these phases together with $\delta$, are not presented here for simplicity. Including these relations results in the sum rule for $\cos\delta$ in terms of $\theta^\nu_{12}$, $\theta^\nu_{23}$, $\delta^e_{13}$, and the measured lepton mixing angles, as follows:
\begin{equation}\label{eq:cosdelformula1213}
\begin{aligned}
	\cos\delta = {}&
	\frac{2}{s'_{12} s_{13} s'_{23}} \big\{ (c^2_{13}c^2_{23}+1)(s^2_{12}-(c^\nu_{12})^2)+(2c^2_{12}-1)c^2_{23}\\
	&+(2(c^\nu_{12})^2-1)(c_{13}c_{23}sc^\nu_{23})^2
	-2\cos(\delta^e_{13})c^\nu_{12}s^\nu_{12}t^\nu_{23}\sqrt{1-(c_{13}c_{23}sc^\nu_{23})^2}\vert c_{13}c_{23}\vert \big \},
\end{aligned}
\end{equation}
as well as the following relations for the lepton mixing angles in this scenario:
\begin{equation}\label{eq:sin131213}
\sin^{2}(\theta_{13})=(s_{12}^{e})^{2}(s_{23}^{\nu})^{2}+c_{12}^{e}s_{13}^{e}\left[c_{12}^{e}(c_{23}^{\nu})^{2}s_{13}^{e}+\cos(\delta_{12}^{e}-\delta_{13}^{e})s_{12}^{e}s^{\prime \nu}_{23}\right],
\end{equation}
\begin{equation}\label{eq:sin231213}
\sin^{2}(\theta_{23})=\frac{(c_{23}^{\nu})^{2}(s_{12}^{e})^{2}(s_{13}^{e})^{2}-\cos(\delta_{12}^{e}-\delta_{13}^{e})c_{23}^{\nu}s_{12}^{\prime e}s_{13}^{e}s_{23}^{\nu}+(c_{12}^{e})^{2}(s_{23}^{\nu})^{2}}{1-\sin^{2}\theta_{13}},
\end{equation}
\begin{equation}\label{eq:sin121213}
\sin^{2}(\theta_{12})=\frac{\Xi_{12}}{1-\sin^{2}\theta_{13}},
\end{equation}
in which $\Xi_{12}$ is given by
\begin{equation}
\begin{split}
	\Xi_{12} = {}& (c_{12}^{e})^{2}(c_{13}^{e})^{2}(s_{12}^{\nu})^{2}+
		2c_{12}^{e}c_{12}^{\nu}c_{13}^{e}s_{12}^{\nu} \big[
			\cos\delta_{13}^{e}c_{12}^{e}s_{13}^{e}s_{23}^{\nu}
			-\cos\delta_{12}^{e}c_{23}^{\nu}s_{12}^{e} 
		\big]\\
		& +(c_{12}^{\nu})^{2}\left[
			(c_{23}^{\nu})^{2}(s_{12}^{e})^{2}
			-\cos(\delta_{12}^{e}-\delta_{13}^{e})
			 c_{23}^{\nu}s_{12}^{\prime e}s_{13}^{e}s_{23}^{\nu}
			+(c_{12}^{e})^{2}(s_{13}^{e})^{2}(s_{23}^{\nu})^{2} \right].
\end{split}
\end{equation}
We can of course re-express Eq.~(\ref{eq:cosdelformula1213}) by incorporating either some or all of the angle relations as given in  Eqs.~(\ref{eq:sin131213})--(\ref{eq:sin121213}), as in previous sections. However, this necessarily reintroduces additional model parameters, unlike in the previous cases.


\section{Numerical analysis of sum rules} \label{sec:NA}


In this section, we present a numerical analysis of the predictions for $\cos
\delta$ and $\sin \delta$ as functions of the model parameters
$\theta_{ij}^{e}$ and $\delta_{ij}^{e}$ with $ij=\{12,13,23\}$ for scenarios
with TBM, BM, HEX, GR1, and GR2 mixing angles in the mixing matrix of the neutrino sector, $U_\nu$.  We recall that each of these scenarios has been taken to have $\theta^\nu_{23}=\pi/4$. The values for the parameter of interest $y=\sin^2(\theta^\nu_{12})\equiv (s^\nu_{12})^2$ are given in Table~\ref{tab:2}; they are $0.25$ (HEX), $0.28$ (GR1), $0.33$ (TBM), $0.35$ (GR2), and $0.5$ (BM).  
\begin{table}[H]
\centering
\begin{tabular}{|c|c|c|c|c|c|}
\hline 
 & BM & TBM  & HEX & GR1 & GR2 \\
\hline 
$y=(s^\nu_{12})^2$ & $1/2$ & $1/3$ & $1/4$ & $(5-\sqrt{5})/10$ & $ (5-\sqrt{5})/8$ \\
\hline 
\end{tabular}
\caption{The values of $\sin^2(\theta^\nu_{12})\equiv (s^\nu_{12})^2$ for the theoretical scenarios under consideration.  All scenarios have been taken to have $\theta^\nu_{23}=\pi/4$.}
\label{tab:2}
\end{table}

In Section~\ref{subsec:easyr}, we consider single rotations in $U_e$ that result in nontrivial predictions for $\theta_{13}$, while in
Section~\ref{subsec:harderr}, we consider the cases with two rotations in
$U_e$, as previously discussed.   Here we note that numerical
analyses of these sum rules have appeared in the
literature~\cite{atm5,sum1,sum2a,sum2,sum3,sum4,sum5,sum6,sum7,sum8,sum9,sum10}. The focus in this work is on the allowed parameter space of the model parameters for each scenario of interest, in light of updated global fits.

In this analysis we use the allowed mixing parameter ranges at 3$\sigma$ as reported in the most recent global fit of \cite{c1} and summarized in Table~\ref{tab:1}.  As the sum rules for $\cos\delta$ do not explicitly depend on the mass ordering, the only differences between the normal ordering (NO) and inverted ordering (IO) cases arise due to the slight differences in the global fit values for the measured lepton mixing angles, as seen in Table~\ref{tab:1}. Hence, here we will restrict ourselves just to the NO values and not the IO values, as the differences between the global fit data in the two cases are at sub-percent levels.


\subsection{Single rotations}
\label{subsec:easyr}
Here we will begin with the cases that have a single rotation in $U_e$, and show the dependence of $\cos\delta$ and $\sin\delta$ on the model parameters, using a colorscale to denote parameter values.  It will immediately be evident that the patterns for $\cos\delta$ and $\sin\delta$ repeat as a function of model parameters.  This results from the intrinsic degeneracy for $\delta_{ij}^{\prime e}\rightarrow \delta_{ij}^{e}-\pi$ and
$\theta_{ij}^{\prime e}\rightarrow \theta_{ij}^{e}-\pi /2$, which provide identical predictions up to unobservable sign flips.\\


\subsubsection*{Case 1: One rotation in the $1-2$ sector ($U_{e}=U^{e}_{12}(\theta_{12}^{e},\delta_{12}^{e})$)} 

For the case in which $U_e$ is given by a single rotation of the form $U_{e}=U^{e}_{12}(\theta_{12}^{e},\delta_{12}^{e})$, we will begin by writing the mixing angle conditions as given in Eq.~(\ref{eq:sij2}) for the models considered here, which all have $\theta^\nu_{23}=\pi/4$.  Eq.~(\ref{eq:sij2}) already indicates that the reactor and atmospheric mixing angles are in fact independent of $\theta^\nu_{12}$ and $\delta^e_{12}$, and as such are precisely correlated with each other. The reactor mixing angle takes the very simple form
\begin{equation}\label{eq:reactor12}
\sin^2(\theta_{13})\equiv s^2_{13}= (s^e_{12})^2/2.
\end{equation}
Hence, we see that fixing the reactor angle to within its quite precisely measured $3\sigma$ range fixes the model parameter $\theta^e_{12}$ to a strict range,  $0.0378\leq (s^e_{12})^2\leq 0.0478$, which shows that $\theta^e_{12}$ is roughly of the order of the Cabibbo angle $\theta_c\simeq 0.22$.  This also clearly indicates that these parameters can be traded for each other in the analysis.  The atmospheric mixing angle is given by
\begin{equation}\label{eq:atm12}
\sin^2(\theta_{23})\equiv s^2_{23}= \frac{(c^e_{12})^2/2}{(1-(s^e_{12})^2/2)} = \frac{\;\; 1-2s^2_{13}}{2(1-s^2_{13})}.
\end{equation}
The atmospheric mixing angle is thus precisely determined by the reactor mixing angle to fall into a much smaller range than what is allowed experimentally at $3\sigma$, as is also evident from Eq.~(\ref{eq:sij2}); it is further evident that $s^2_{23} <1/2$.  More precisely, the reactor angle bounds result in the range $0.4878 \leq s^2_{23}\leq 0.4904.$

The solar mixing angle, however, also depends on the model parameter $\delta^e_{12}$, as well as the parameter $y=(s^\nu_{12})^2$, which varies among the scenarios under consideration.  The expression for $\sin^2(\theta_{12})$ takes the form
\begin{equation}\label{eq:solar12}
\sin^2(\theta_{12}) \equiv s^2_{12}=\frac{(c^\nu_{12})^2(s^e_{12})^2/2+(c^e_{12})^2(s^\nu_{12})^2-\sqrt{2}c^e_{12}s^e_{12}c^\nu_{12}s^\nu_{12}\cos(\delta^e_{12})}{1-s^2_{13}},
\end{equation}
which can be expressed in term of $s^2_{13}$ and $y$ as follows:
\begin{equation}\label{eq:s12sq12_sec4}
s^2_{12}=\frac{y-3y s^2_{13}+s^2_{13}-2 \cos(\delta^e_{12}) \sqrt{(1-2s^2_{13})s^2_{13}}\sqrt{y(1-y)}}
{1-s^2_{13}}\equiv \frac{\tilde{\alpha}^{(12)}}{1-s^2_{13}},
\end{equation}
in which we have introduced the parameter $\tilde{\alpha}^{(12)}$ for later convenience. We see from Eq.~(\ref{eq:s12sq12_sec4}) that for a given theoretical scenario as given by a fixed value of $y$, there is a possible range of allowed values of $s^2_{12}$, which may or may not cover the full experimentally allowed range at $3\sigma$.  The predicted range of $\cos\delta$ also depends on these model parameters, and thus on the possible values of $s^2_{12}$, as well as the tight predictions of the atmospheric and reactor mixing angles that result for this general form of $U_{e}$.  For this set of models, it is straightforward to see from Eqs.~(\ref{eq:sumruleorig}), (\ref{eq:cdlistredcompact}), and (\ref{eq:cdlistred13}) that $\cos\delta$ takes the form
\begin{equation}\label{eq:cosdel12num}
\cos\delta = \frac{s^2_{12}+s^2_{13}-3 s^2_{12}s^2_{13}-y(1-s^2_{13})}
{2\sqrt{s^2_{12}(1-s^2_{12})}\sqrt{s^2_{13}(1-2s^2_{13})}}.
\end{equation}
 Eq.~(\ref{eq:cosdel12num}), taken together with Eq.~(\ref{eq:s12sq12_sec4}), can be used to express $\cos\delta$ as a function of $s^2_{13}$, $y$, and $\cos(\delta^e_{12})$, as follows:
 \begin{equation}\label{eq:cosdelta12f}
\cos\delta=\frac{\tilde{\alpha}^{(12)}(1-3s^2_{13})+s^2_{13}(1-s^2_{13})-y(1-s^2_{13})^2}{2\sqrt{s^2_{13}(1-2s^2_{13})}\sqrt{\tilde{\alpha}^{(12)}(1-s^2_{13}-\tilde{\alpha}^{(12)})}},
 \end{equation}  
 and we can obtain an analogous expression for $\sin\delta$ from this result, which takes the form
 \begin{equation}\label{eq:sindelta12f}
 \sin\delta =\pm \frac{\sin(\delta^e_{12})(1-s^2_{13})\sqrt{y(1-y)}}{\sqrt{\tilde{\alpha}^{(12)}(1-s^2_{13}-\tilde{\alpha}^{(12)})}},
 \end{equation}
where the sign is taken carefully according to the specified conventions. Note from Eq.~(\ref{eq:sindelta12f}) that as $\delta^e_{12}\rightarrow 0$, $\sin\delta \rightarrow 0$, as expected.
  In this quite predictive set of scenarios, therefore, we see that there is a strong correlation between the allowed values of $s^2_{12}$ and the predicted range for $\cos\delta$ and $\sin\delta$. 


	\begin{figure}[H]
		\begin{subfigure}[b]{0.475\textwidth}
			\caption{$\cos \delta$($\theta_{12}^e{},\delta_{12}^{e}$) }
			\includegraphics[width=\textwidth]{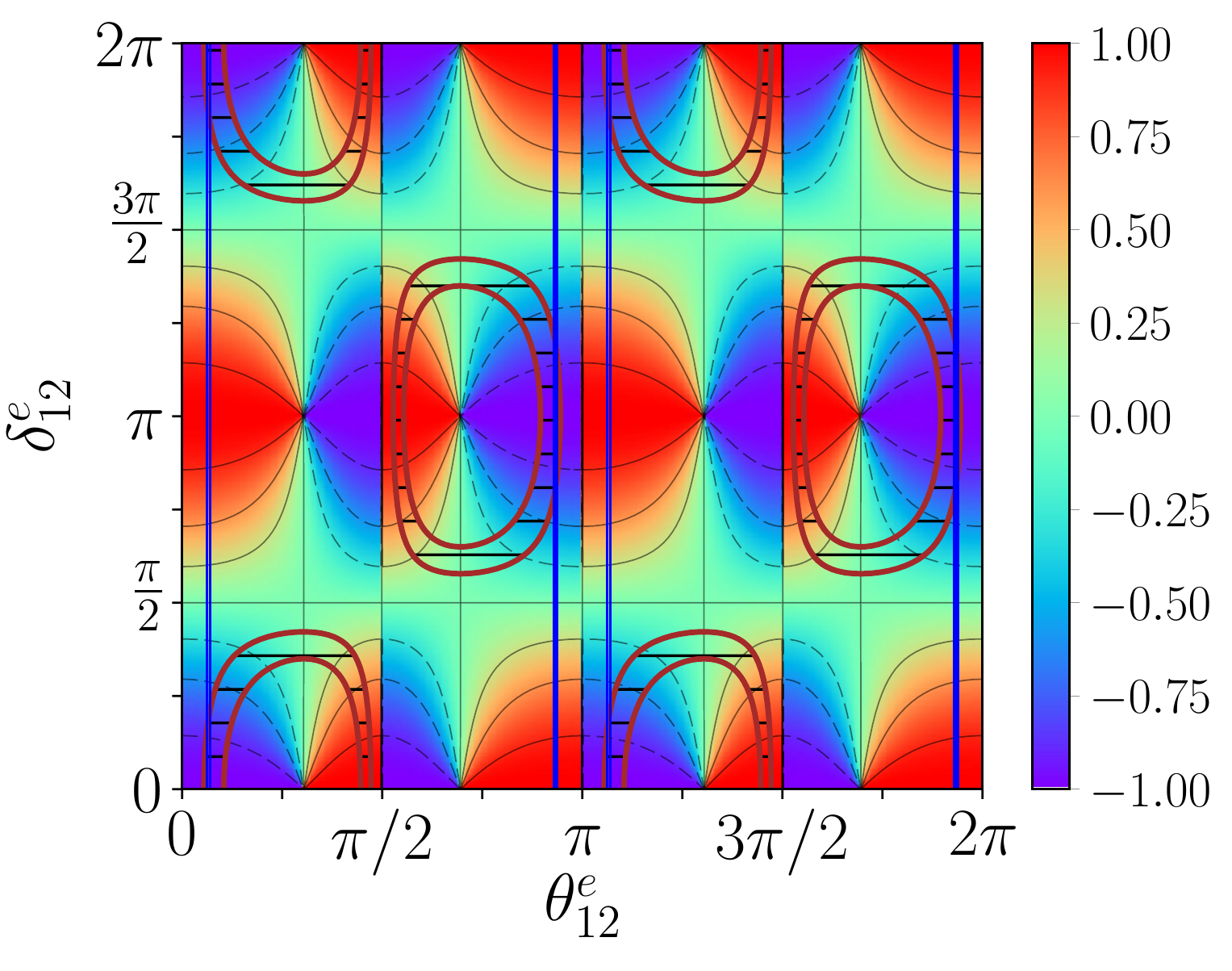}
		\end{subfigure}
		\hfill
		\begin{subfigure}[b]{0.475\textwidth}
			\caption{$\sin \delta$($\theta_{12}^e{},\delta_{12}^{e}$)}
			\includegraphics[width=\textwidth]{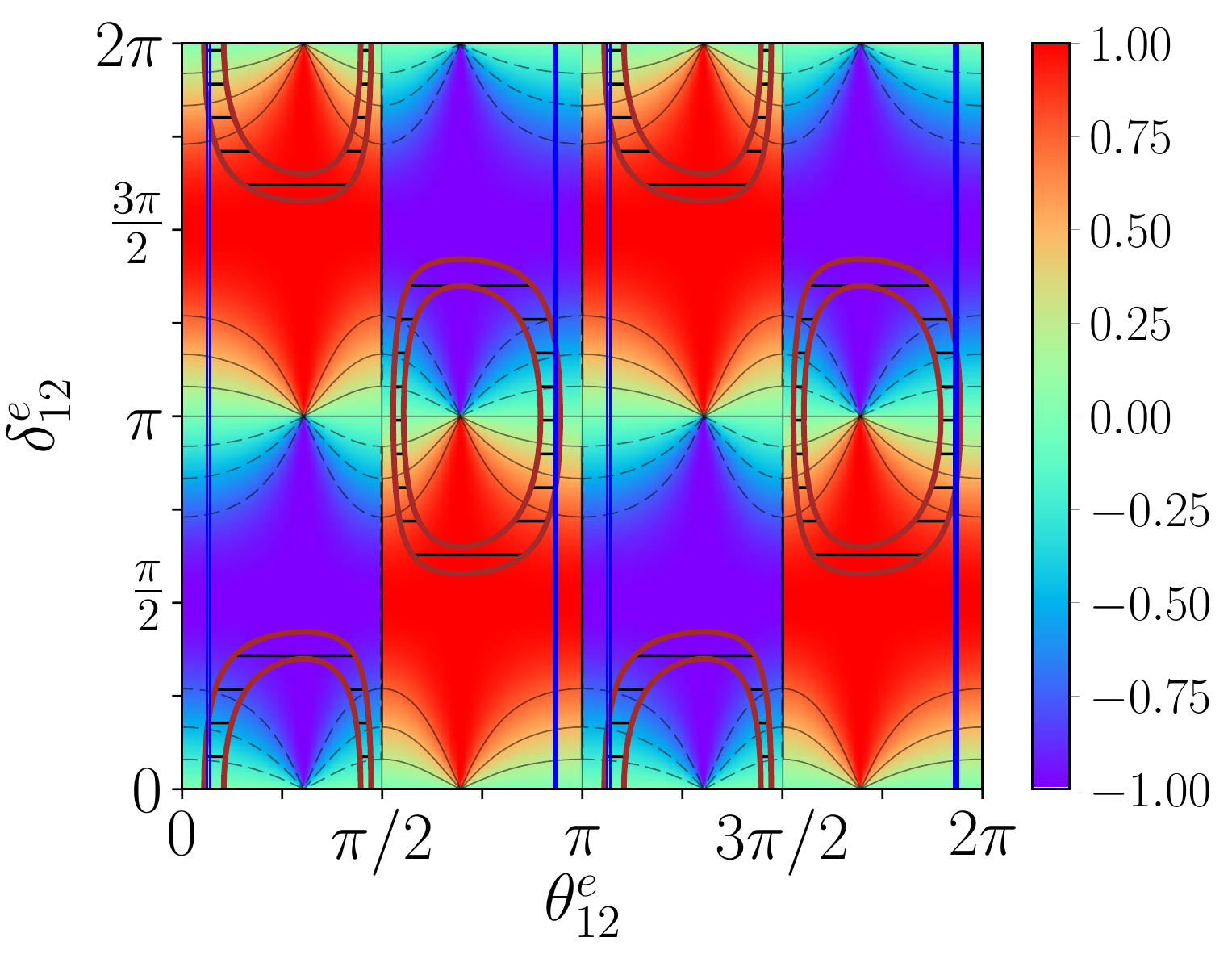}
		\end{subfigure}
		\caption{The predictions for $\cos \delta$ and $\sin \delta$ as a
		         function of $\theta_{12}^{e}$ and $\delta_{12}^{e}$ for BM
			 mixing. The blue band and the region between the dark-red
		 contours represent the regions allowed by $\sin^2(\theta_{13})$ and
	 $\sin^2(\theta_{12})$ at $3\sigma$, respectively.}
	 	 \label{fig:fullspace12bm}
	\end{figure}
We will begin our study of individual scenarios with the case of BM mixing, for which $y=1/2$.  As is well known in the literature, the experimental constraints on $s^2_{12}$ are quite stringent for BM mixing since the value of the solar angle in the absence of charged lepton corrections is maximal, and thus quite far from the experimentally allowed range.   This results in a very focused range of predictions for $\cos\delta$.  This can be seen in the full range of model parameter space for $\cos\delta$ and $\sin\delta$ as shown in Figure~\ref{fig:fullspace12bm}. 
	\begin{figure}[H]
		\begin{subfigure}[b]{0.475\textwidth}
			\caption{$\cos \delta$($\theta_{12}^e{},\delta_{12}^{e}$) }
			\includegraphics[width=\textwidth]{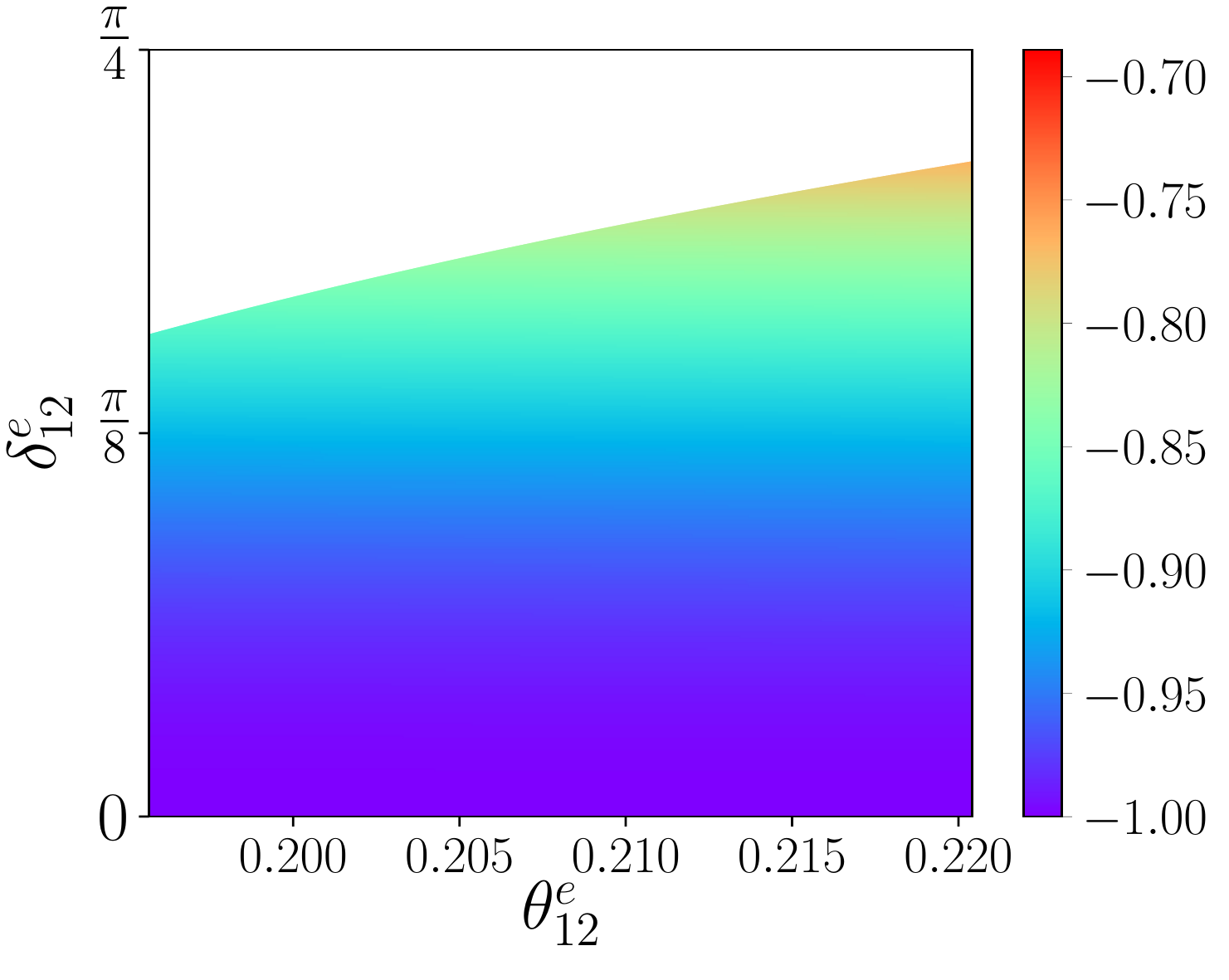}
		\end{subfigure}
		\hfill
		\begin{subfigure}[b]{0.475\textwidth}
			\caption{$\sin \delta$($\theta_{12}^e{},\delta_{12}^{e}$)}
			\includegraphics[width=\textwidth]{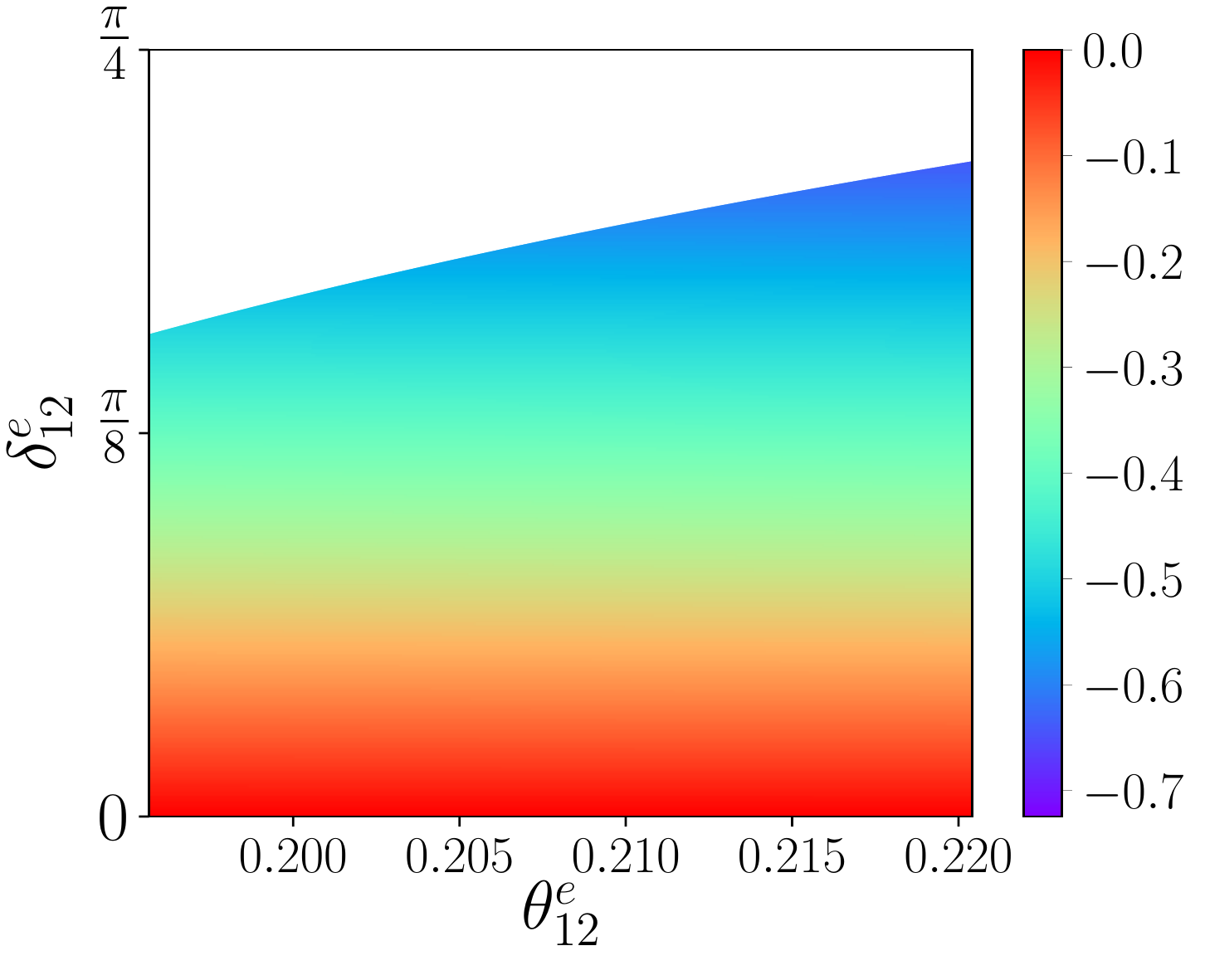}
		\end{subfigure}
		\caption{A close-up view of one of the allowed parameter regions in $\theta^e_{12}$ and $\delta^e_{12}$ and the predictions for $\cos\delta$ and $\sin\delta$ for BM mixing, for $U_e=U^e_{12}(\theta^e_{12},\delta^e_{12})$.}
			 \label{fig:zoomedin12bm}
	\end{figure}
	More precisely, we see that Figure~\ref{fig:fullspace12bm} shows the very limited regions of parameter space for $\theta^e_{12}$ where the reactor and solar mixing angle constraints overlap.  Upon taking a closer look at one of these regions in Figure~\ref{fig:zoomedin12bm}, we see the quite restrictive range of possible values of $\cos\delta$ in this scenario, which is characterized by a most probable value of $\cos\delta \sim -1$. 

	\begin{figure}[H]
		\begin{subfigure}[b]{0.475\textwidth}
  \caption{$\cos \delta$($\theta_{12}^e{},\delta_{12}^{e}$) }
    \includegraphics[width=\textwidth]{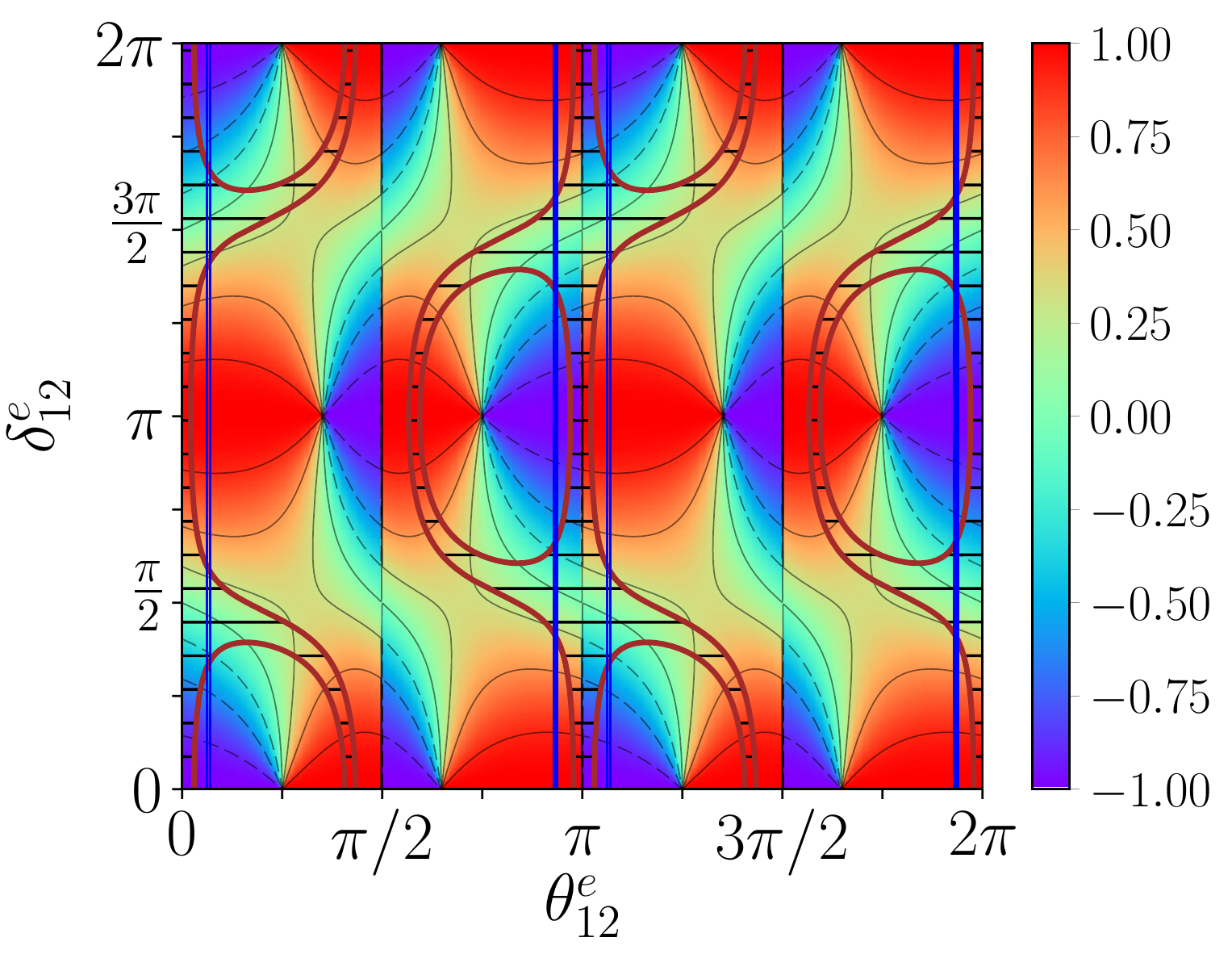}
  \end{subfigure}
  \hfill
  \begin{subfigure}[b]{0.475\textwidth}
  \caption{$\sin \delta$($\theta_{12}^e{},\delta_{12}^{e}$) }
    \includegraphics[width=\textwidth]{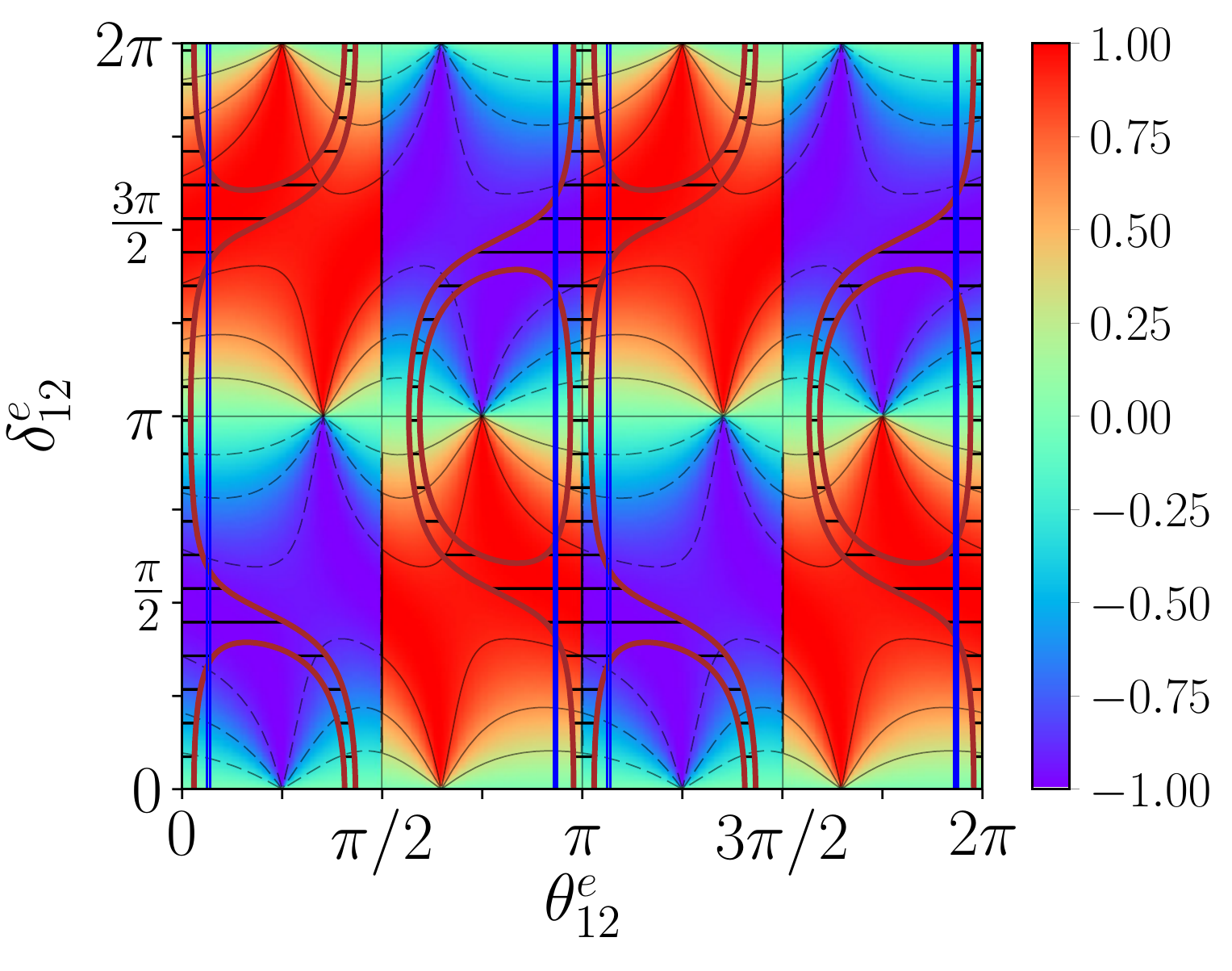}
  \end{subfigure}
  \caption{The predictions for $\cos \delta$ and $\sin \delta$ as a
  function of $\theta_{12}^{e}$ and $\delta_{12}^{e}$ for TBM mixing, in the case that $U_e=U^e_{12}(\theta^e_{12},\delta^e_{12})$. The blue band and the region between the dark-red
		 contours represent the regions allowed by $\sin^2(\theta_{13})$ and
	 $\sin^2(\theta_{12})$ at $3\sigma$, respectively.}
    \label{fig:fullspace12tbm}
\end{figure}
	\begin{figure}[H]
		\begin{subfigure}[b]{0.475\textwidth}
  \caption{$\cos \delta$($\theta_{12}^e{},\delta_{12}^{e}$) }
    \includegraphics[width=\textwidth]{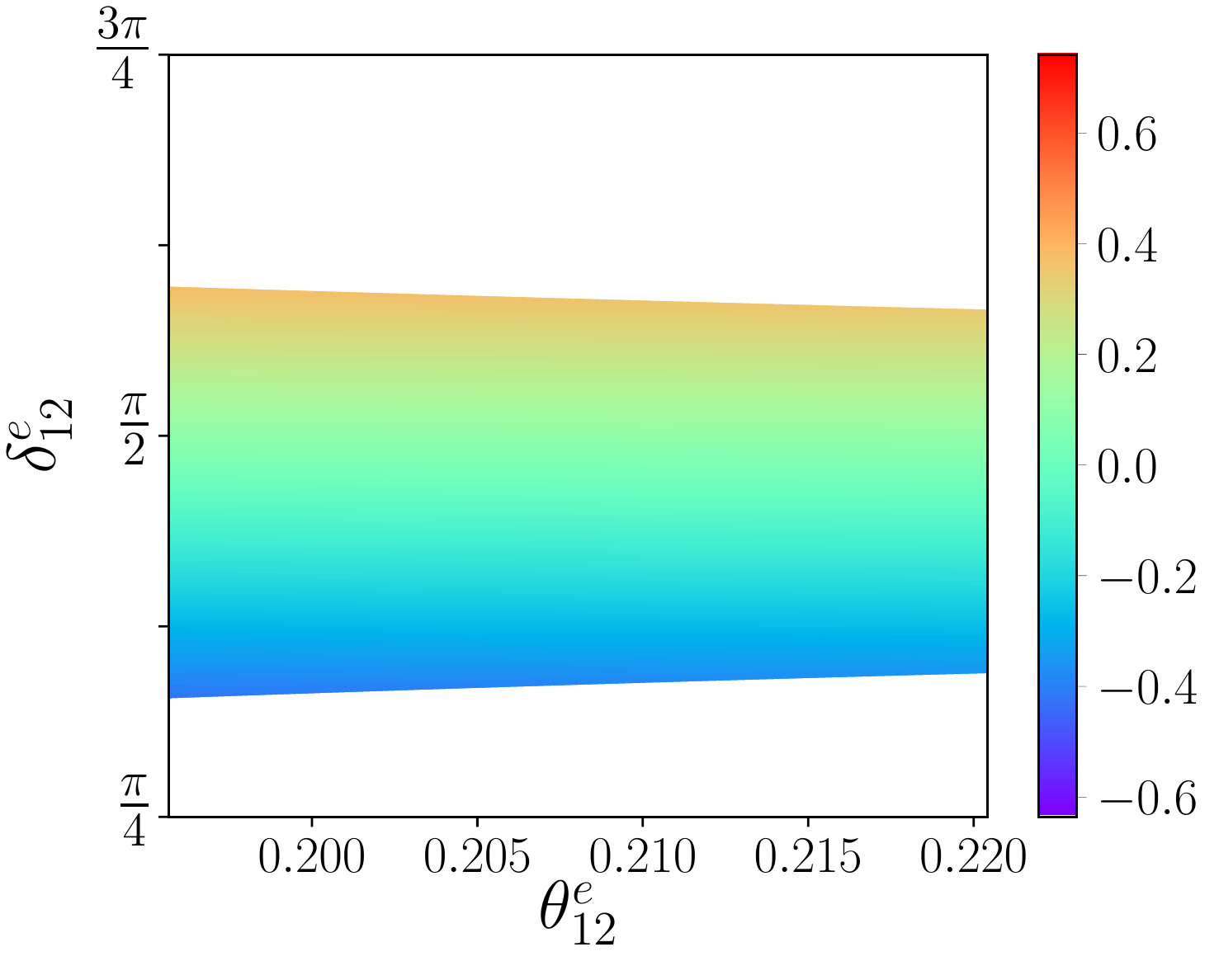}
  \end{subfigure}
  \hfill
  \begin{subfigure}[b]{0.475\textwidth}
  \caption{$\sin \delta$($\theta_{12}^e{},\delta_{12}^{e}$) }
    \includegraphics[width=\textwidth]{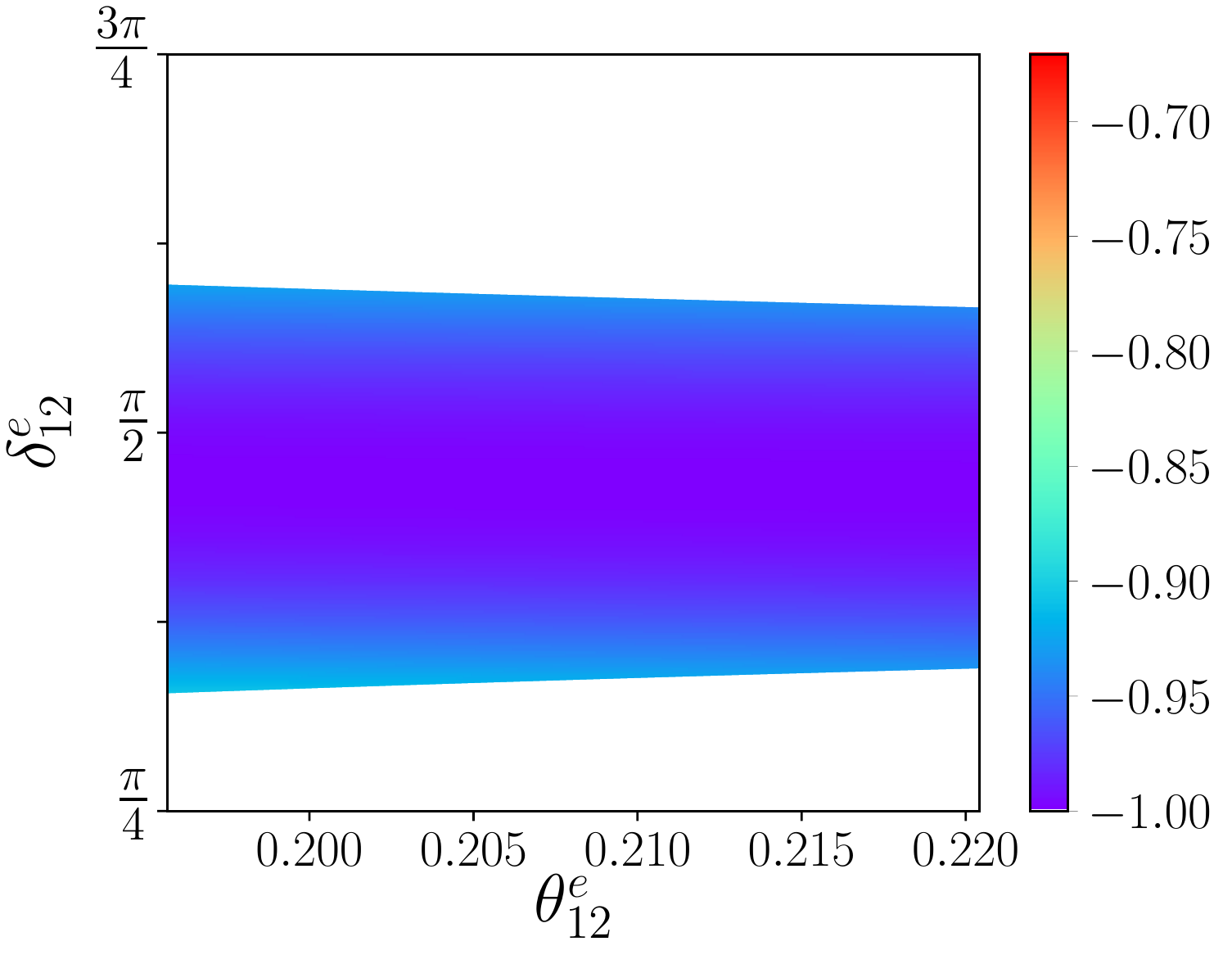}
  \end{subfigure}
		\caption{A close-up view of one of the allowed parameter regions in $\theta^e_{12}$ and $\delta^e_{12}$ and the predictions for $\cos\delta$ and $\sin\delta$ for TBM mixing, for $U_e=U^e_{12}(\theta^e_{12},\delta^e_{12})$.}
		 \label{fig:zoomedin12tbm}
\end{figure}

In the other scenarios, the smaller values of $y$ in each case allow for greater ease in obtaining $s^2_{12}$ in the allowed range, and hence a significantly wider region of parameter space for which the reactor and solar angle constraints can be satisfied simultaneously than what occurs in the BM mixing case. Using TBM mixing as a representative example, we show the full parameter space in Figure~\ref{fig:fullspace12tbm}, and a closer view of one of the allowed parameter regions in Figure~\ref{fig:zoomedin12tbm}. The preferred range for $\delta^e_{12}$ includes maximal values for the MNSP phase $\delta$.  An inspection of Figure~\ref{fig:fullspace12tbm} also shows that in the allowed parameter regions in which $\theta_{12}^e$ is roughly shifted to $\pi-\theta^e_{12}$, the predicted values of $\sin\delta$ change sign, accordingly.

The remaining scenarios of HEX mixing and GR1/GR2 are similar to that of TBM mixing. Both HEX and GR1 have values of $\theta^\nu_{12}$ that are smaller than that of TBM, while in the GR2 case $\theta^\nu_{12}$ is larger, resulting in characteristic allowed regions of the remaining model parameters.   Therefore, for simplicity for these cases we highlight here specific portions of the allowed parameter space, and their predictions for $\cos\delta$ and $\sin\delta$.  
\begin{figure}[H]
  \begin{subfigure}[b]{0.475\textwidth}
  \caption{$\cos \delta$($\theta_{12}^e{},\delta_{12}^{e}$) }
    \includegraphics[width=\textwidth]{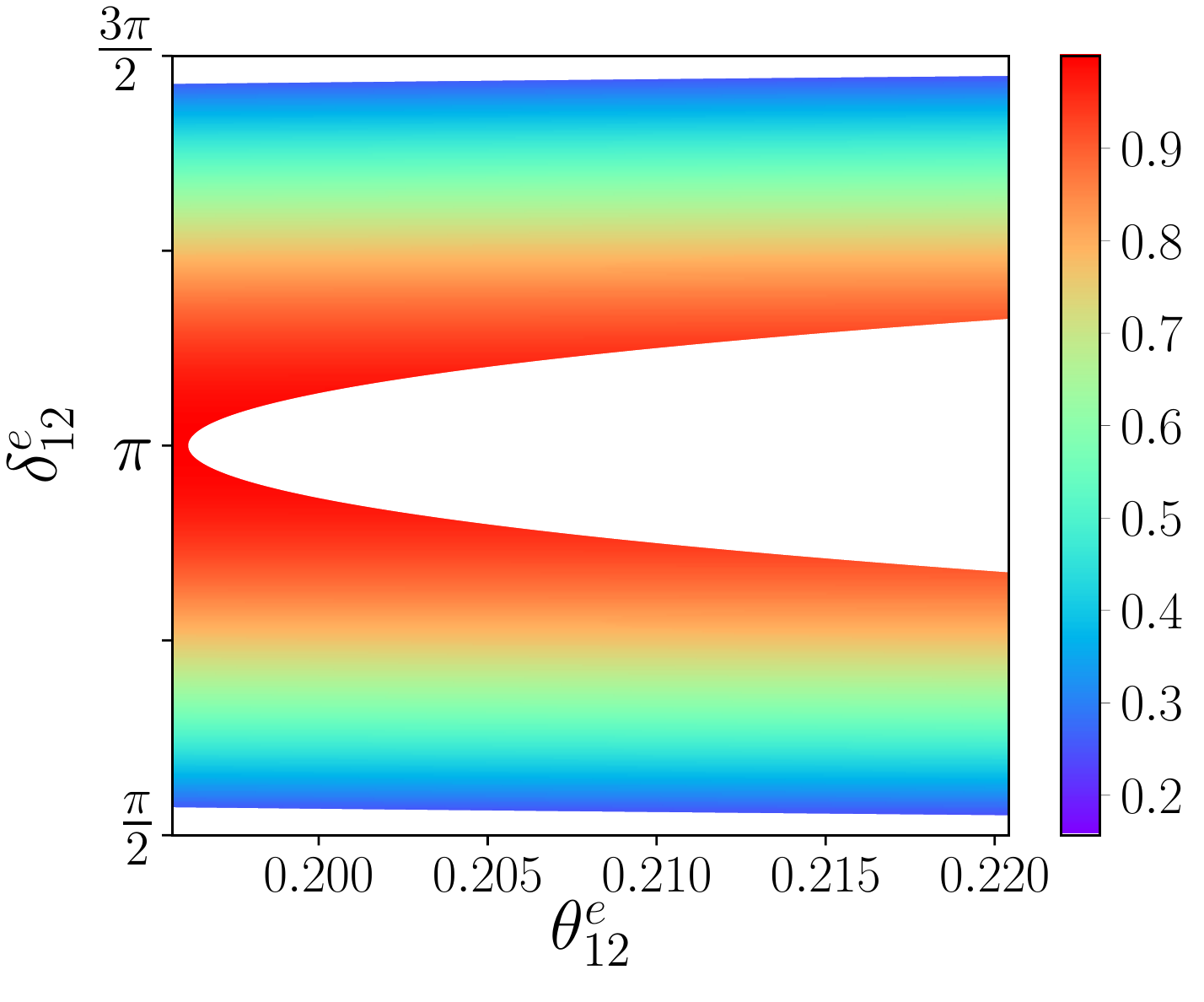}
  \end{subfigure}
  \hfill
  \begin{subfigure}[b]{0.475\textwidth}
  \caption{$\sin \delta$($\theta_{12}^e{},\delta_{12}^{e}$) }
    \includegraphics[width=\textwidth]{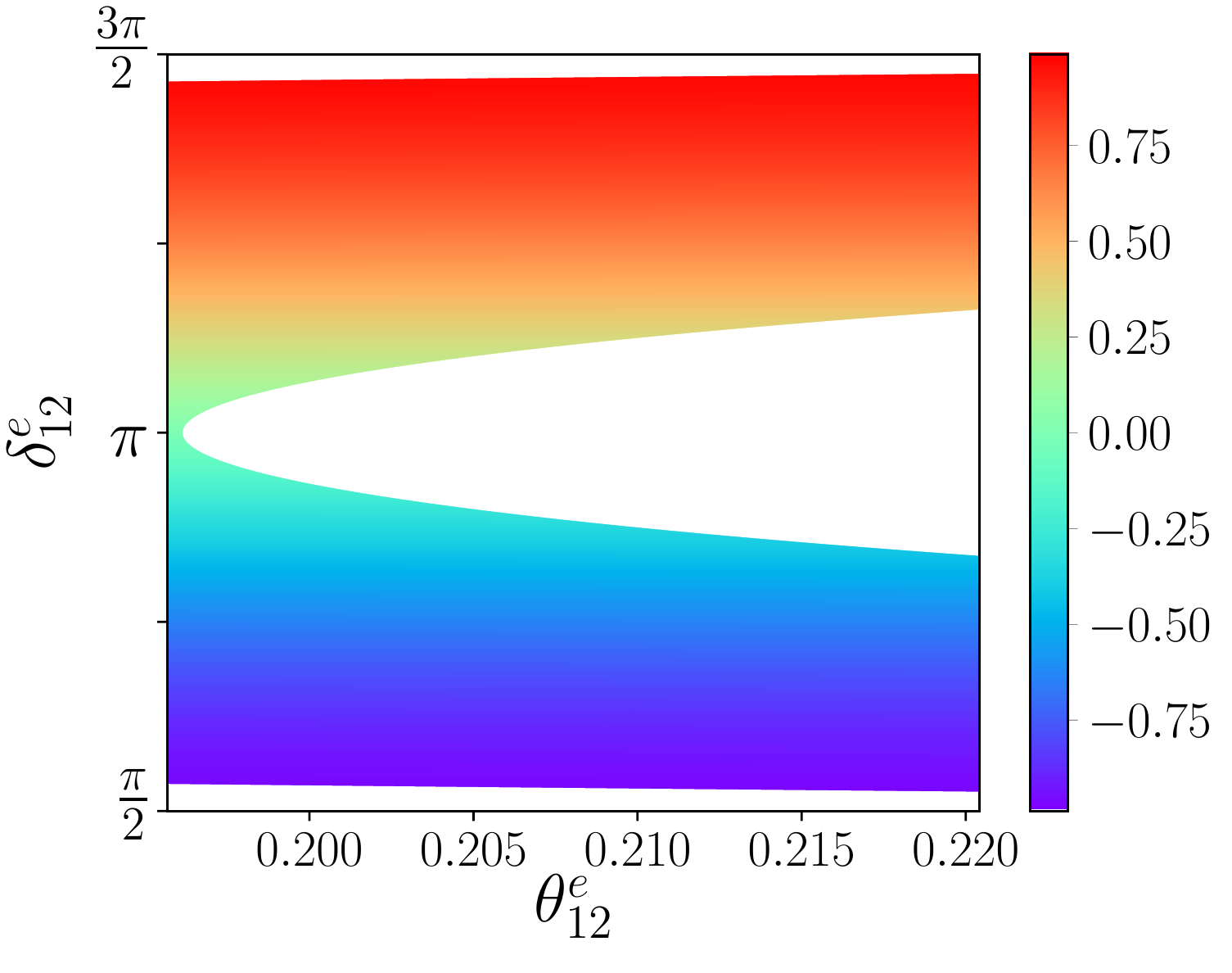}
  \end{subfigure}
		\caption{A close-up view of the allowed parameter region in $\theta^e_{12}$ and $\delta^e_{12}$ and the predictions for $\cos\delta$ and $\sin\delta$ for HEX mixing, for $U_e=U^e_{12}(\theta^e_{12},\delta^e_{12})$.}
		\label{fig:zoomedin12hex}
\end{figure}
\begin{figure}[H]
  \begin{subfigure}[b]{0.475\textwidth}
  \caption{$\cos \delta$($\theta_{12}^e{},\delta_{12}^{e}$) }
    \includegraphics[width=\textwidth]{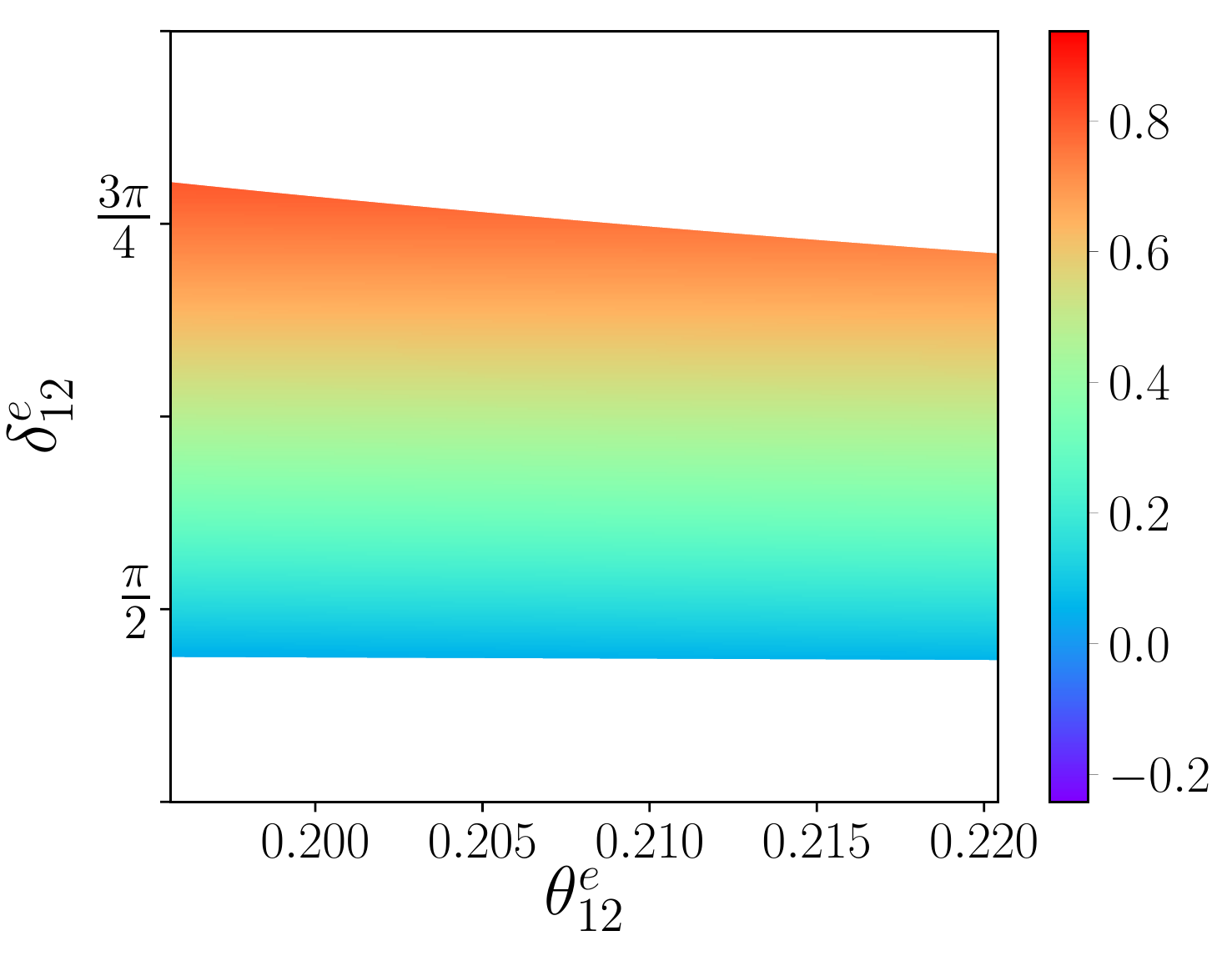}
  \end{subfigure}
  \hfill
  \begin{subfigure}[b]{0.475\textwidth}
  \caption{$\sin \delta$($\theta_{12}^e{},\delta_{12}^{e}$) }
    \includegraphics[width=\textwidth]{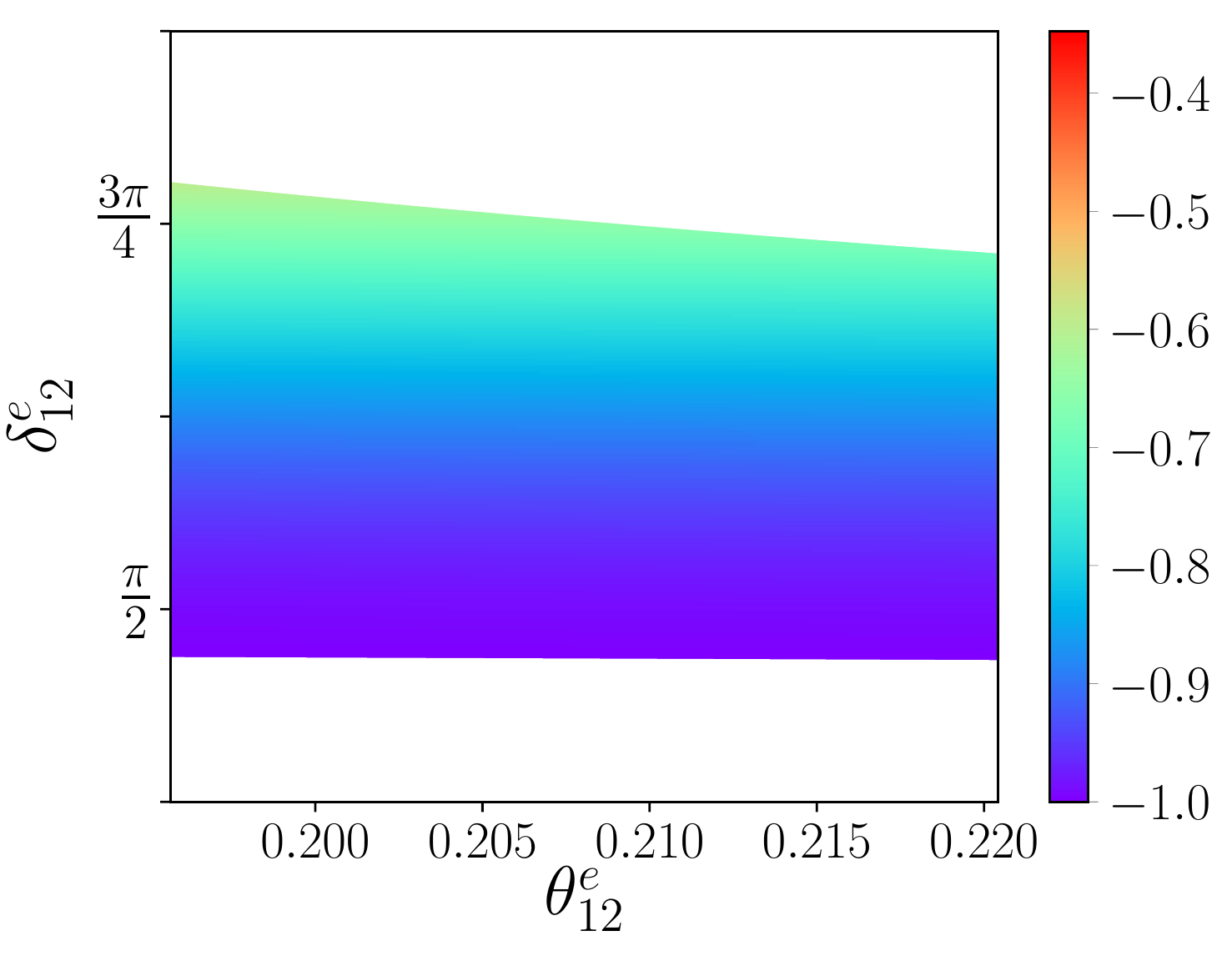}
  \end{subfigure}
		\caption{A close-up view of the allowed parameter region in $\theta^e_{12}$ and $\delta^e_{12}$ and the predictions for $\cos\delta$ and $\sin\delta$ for GR1 mixing, for $U_e=U^e_{12}(\theta^e_{12},\delta^e_{12})$.}
		\label{fig:zoomedin12gr1}
\end{figure}
\begin{figure}[H]
  \begin{subfigure}[b]{0.475\textwidth}
  \caption{$\cos \delta$($\theta_{12}^e{},\delta_{12}^{e}$) }
    \includegraphics[width=\textwidth]{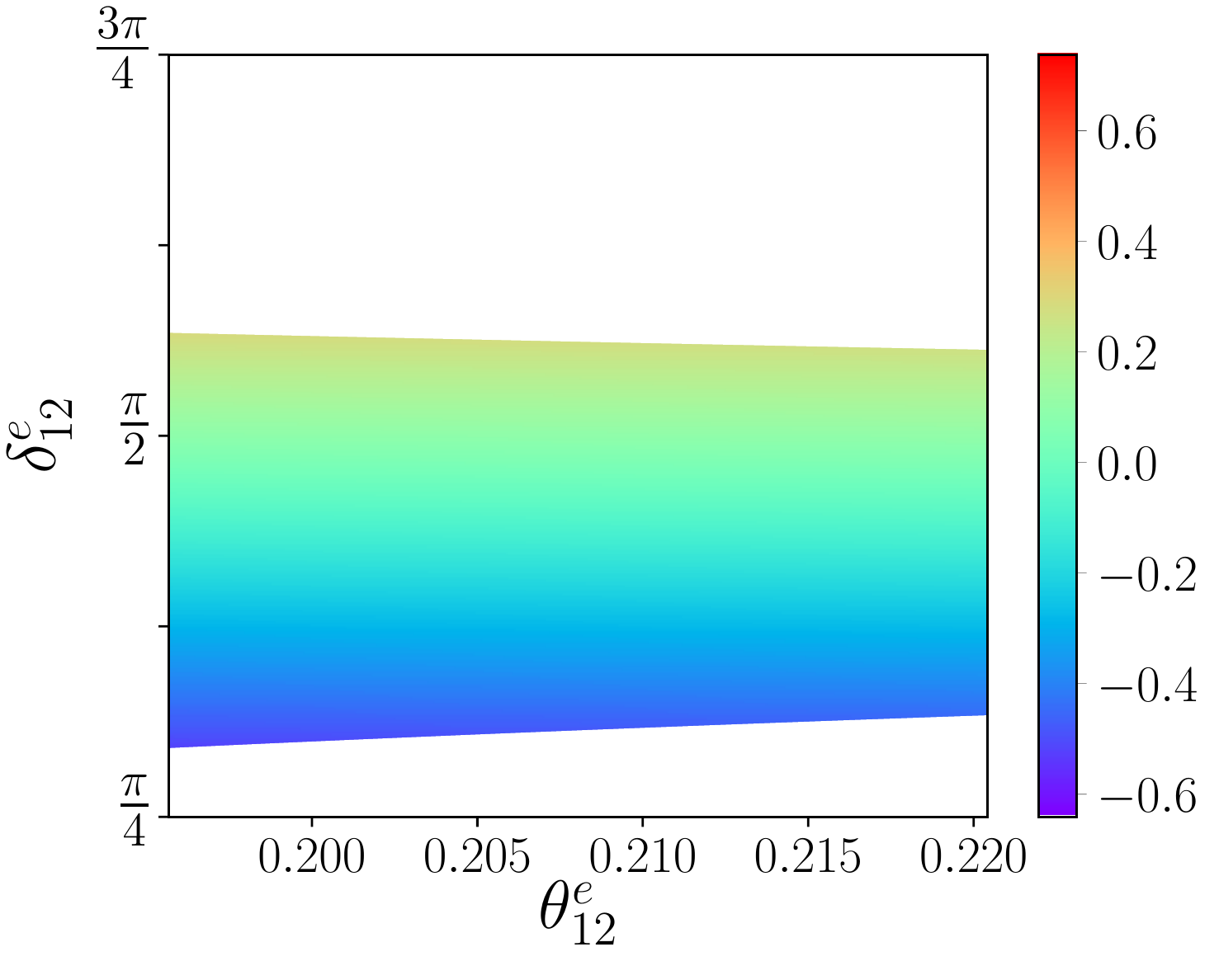}
  \end{subfigure}
  \hfill
  \begin{subfigure}[b]{0.475\textwidth}
  \caption{$\sin \delta$($\theta_{12}^e{},\delta_{12}^{e}$) }
    \includegraphics[width=\textwidth]{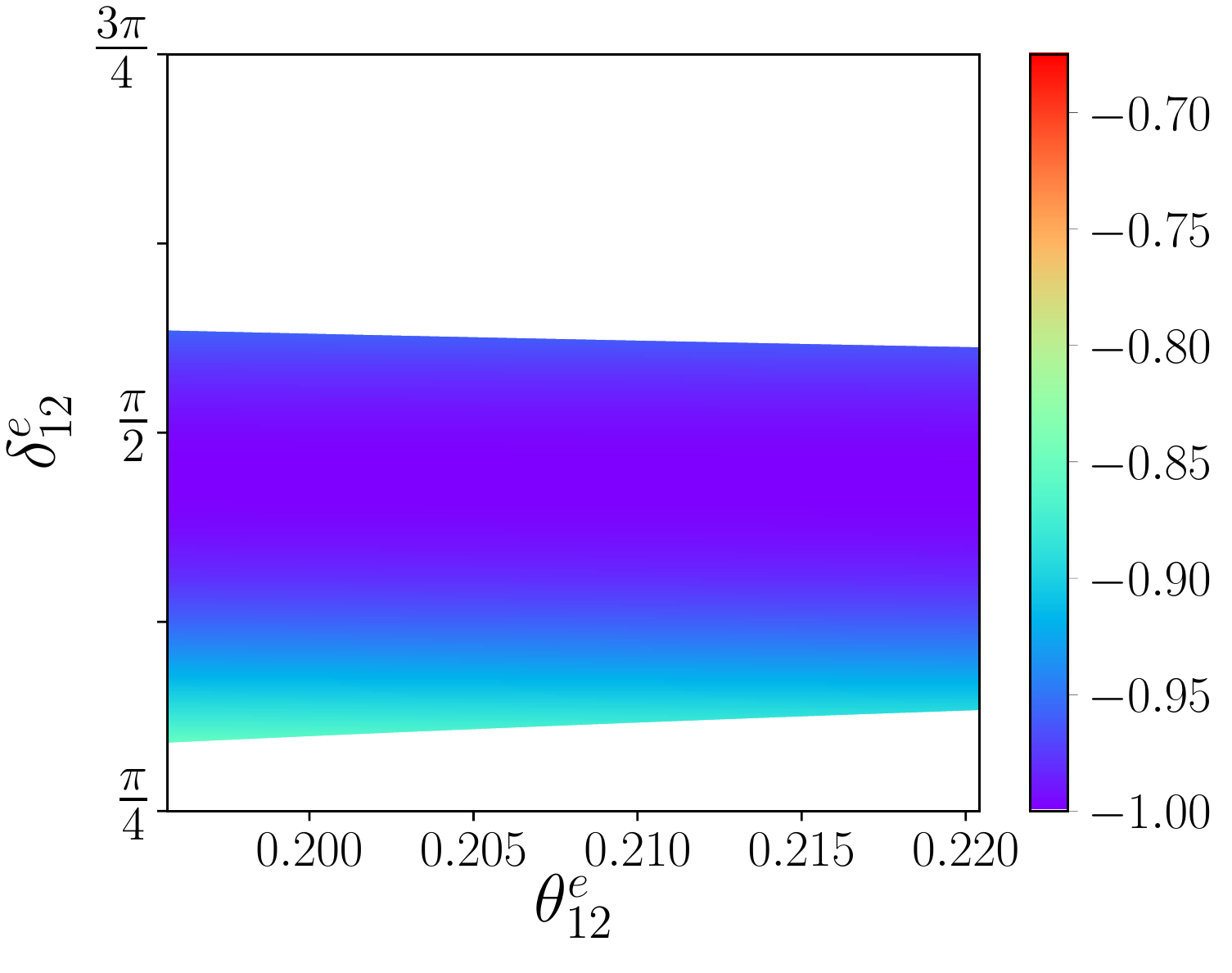}
  \end{subfigure}
		\caption{A close-up view of the allowed parameter region in $\theta^e_{12}$ and $\delta^e_{12}$ and the predictions for $\cos\delta$ and $\sin\delta$ for GR2 mixing, for $U_e=U^e_{12}(\theta^e_{12},\delta^e_{12})$.}
		\label{fig:zoomedin12gr2}
\end{figure}

These results are shown for the case of HEX mixing in Figure~\ref{fig:zoomedin12hex}. In this case, there is a disallowed parameter region focused around $\delta^e_{12}=\pi$, and allowed bands both below and above this region between $\pi/2$ and $3\pi/2$ that are characterized by the sign of $\sin\delta$ and a magnitude of $\sin\delta$ that most often falls in the range $\pm 0.75$.  

The results for GR1 and GR2 are shown in  Figure~\ref{fig:zoomedin12gr1} and  Figure~\ref{fig:zoomedin12gr2}.  In the case of GR1 mixing, we see for one set of allowed values of $\delta^e_{12}$ that fall into a band between approximately $\pi/2$ and $3\pi/4$, the prediction is that $\sin\delta<0$ throughout the parameter region, while for GR2, the analogous band of allowed values for $\theta^e_{12}$ ranges between just above $\pi/4$ and just above $\pi/2$, with $\sin\delta$ close to $-1$ for the  majority of this parameter space.  We also note that as in the TBM case, for each of the two GR scenarios there is another band of allowed values of $\theta^e_{12}$ near $\pi$ with similar features, but with a change in sign in the values of $\sin\delta$.


\subsubsection*{Case 2: One rotation in the $1-3$ sector $(U_{e}=U^{e}_{13}(\theta_{13}^{e},\delta_{13}^{e}))$} 
This set of perturbations resembles the previous case in that once again, the reactor and atmospheric angles are independent of $\theta^\nu_{12}$ and the phase angle $\delta^e_{13}$, while the solar angle (and thus also $\cos\delta$) also depends on $\theta^\nu_{12}$ and $\delta^e_{13}$, as seen in Eq.~(\ref{eq:sij213}). More precisely, with $\theta^\nu_{23}=\pi/4$, $s^2_{13}$ now takes the form
\begin{equation}\label{eq:reactor13}
s_{13}^2 = (s^e_{13})^2/2,
\end{equation}
(compare Eq.~(\ref{eq:reactor12})).  Therefore, the parameter $(s^e_{13})^2$ takes an identical range as did $(s^e_{12})^2$ for the $12$ rotations, and again we can trade the model parameter (here $(s^e_{13})^2$) for $s^2_{13}$.  The atmospheric angle constraint now is predicted to be
\begin{equation}\label{eq:atm13}
s^2_{23} = \frac{1}{2(1-s^2_{13})},
\end{equation}
such that once again it is precisely determined by the reactor angle bounds to fall into a much smaller range than its $3\sigma$ allowed region.  In this case, however, the reactor angle now always satisfies $s^2_{23}>1/2$, with the range $0.5096<s^2_{23}<0.5122$.  

From Eq.~(\ref{eq:reactor13}), we see that the solar mixing angle takes the form
\begin{equation}\label{eq:solar13}
s^2_{12} = \frac{s^2_{13}+y-3 s^2_{13}y +2\cos(\delta^e_{13})\sqrt{(1-2s^2_{13})s^2_{13}}\sqrt{y(1-y)}}{1-s^2_{13}} \equiv \frac{\tilde{\alpha}^{(13)}}{1-s^2_{13}},
\end{equation}
in which we have defined $\tilde{\alpha}^{(13)}$ for later convenience, as we recall that $y=\sin^2(\theta^\nu_{12})$.  Eq.~(\ref{eq:solar13}) should be compared with Eq.~(\ref{eq:solar12}), as these expressions share many similar features.  Once again, there is an interplay between the allowed values of the phase angle $\delta^e_{13}$ and the allowed range for $s^2_{12}$ for a given value of $y$.  In this scenario, the expression for $\cos\delta$, in terms of $s^2_{12}$, $s^2_{13}$, and $y$, is given by
\begin{equation}\label{eq:cosdel13num}
\cos\delta = \frac{(1-s^2_{13})y+3s^2_{12}s^2_{13}-s^2_{12}-s^2_{13}}{2 \sqrt{(1-s^2_{12})s^2_{12}}\sqrt{s^2_{13}(1-2s^2_{13})}},
\end{equation}
which is just the opposite of Eq.~(\ref{eq:cosdel12num}), as is known in the literature.
Together with Eq.~(\ref{eq:solar13}), we can use Eq.~(\ref{eq:cosdel13num}) to obtain $\cos\delta$ and $\sin\delta$ in terms of $s^2_{13}$, $y$, and $\delta^e_{13}$, as follows:
\begin{equation}\label{eq:cosdelta13f}
\cos\delta = \frac{y(1-s^2_{13})^2-s^2_{13}(1-s^2_{13})-\tilde{\alpha}^{(13)}(1-3s^2_{13})}{2\sqrt{s^2_{13}(1-2s^2_{13})}\sqrt{\tilde{\alpha}^{(13)}(1-s^2_{13}-\tilde{\alpha}^{(13)})}},
\end{equation}
which should be compared with Eq.~(\ref{eq:cosdelta12f}), its counterpart for the $1-2$ perturbations, and 
\begin{equation}\label{eq:sindelta13f}
\sin\delta = \pm \frac{\sin\delta^e_{13}(1-s^2_{13})\sqrt{y(1-y)}}{\sqrt{\tilde{\alpha}^{(13)}(1-s^2_{13}-\tilde{\alpha}^{(13)})}},
\end{equation}
in which the sign is to be taken carefully to ensure conformity with the chosen conventions.  Once again, it is instructive to compare this result to the very similar form of Eq.~(\ref{eq:sindelta12f}).

Hence, the structural forms of the charged lepton corrections in the $1-2$ sector and the $1-3$ sector are quite similar, but with slightly different correlations between the predicted range of $\sin^2(\theta_{12})$ and the values for $\cos\delta$ and $\sin\delta$ between the two cases.  For this reason, we will now present the identical set of figures for the $1-3$ sector perturbations.

\begin{figure}[H]
  \begin{subfigure}[b]{0.475\textwidth}
  \caption{$\cos \delta$($\theta_{13}^{e}$, $\delta_{13}^{e}$) }
    \includegraphics[width=\textwidth]{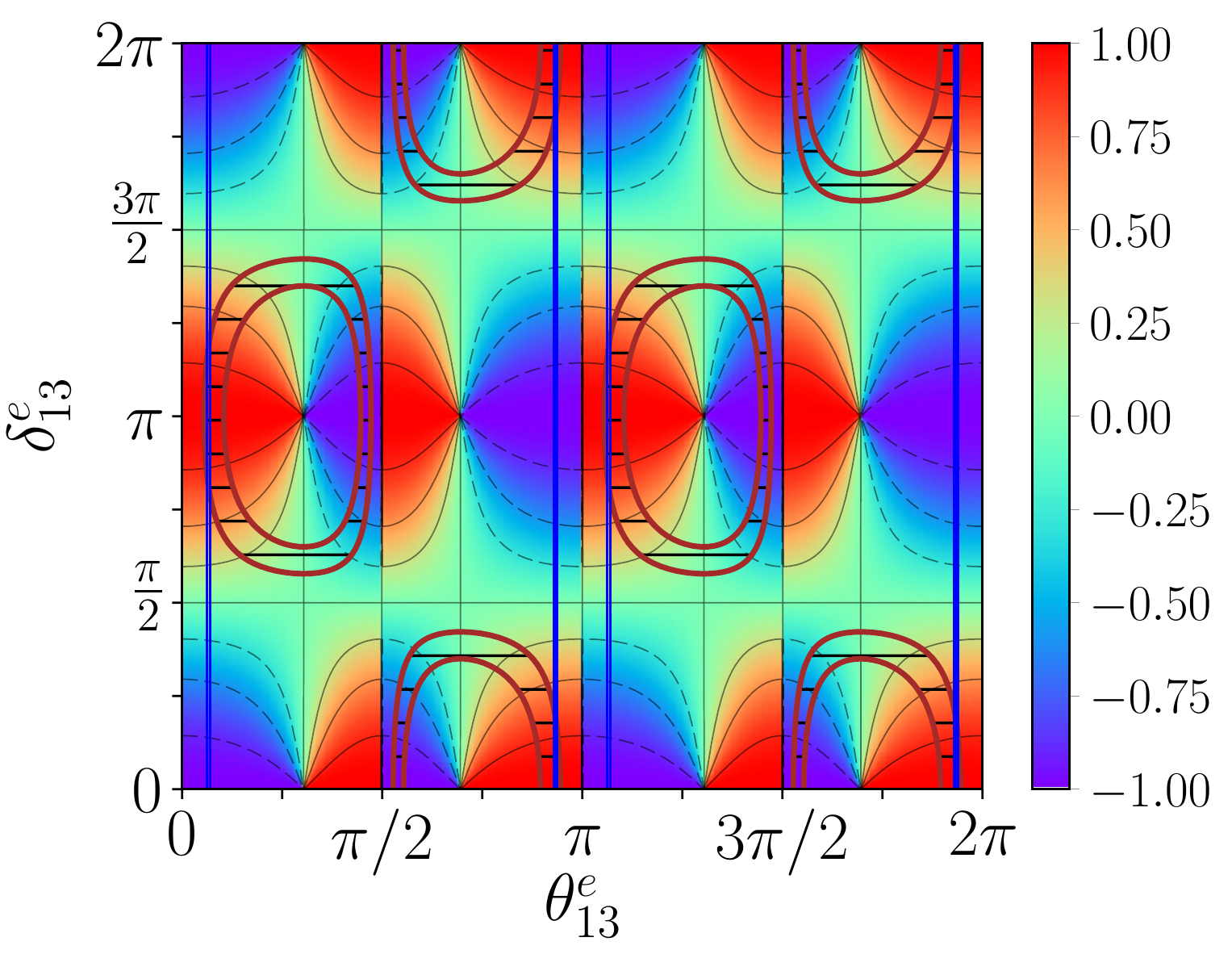}
  \end{subfigure}
  \hfill
  \begin{subfigure}[b]{0.475\textwidth}
  \caption{$\sin \delta$($\theta_{13}^{e}$, $\delta_{13}^{e}$) }
    \includegraphics[width=\textwidth]{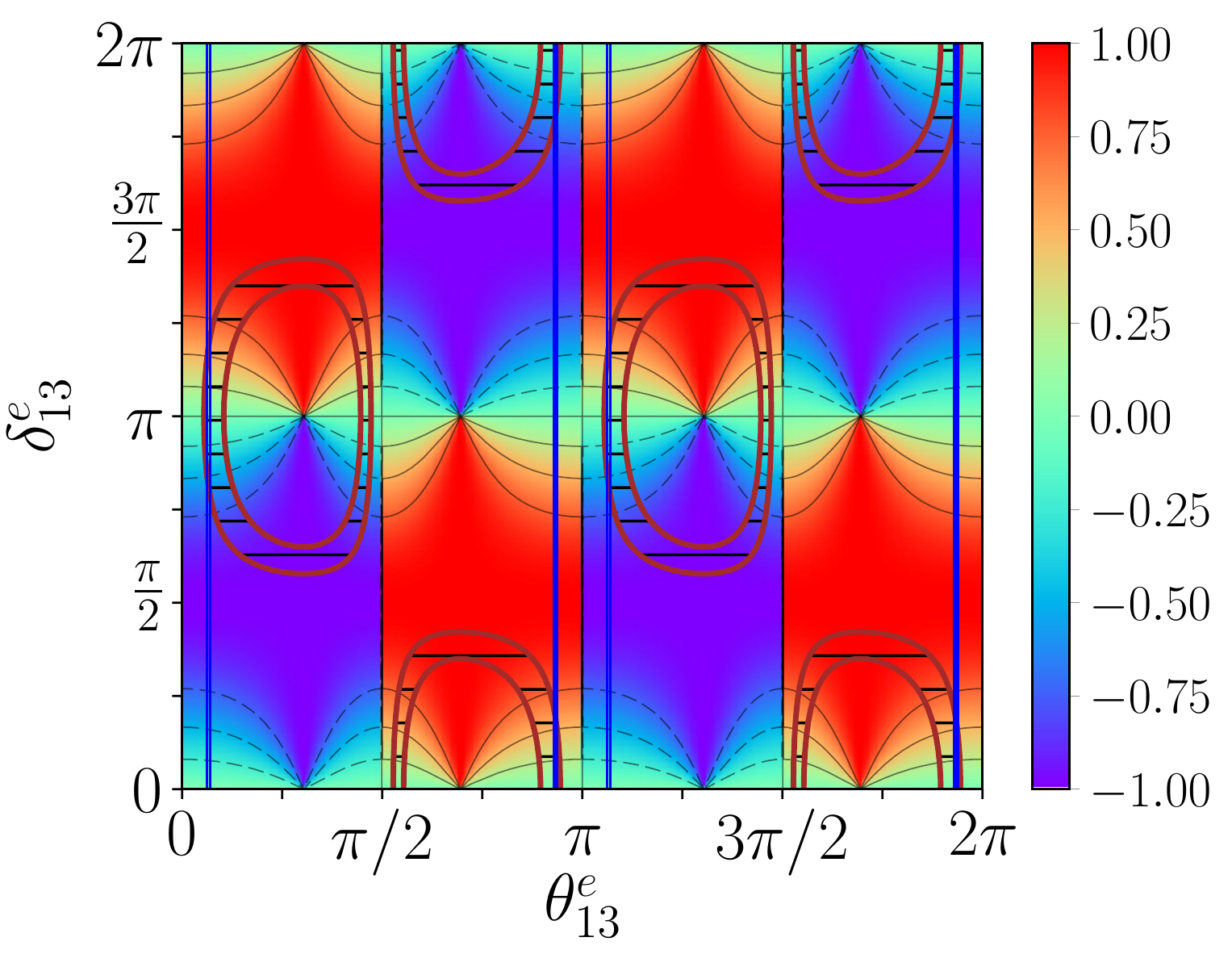}
  \end{subfigure}
  \caption{The predictions for $\cos \delta$ and $\sin \delta$ as a
		         function of $\theta_{13}^{e}$ and $\delta_{13}^{e}$ for BM
			 mixing. The blue band and the region between the dark-red
		 contours represent the regions allowed by $\sin^2(\theta_{13})$ and
	 $\sin^2(\theta_{12})$ at $3\sigma$, respectively.}
  \label{fig:fullspace13bm}
\end{figure}
\begin{figure}[H]
  \begin{subfigure}[b]{0.475\textwidth}
  \caption{$\cos \delta$($\theta_{13}^{e}$, $\delta_{13}^{e}$) }
    \includegraphics[width=\textwidth]{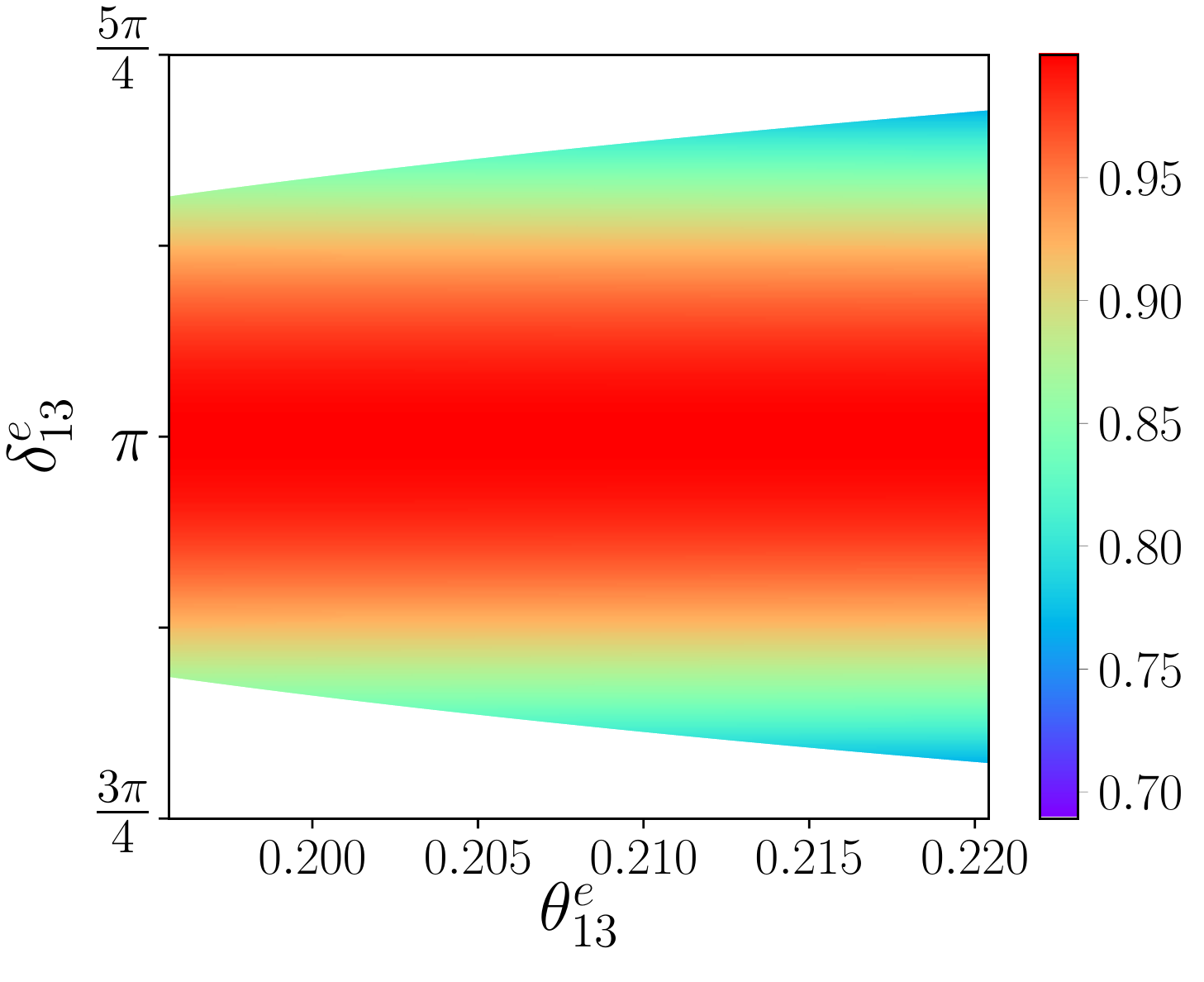}
  \end{subfigure}
  \hfill
  \begin{subfigure}[b]{0.475\textwidth}
  \caption{$\sin \delta$($\theta_{13}^{e}$, $\delta_{13}^{e}$) }
    \includegraphics[width=\textwidth]{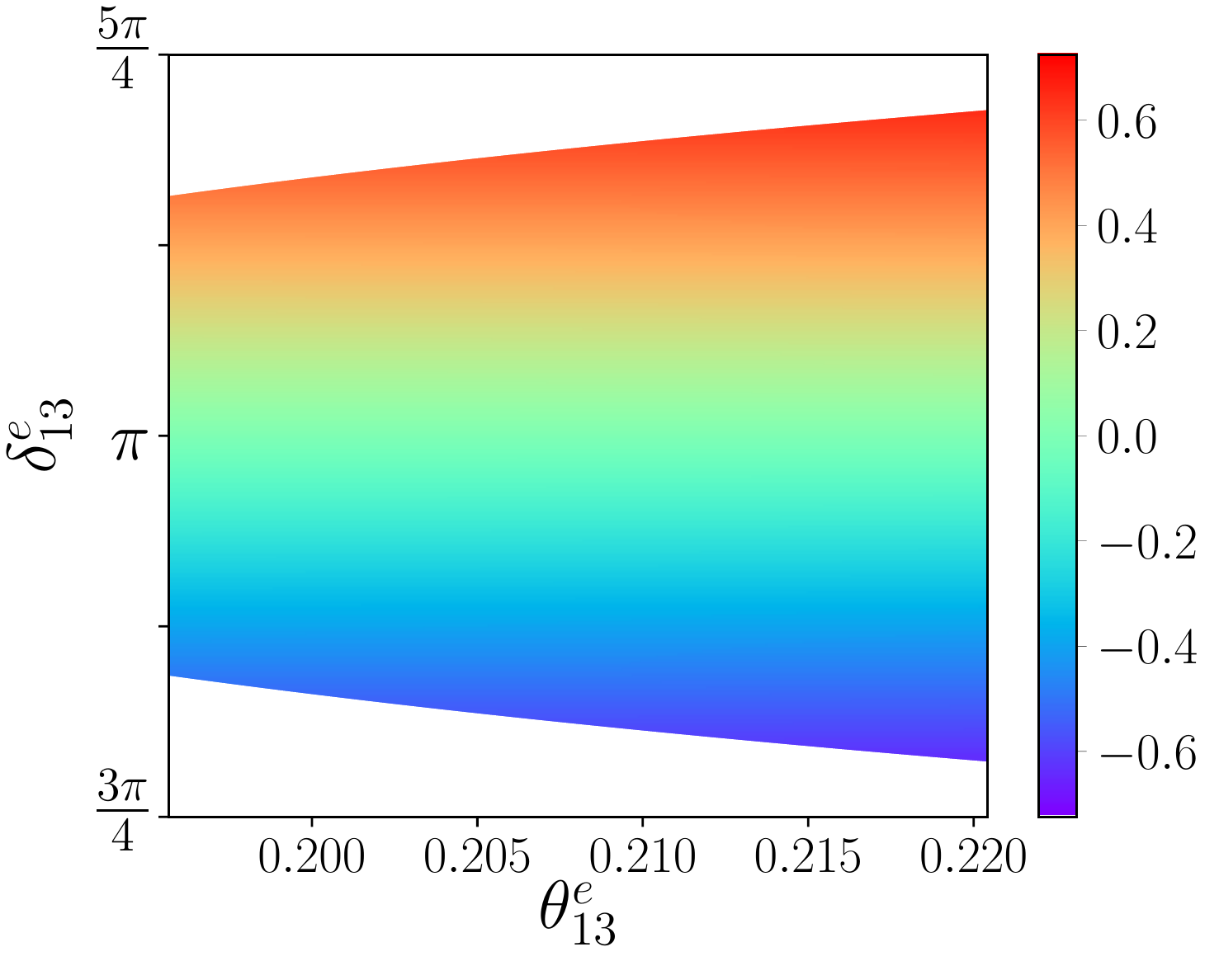}
  \end{subfigure}
		\caption{A close-up view of one of the allowed parameter region in $\theta^e_{13}$ and $\delta^e_{13}$ and the predictions for $\cos\delta$ and $\sin\delta$ for BM mixing, for $U_e=U^e_{13}(\theta^e_{13},\delta^e_{13})$.}
		\label{fig:zoomedin13bm}
\end{figure}
\begin{figure}[H]
  \begin{subfigure}[b]{0.475\textwidth}
  \caption{$\cos \delta$($\theta_{13}^{e}$, $\delta_{13}^{e}$) }
    \includegraphics[width=\textwidth]{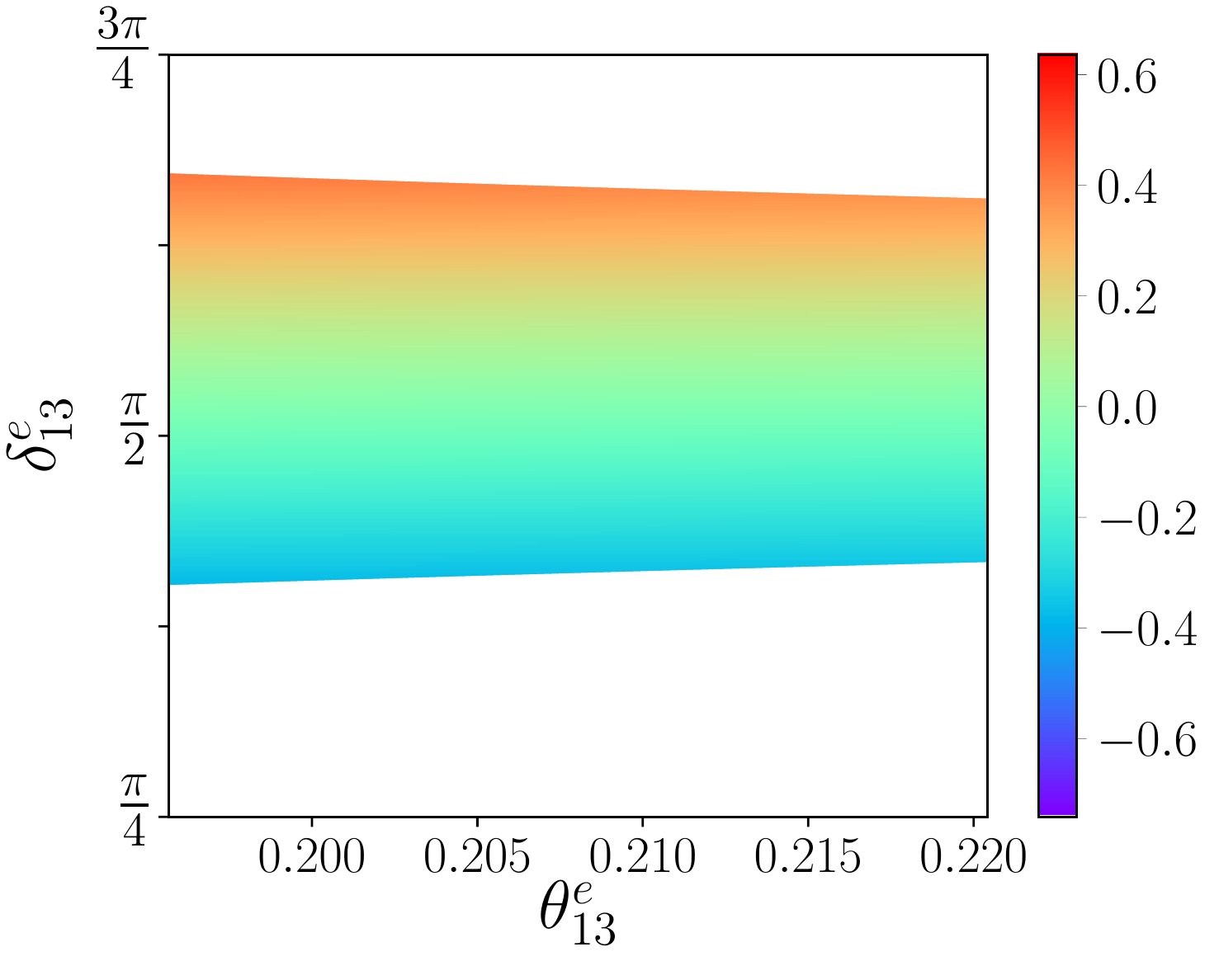}
  \end{subfigure}
  \hfill
  \begin{subfigure}[b]{0.475\textwidth}
  \caption{$\sin \delta$($\theta_{13}^{e}$, $\delta_{13}^{e}$) }
    \includegraphics[width=\textwidth]{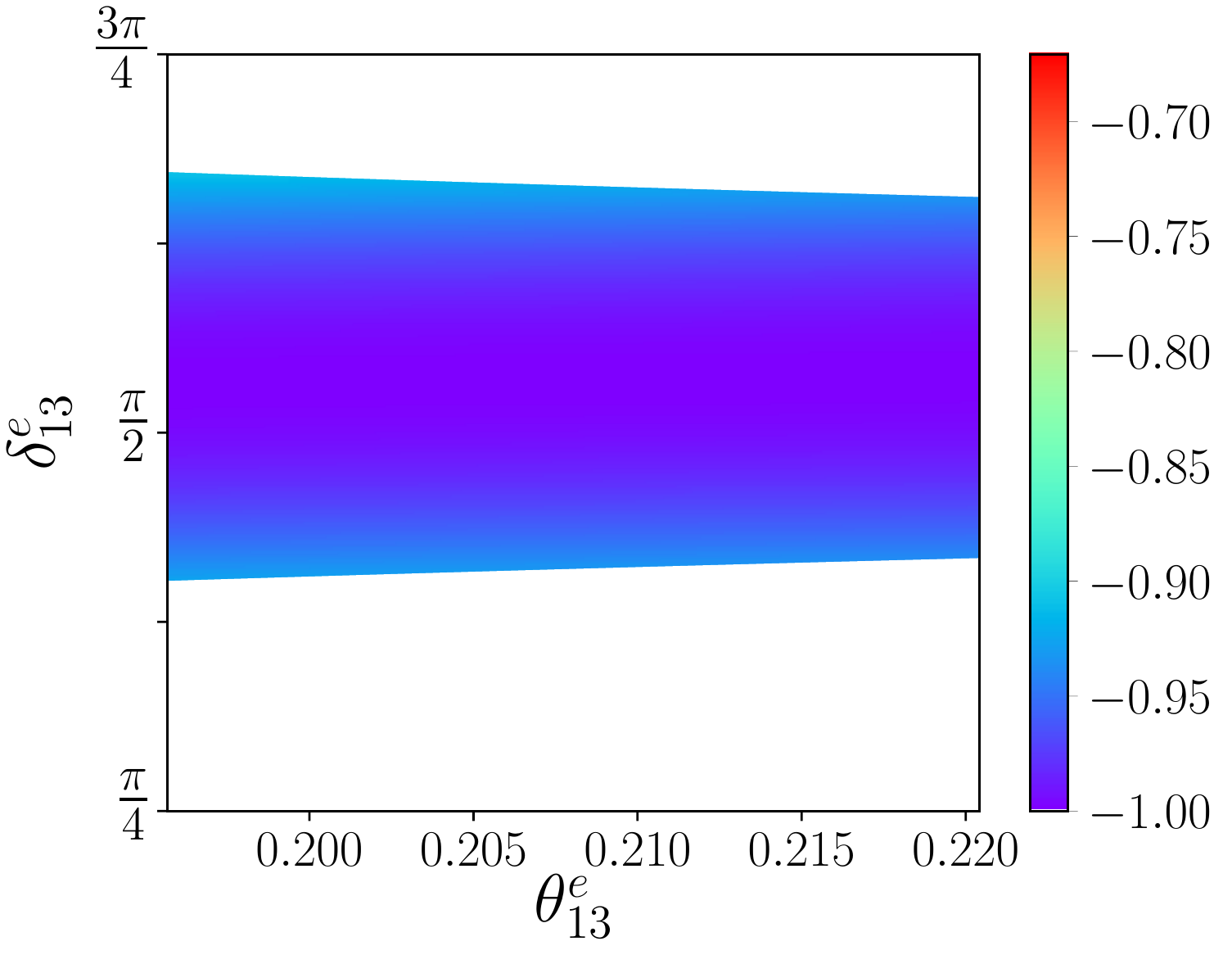}
  \end{subfigure}
		\caption{A close-up view of one of the allowed parameter regions in $\theta^e_{13}$ and $\delta^e_{13}$ and the predictions for $\cos\delta$ and $\sin\delta$ for TBM mixing, for $U_e=U^e_{13}(\theta^e_{13},\delta^e_{13})$.}
		\label{fig:zoomedin13tbm}
\end{figure}
\begin{figure}[H]
  \begin{subfigure}[b]{0.475\textwidth}
  \caption{$\cos \delta$($\theta_{13}^{e}$, $\delta_{13}^{e}$)}
    \includegraphics[width=\textwidth]{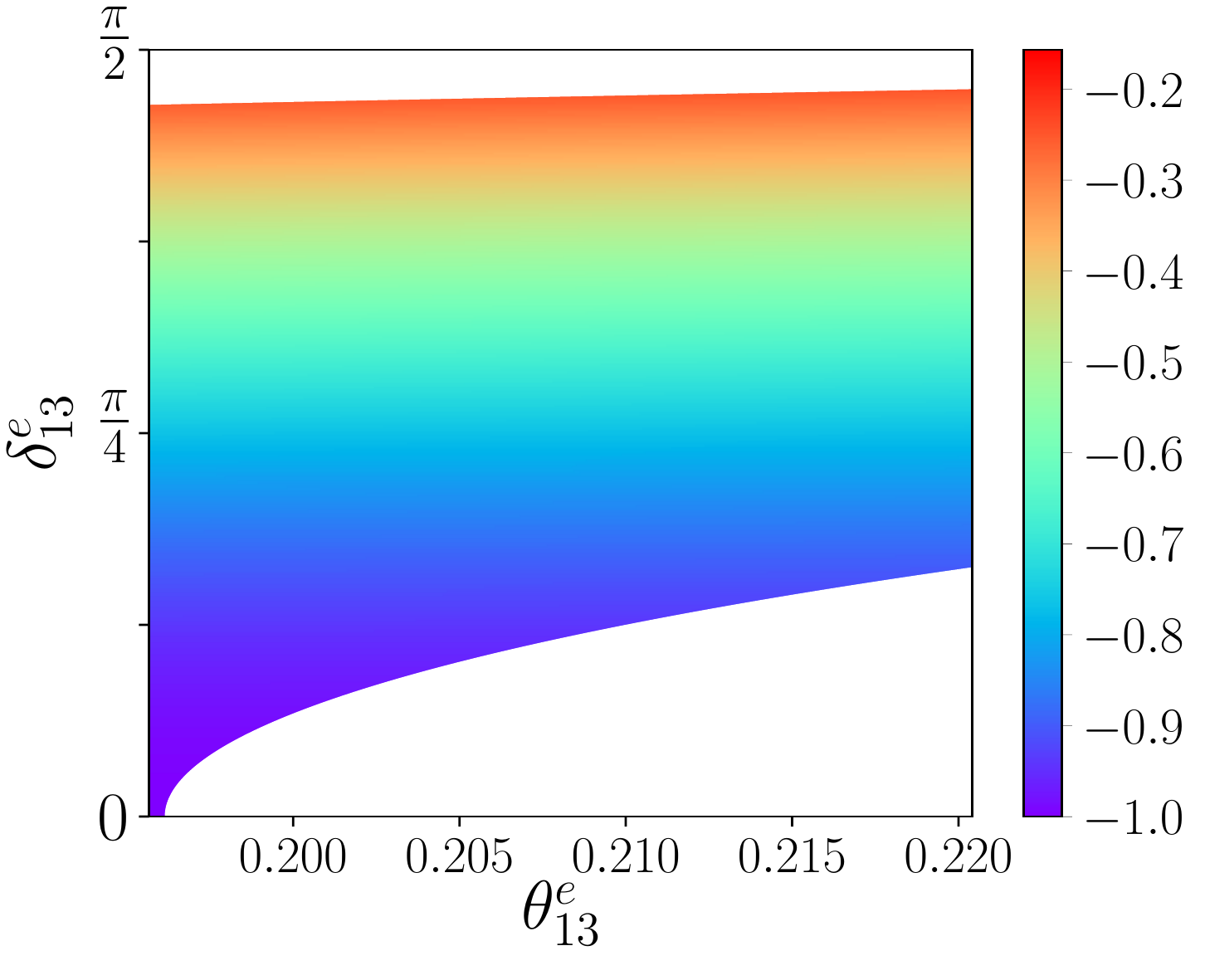}
  \end{subfigure}
  \hfill
  \begin{subfigure}[b]{0.475\textwidth}
  \caption{$\sin \delta$($\theta_{13}^{e}$, $\delta_{13}^{e}$) }
    \includegraphics[width=\textwidth]{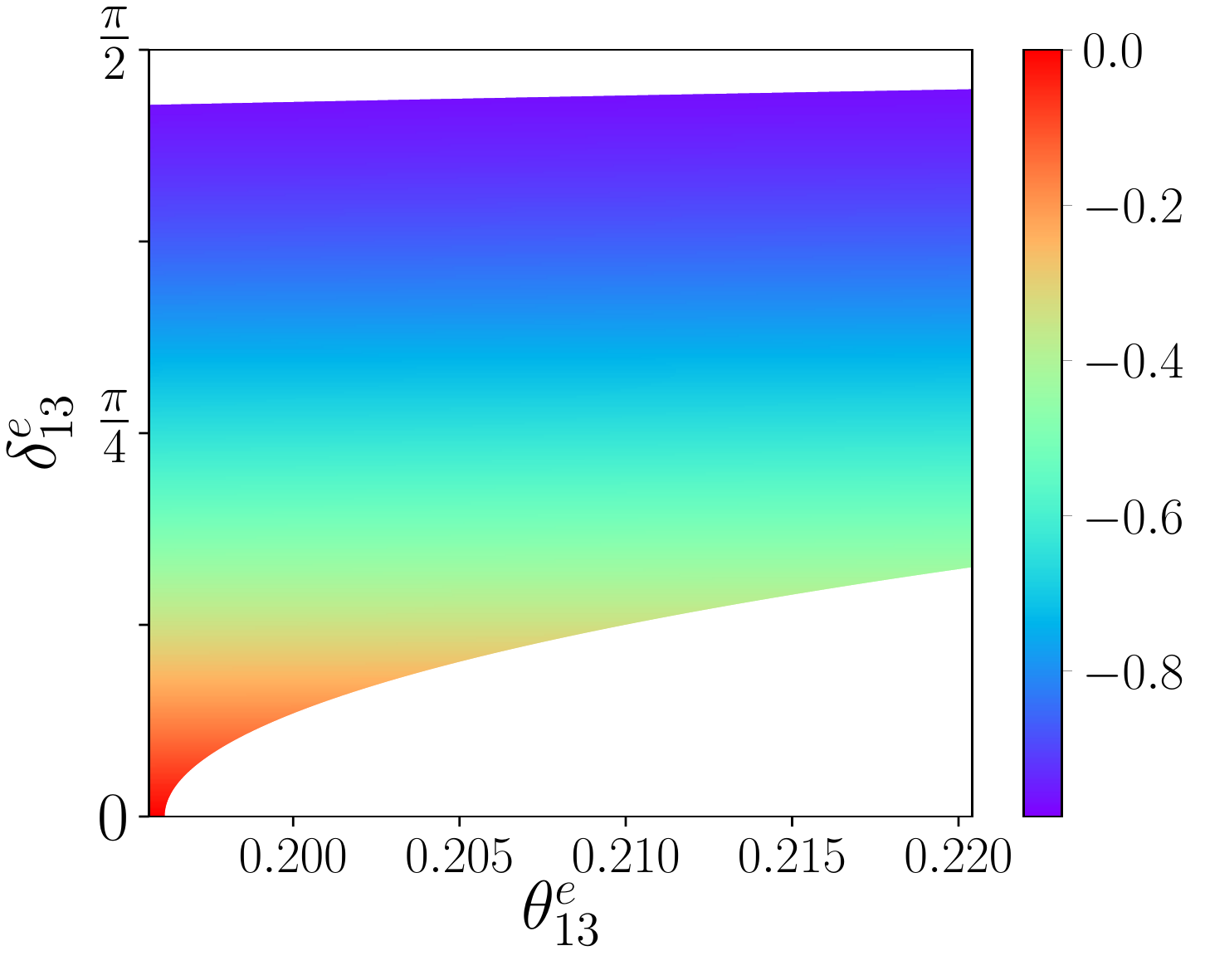}
  \end{subfigure}
		\caption{A close-up view of one of the allowed parameter regions in $\theta^e_{13}$ and $\delta^e_{13}$ and the predictions for $\cos\delta$ and $\sin\delta$ for HEX mixing, for $U_e=U^e_{13}(\theta^e_{13},\delta^e_{13})$.}
		\label{fig:zoomedin13hex}
\end{figure}
\begin{figure}[H]
  \begin{subfigure}[b]{0.475\textwidth}
  \caption{$\cos \delta$($\theta_{13}^{e}$, $\delta_{13}^{e}$)}
    \includegraphics[width=\textwidth]{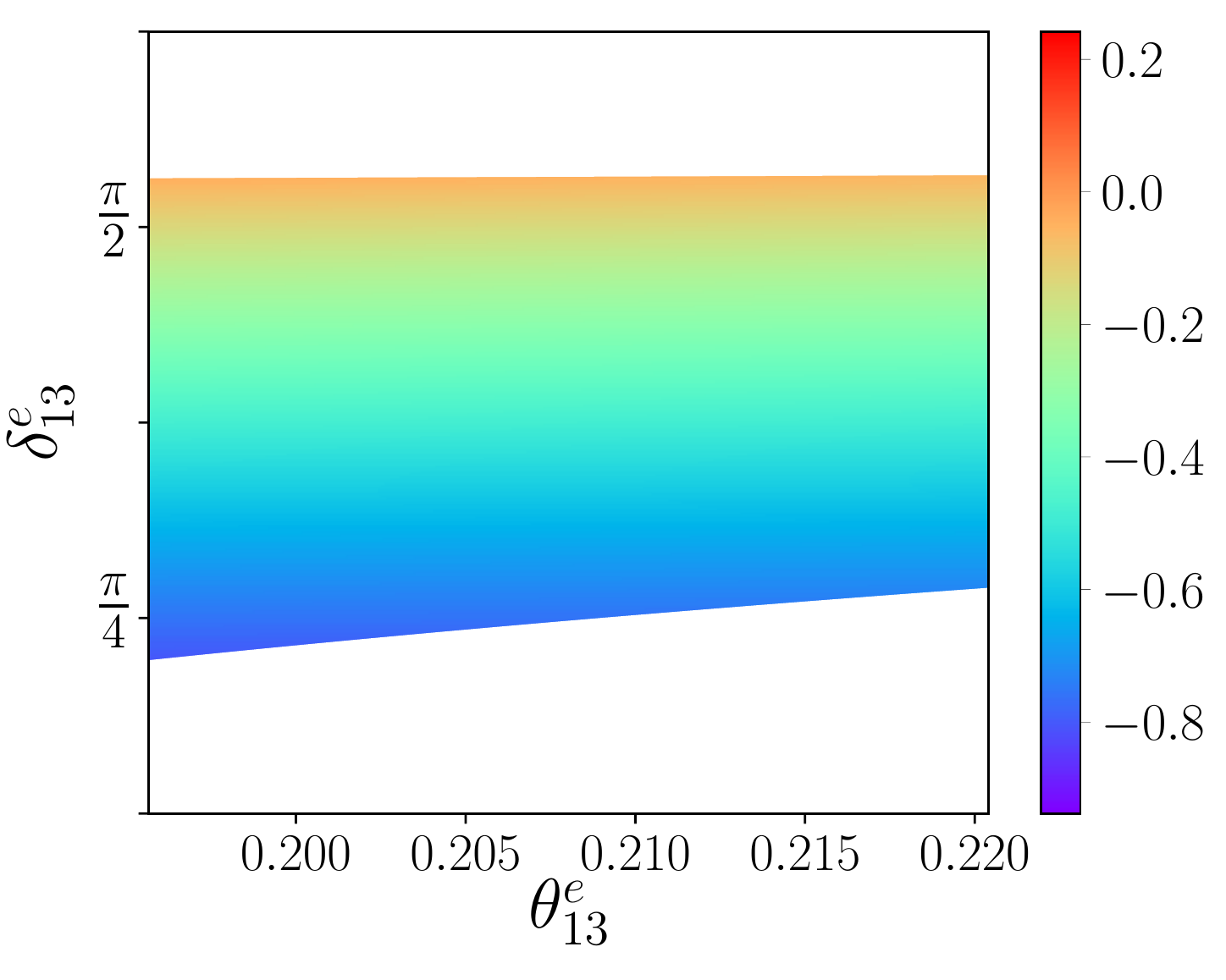}
  \end{subfigure}
  \hfill
  \begin{subfigure}[b]{0.475\textwidth}
  \caption{$\sin \delta$($\theta_{13}^{e}$, $\delta_{13}^{e}$) }
    \includegraphics[width=\textwidth]{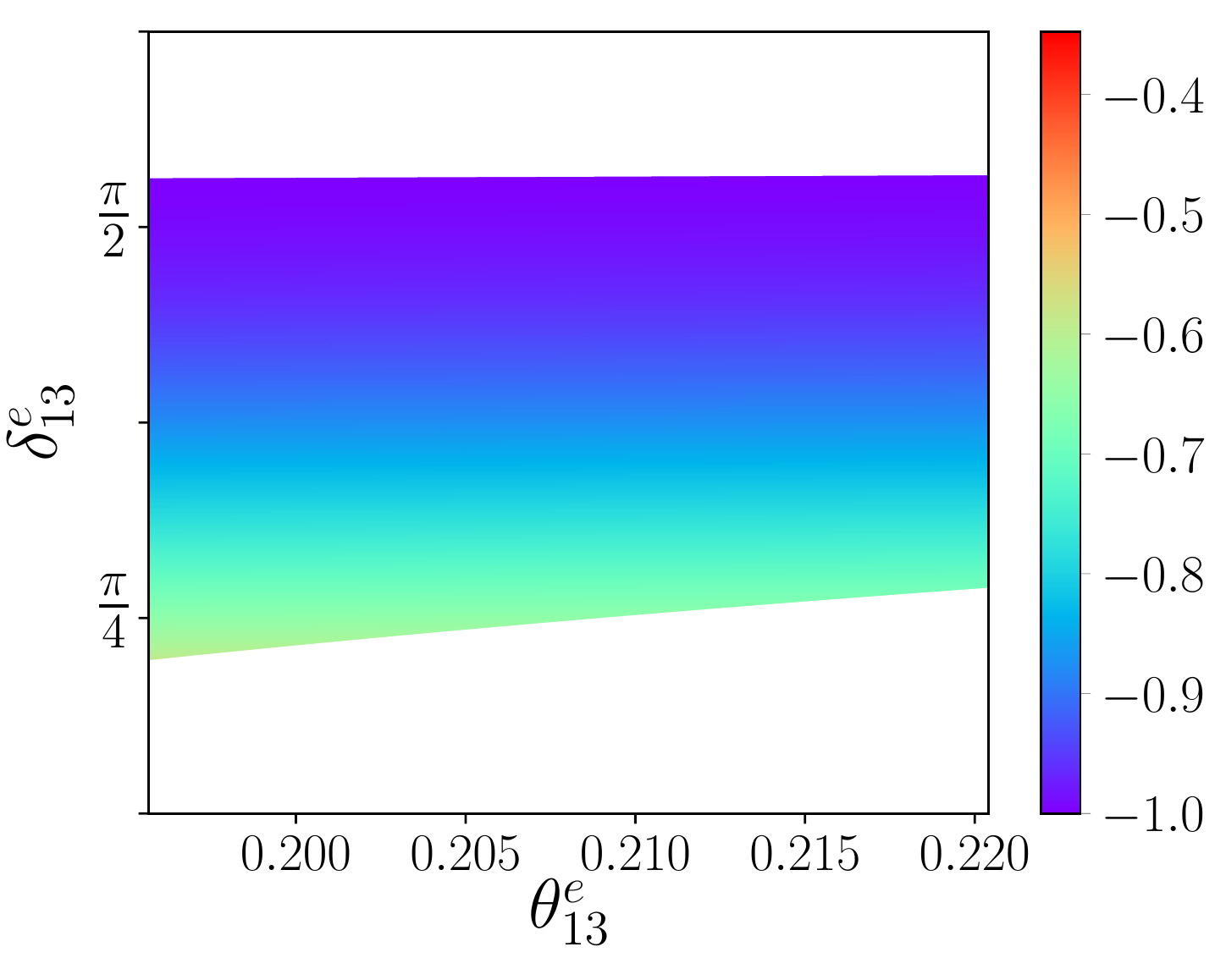}
  \end{subfigure}
		\caption{A close-up view of one of the allowed parameter regions in $\theta^e_{13}$ and $\delta^e_{13}$ and the predictions for $\cos\delta$ and $\sin\delta$ for GR1 mixing, for $U_e=U^e_{13}(\theta^e_{13},\delta^e_{13})$.}
		\label{fig:zoomedin13gr1}
\end{figure}
\begin{figure}[H]
  \begin{subfigure}[b]{0.475\textwidth}
  \caption{$\cos \delta$($\theta_{13}^{e}$, $\delta_{13}^{e}$)}
    \includegraphics[width=\textwidth]{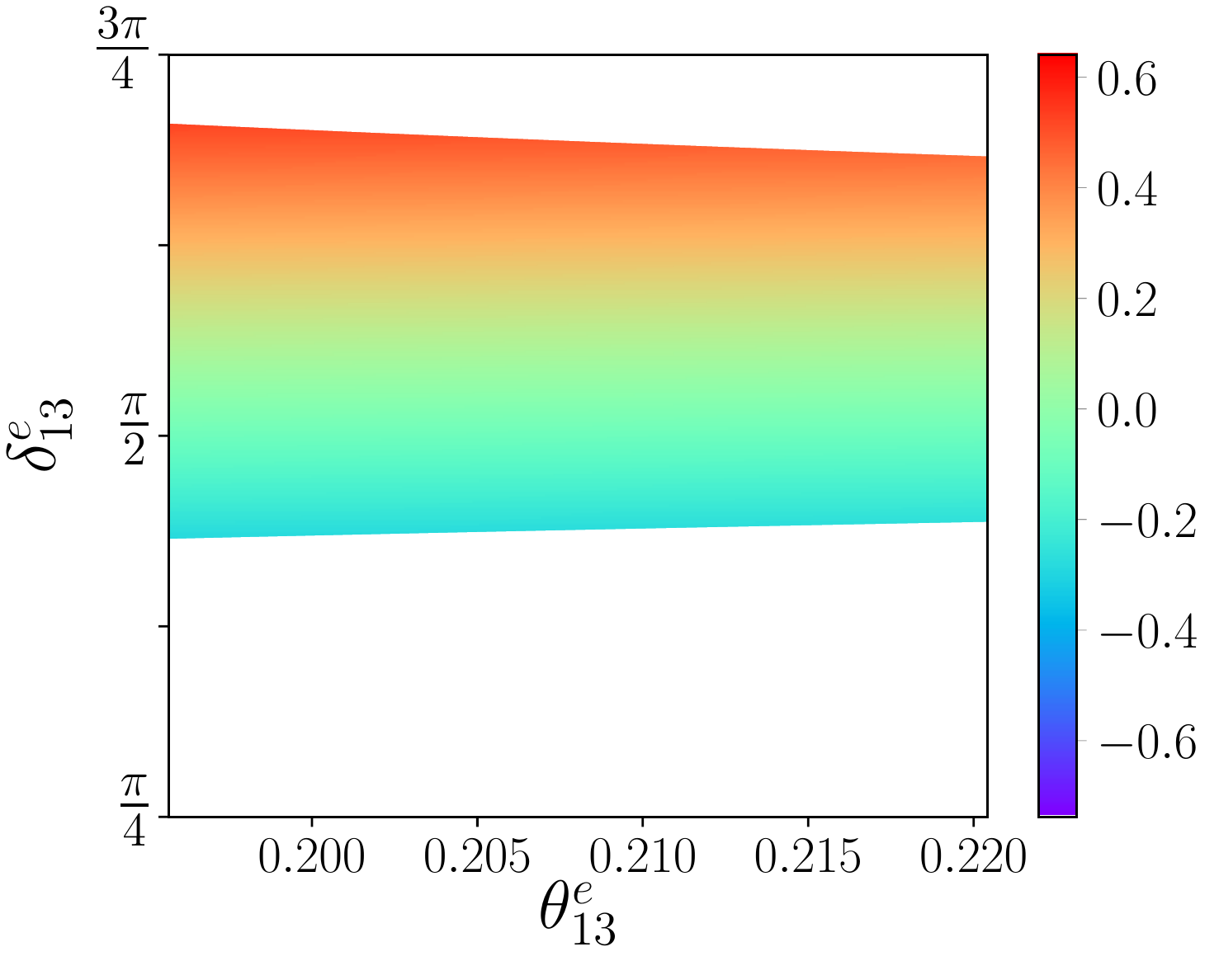}
  \end{subfigure}
  \hfill
  \begin{subfigure}[b]{0.475\textwidth}
  \caption{$\sin \delta$($\theta_{13}^{e}$, $\delta_{13}^{e}$) }
    \includegraphics[width=\textwidth]{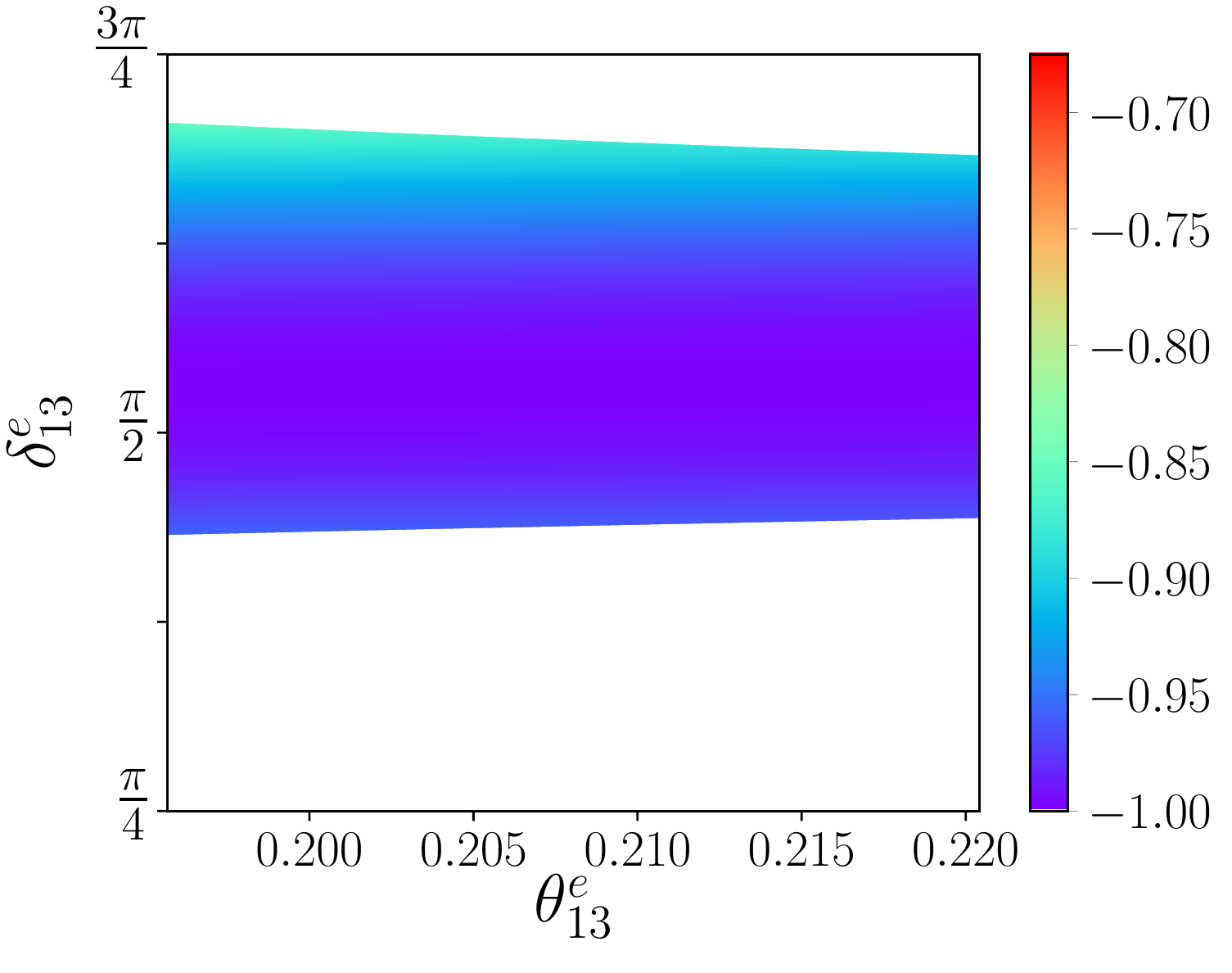}
  \end{subfigure}
		\caption{A close-up view of one of the allowed parameter regions in $\theta^e_{13}$ and $\delta^e_{13}$ and the predictions for $\cos\delta$ and $\sin\delta$ for GR2 mixing, for $U_e=U^e_{13}(\theta^e_{13},\delta^e_{13})$.}
		\label{fig:zoomedin13gr2}
\end{figure}

We start once again with the case of BM mixing, as shown in Figures~\ref{fig:fullspace13bm} and \ref{fig:zoomedin13bm}. Here once again we have tight constraints on the allowed parameter space resulting from the combined constraints on $s^2_{12}$ and $s^2_{13}$, given the large value of the solar angle in absence of the charged lepton sector perturbations. We note that the preferred value for $\cos\delta$ is now $+1$, in accordance with the sign flip as expected from  Eq.~(\ref{eq:cosdel13num}).
\begin{figure}[H]
  \begin{subfigure}[b]{0.475\textwidth}
  \caption{$\cos \delta$($\theta_{13}^{e}$, $\delta_{13}^{e}$) }
    \includegraphics[width=\textwidth]{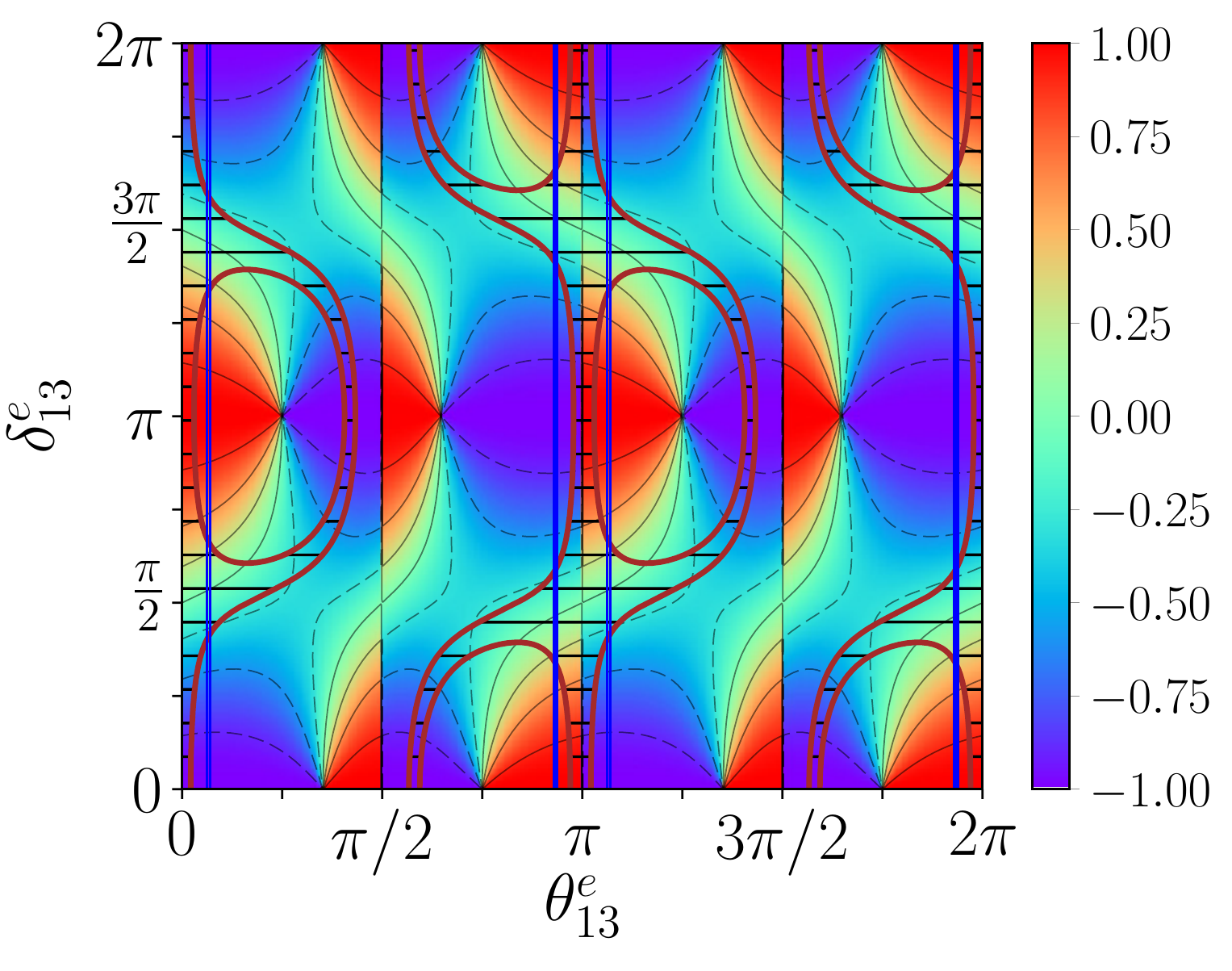}
  \end{subfigure}
  \hfill
  \begin{subfigure}[b]{0.475\textwidth}
  \caption{$\sin \delta$($\theta_{13}^{e}$, $\delta_{13}^{e}$) }
    \includegraphics[width=\textwidth]{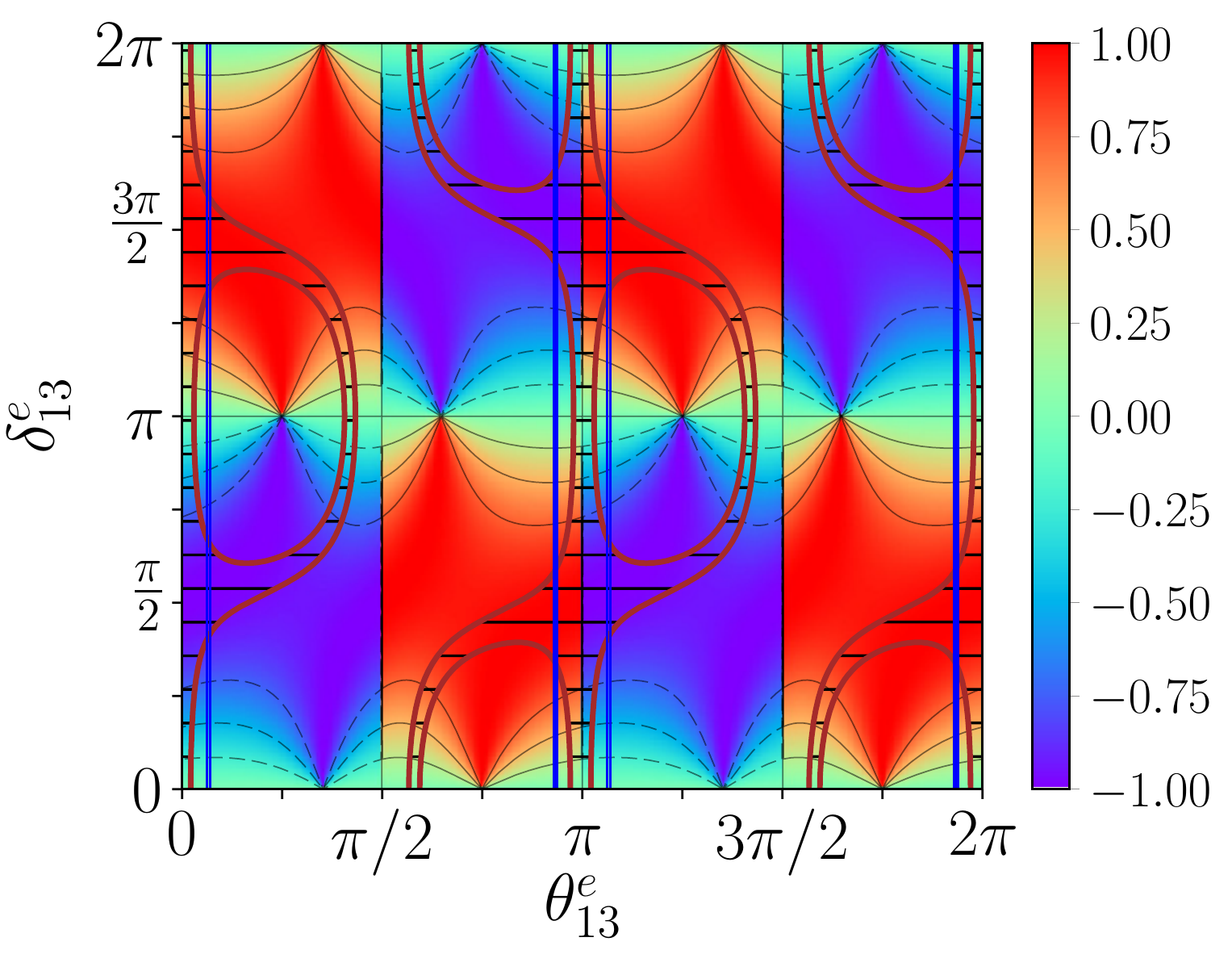}
  \end{subfigure}
  \caption{The predictions of $\cos \delta$ and $\sin \delta$ as a
  function of $\theta_{13}^{e}$ and $\delta_{13}^{e}$ for TBM mixing. The blue band and the region between the dark-red
		 contours represent the regions allowed by $\sin^2(\theta_{13})$ and
	 $\sin^2(\theta_{12})$ at $3\sigma$, respectively.}
  \label{fig:fullspace13tbm}
\end{figure}
We show the analogous results for TBM mixing in Figures~\ref{fig:fullspace13tbm} and \ref{fig:zoomedin13tbm}.  Comparing these results to their counterparts for the $1-2$ perturbations in Figures~\ref{fig:fullspace12tbm} and \ref{fig:zoomedin12tbm}, we see a similar overlap of allowed regions for the reactor and solar mixing angles, and a slightly shifted range of preferred values of $\delta^e_{13}$ compared to the preferred range for $\delta^e_{12}$.
For the remaining cases of HEX, GR1, and GR2 mixing, we follow our previous procedure and show specific zoomed-in regions of allowed parameter space in $\theta^e_{13}$ and $\delta^e_{13}$, which are given in Figures~\ref{fig:zoomedin13hex}, \ref{fig:zoomedin13gr1}, and \ref{fig:zoomedin13gr2}, respectively.  These results, when compared to their counterparts in Figures~\ref{fig:zoomedin12hex}--\ref{fig:zoomedin12gr2}, show the demonstrated sign flip in $\cos\delta$ and the corresponding shifts in the allowed regions for $\delta^e_{13}$ based on the solar mixing angle constraint.
 In particular, we see that for HEX mixing, the allowed range for $\delta^e_{13}$ that corresponds to this particular slice of $\theta^e_{13}$ parameter values now consists of one band that is centered upon $\pi/4$ and ends at values just below $\pi/2$, with negative values of $\sin\delta$.  For GR1 mixing, the $\delta^e_{13}$ range ranges from $\pi/4$ to just above $\pi/2$, with values of $\sin\delta$ that are also negative and tend to be larger in magnitude.  The GR2 mixing case is also dominated by values of $\sin\delta$ that tend to $-1$, with a corresponding range in $\delta^e_{13}$ from slightly below $\pi/2$ to just below $3\pi/4$.  Hence, we see from comparing with the $1-2$ mixing counterparts that together with the sign flip in $\cos\delta$, the predictions for $\sin\delta$ are similar in the perturbation schemes arising from single rotations in the $1-2$ and $1-3$ sectors. 


\subsection{Two rotations}
\label{subsec:harderr}

For the cases in which $U_e$ involves two successive non-commuting rotations, the parameter space involves two mixing angles and two phases.  For the numerical analysis, we have chosen to project to a specific two-dimensional parameter space that focuses on the two phase angles, and uses the experimental constraints on the MNSP mixing angles to fix the two theoretical mixing angle parameters. 

Our approach is as follows.  We evaluate $\sin^2(\theta_{ij})$ as functions of the model parameters using a four-dimensional grid, and remove any combination of parameters that falls outside the 3$\sigma$ bounds reported in \cite{c1} and presented in Table~\ref{tab:1}.  We then fix one of the $\theta^e_{jk}$ based on analyzing their distributions over the parameter range from $[0,\pi]$.  Quite generally, and this is especially evident when one of the two rotations is in the $2-3$ plane, while the phase angles $\delta^e_{jk}$ tend to distribute over the full range, at least one of the $\theta^e_{jk}$ tends to prefer values that are centered around either $\theta^e_{jk}=0.23$, which is roughly the value of the Cabibbo angle $\theta_c$, or $\theta^e_{jk}=2.91=\pi-0.23 \sim \pi -\theta_c$.  This is of course not surprising given that generating the observed value of $\theta_{13}$  prefers that at least one mixing angle parameter in $U_e$ is of the same order as the Cabibbo angle, up to numerical factors of order unity.  In our analysis, we have chosen to fix one of the $\theta^e_{jk}$ in each scenario to be equal to $2.91\simeq \pi-\theta_c$. 
 
This now leaves one $\theta^e_{jk}$ and the two $\delta^e_{jk}$.  To reduce the parameter space further, we use the reactor mixing angle measurement as input.  More precisely, we choose to fix $\sin^2(\theta_{13}) = 0.02155$, the central value for the most precisely measured mixing angle as determined in \cite{c1}.  As we will see for each individual set of perturbations, imposing this value for $s^2_{13}$ together with fixing one of the $\theta^e_{jk}$ to a specific value can be used to determine the remaining $\theta^e_{jk}$ for each set of perturbations.  We can then construct a more dense two-dimensional grid with the two phases in the range $[0,2\pi)$. As before, we test this grid against the 3$\sigma$ range for the normal ordering and discard any set of parameters that falls outside the allowed range (note that either the result that $\sin^2(\theta_{13}) = 0.02155$, or a null solution, will automatically be the generated output). This approach thus allows for the selection of a set of coordinates with the global fit in mind and that can be well represented in a two-dimensional format.  It also focuses on the phase parameters, which play critical roles in generating the MNSP phase $\delta$.


\subsubsection*{Case 3: Two rotations in the 1--2 and 2--3 sectors ($U_{e}=U^{e}_{23}(\theta_{23}^{e},\delta_{23}^{e})
U^{e}_{12}(\theta_{12}^{e},\delta_{12}^{e})$)} 
In this scenario with $\theta^\nu_{23}=\pi/4$, we see from Eq.~(\ref{eq:sintilde})  that 
\begin{equation}
\sin^2(\tilde{\theta}_{23}) = \frac{1}{2} \left (1-\cos(\delta^e_{23})s^{\prime e}_{23} \right ) \equiv \frac{1}{2} (1-z),
\end{equation}
in which we have defined the quantity $z=\cos(\delta^e_{23})s^{\prime e}_{23}$.  From this result and Eq.~(\ref{eq:sin131223}),
$\sin^2(\theta_{13})$ takes the form
\begin{equation}
s^2_{13}=\frac{(s^e_{12})^2}{2}(1-z).
\end{equation}
Here we choose to fix $\theta^e_{12}=2.91$ for concreteness.   Imposing this constraint and fixing $s^2_{13}$ to its central value of $0.02155$ yields the result that $z=0.1890$, and thus $\sin^2(\tilde{\theta}_{23})=0.4091$.  Given that $\cos(\delta^e_{23})$ and $s^{\prime e}_{23}$ are both bounded functions, the $z$ constraint thus disallows specific ranges for both $\delta^e_{23}$ and $\theta^e_{23}$. Here we use $z$ to eliminate $\theta^e_{23}$ for $\delta^e_{23}$, and we will see in the numerical results that there are disallowed regions of $\delta^e_{23}$ centered at $\pi/2$ (mod $\pi$).

For the atmospheric mixing angle, Eq.~(\ref{eq:sin231223}) shows that for $\theta^\nu_{23}=\pi/4$, $s^2_{23}$ takes the form
\begin{equation}
s^2_{23}=\frac{(c^e_{12})^2(1-z)}{2-(s^e_{12})^2(1-z)}=\frac{1-z-2s^2_{13}}{2(1-s^2_{13})},
\end{equation}
such that for $\theta^e_{12}=2.91$ and $z=0.1890$, the atmospheric mixing angle is predicted to be given by $s^2_{23}= 0.396$, which is just slightly above the lower $3\sigma$ limit on this quantity.  Hence, in this approach, once $s^2_{13}$ is fixed $s^2_{23}$ is also fixed, just as in the single rotation cases.

Turning now to the solar mixing angle, we have from Eq.~(\ref{eq:sin121223}) that
\begin{equation}\label{eq:solar1223}
s^2_{12} = \frac{(c^e_{12})^2y+\frac{1}{2}(1-y)(1+z)(s^e_{12})^2-s^{\prime e}_{12}\sqrt{\frac{y(1-y)}{2}}(c^e_{23}\cos(\delta^e_{12})+s^e_{23}\cos(\delta^\prime))}{1-(s^e_{12})^2(1-z)/2},
 \end{equation} 
in which we have set $\delta^\prime=\delta^e_{12}+\delta^e_{23}$, and we recall that $\theta^e_{23}$ is determined by the $z$ constraint as described above.  For fixed $\theta^e_{12}$, $z$ (and thus $\theta^e_{23}$ as a function of $\delta^e_{23}$), and $y$, we see that the solar mixing angle constraint fixes a preferred range for the phase angle $\delta^e_{12}$.  From Eq.~(\ref{eq:sumruleorig}) (or equivalently Eq.~(\ref{eq:cddouble1223form1}) or (\ref{eq:cddouble1223form2})), we can write an expression for $\cos\delta$ in terms of $s^2_{13}$, $y$, $z$, and $s^2_{12}$, as follows:
\begin{equation}\label{eq:cosdel1223f}
\cos\delta = \frac{(1-z)s^2_{12}+s^2_{13}(1+z)-s^2_{12}s^2_{13}(z+3)-y(1-s^2_{13})(1-z)}{s^\prime_{12}s_{13}\sqrt{(1+z)(1-z-2s^2_{13})}},
\end{equation}
in which $s^2_{12}$ is given by Eq.~(\ref{eq:solar1223}). As in the single rotation cases, there is an interplay between the allowed values of $s^2_{12}$ and the predicted range of $\cos\delta$ as given in Eq.~(\ref{eq:cosdel1223f}).  We can go one step further and obtain expressions for $\cos\delta$ and $\sin\delta$ in terms of the model parameters, but these results are rather cumbersome so we will refrain from presenting them here.
\begin{figure}[H]
  \begin{subfigure}[b]{0.475\textwidth}
  \caption{$\cos \delta$($\delta_{12}^e{},\delta_{23}^{e}$) }
    \includegraphics[width=\textwidth]{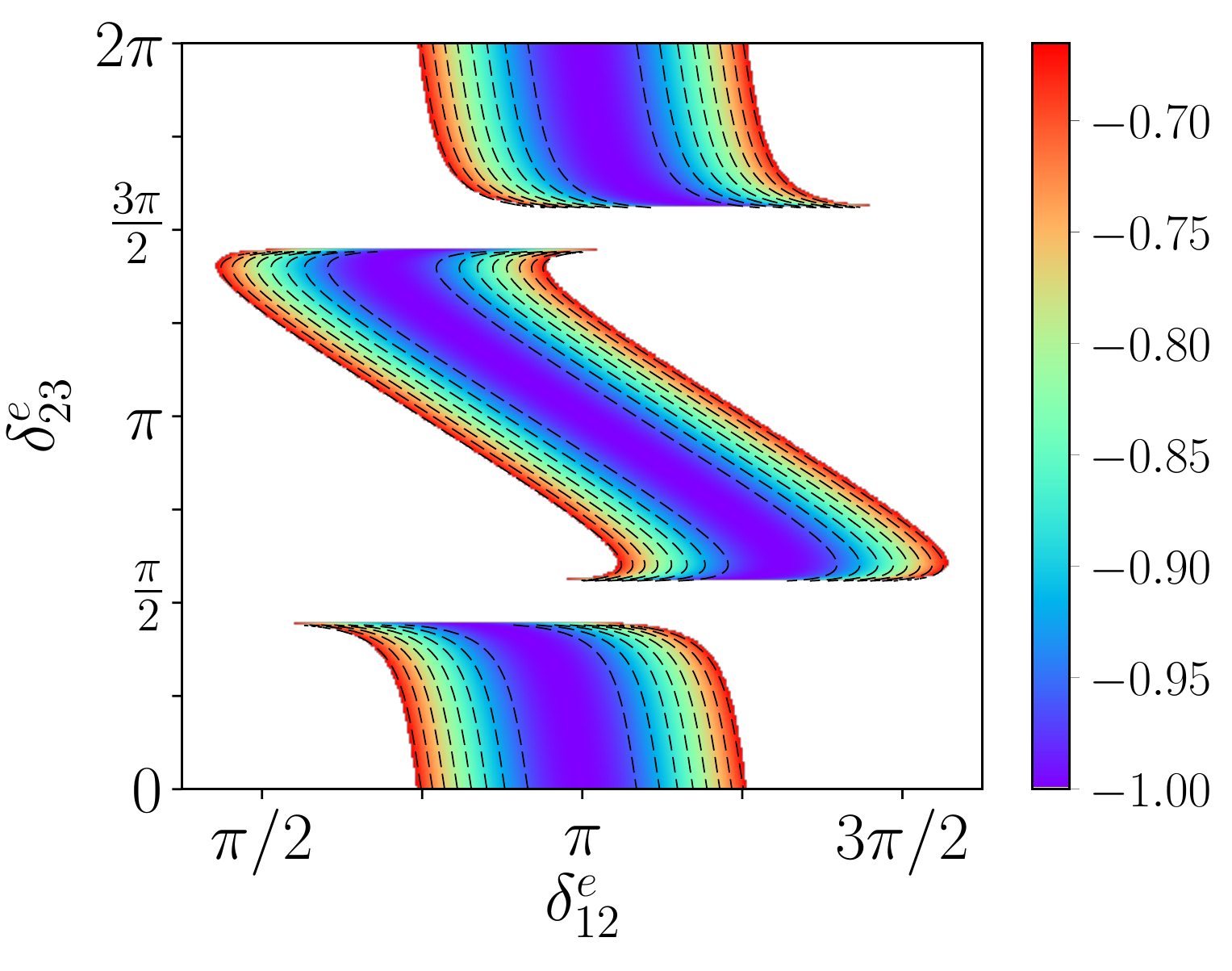}
  \end{subfigure}
  \hfill
  \begin{subfigure}[b]{0.475\textwidth}
  \caption{$\sin \delta$($\delta_{12}^e{},\delta_{23}^{e}$) }
    \includegraphics[width=\textwidth]{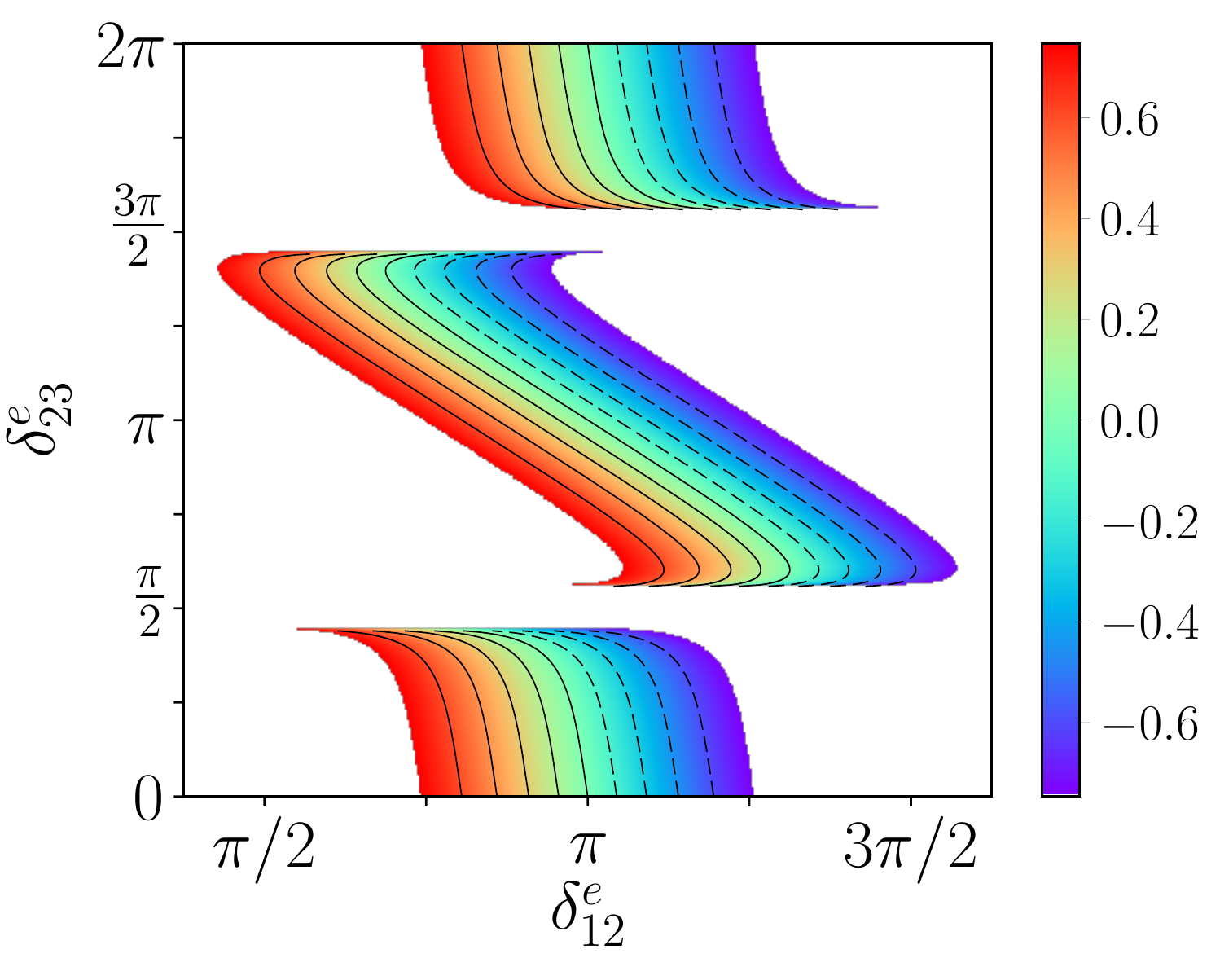}
  \end{subfigure}
  \caption{Distributions of $\cos \delta$ and $\sin\delta$ as a
  function of $\delta_{12}^{e}$ and $\delta_{23}^{e}$ for BM mixing, for the case that $U_{e}=U^{e}_{23}(\theta_{23}^{e},\delta_{23}^{e})
U^{e}_{12}(\theta_{12}^{e},\delta_{12}^{e})$.}
\label{fig:1223bm}
\end{figure}
Starting our numerical analysis of this class of perturbations with the case of BM mixing, we present in Figure~\ref{fig:1223bm} a set of plots for $\cos\delta$ and $\sin\delta$ as a function of the model parameters $\delta^e_{12}$ and $\delta^e_{23}$.  These figures show a wide range for $\delta^e_{23}$, other than the disallowed regions at/near $\pi/2$ (mod $\pi$), which arise from the $z$ constraint discussed above. In contrast, the allowed range of $\delta^e_{12}$ values is more focused (note the differences in the axis labeling).  This is as expected since it is this parameter that is most critical in obtaining an experimentally allowed range of the solar angle.  As we will see throughout this subsection, the shape of these allowed bands in this two-dimensional parameter space is characteristic within this scenario of our approach in which the theoretical mixing angle parameters have been fixed. 

\begin{figure}[H]
  \begin{subfigure}[b]{0.475\textwidth}
  \caption{$\cos \delta$($\delta_{12}^e{},\delta_{23}^{e}$) }
    \includegraphics[width=\textwidth]{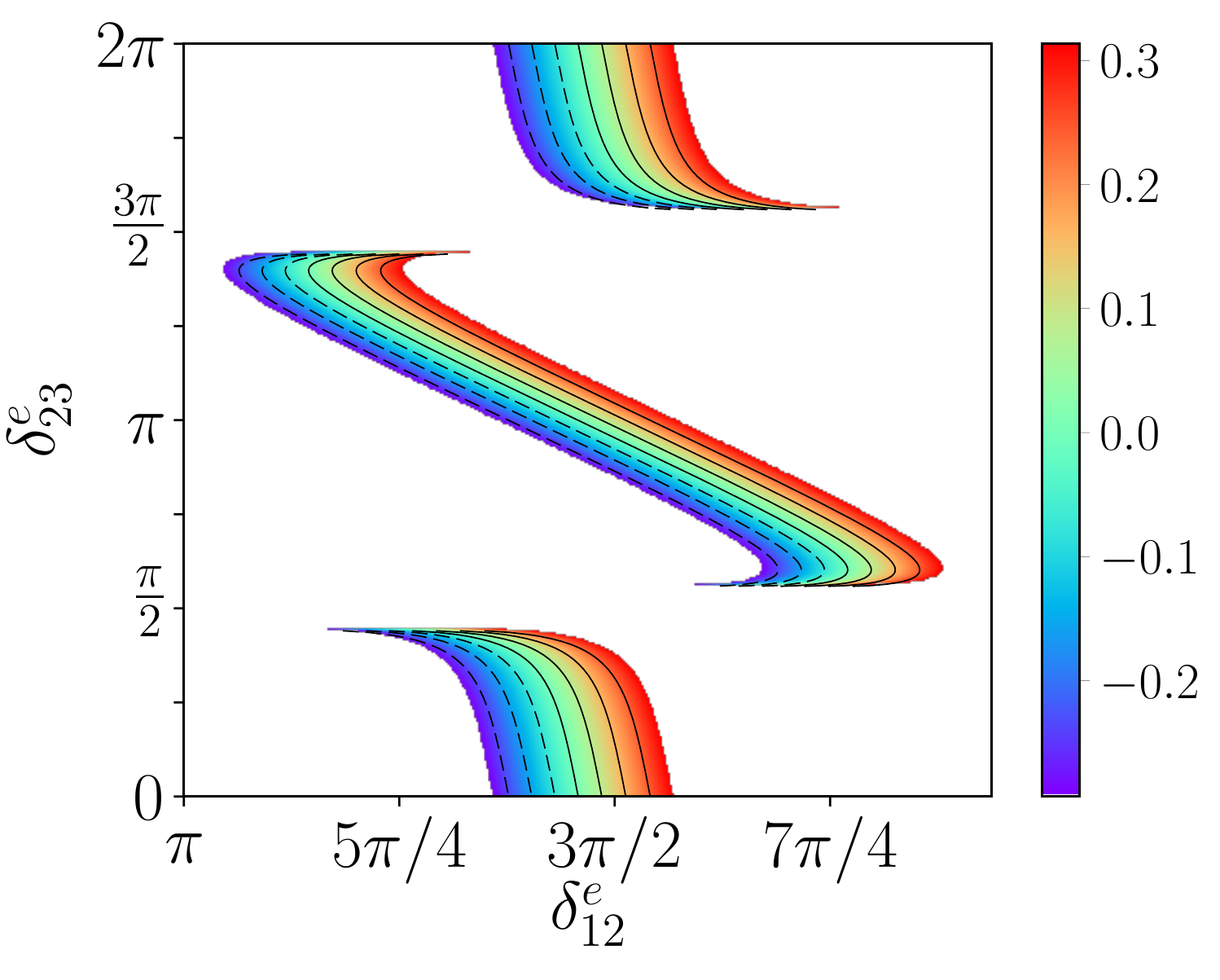}
  \end{subfigure}
  \hfill
  \begin{subfigure}[b]{0.475\textwidth}
  \caption{$\sin \delta$($\delta_{12}^e{},\delta_{23}^{e}$) }
    \includegraphics[width=\textwidth]{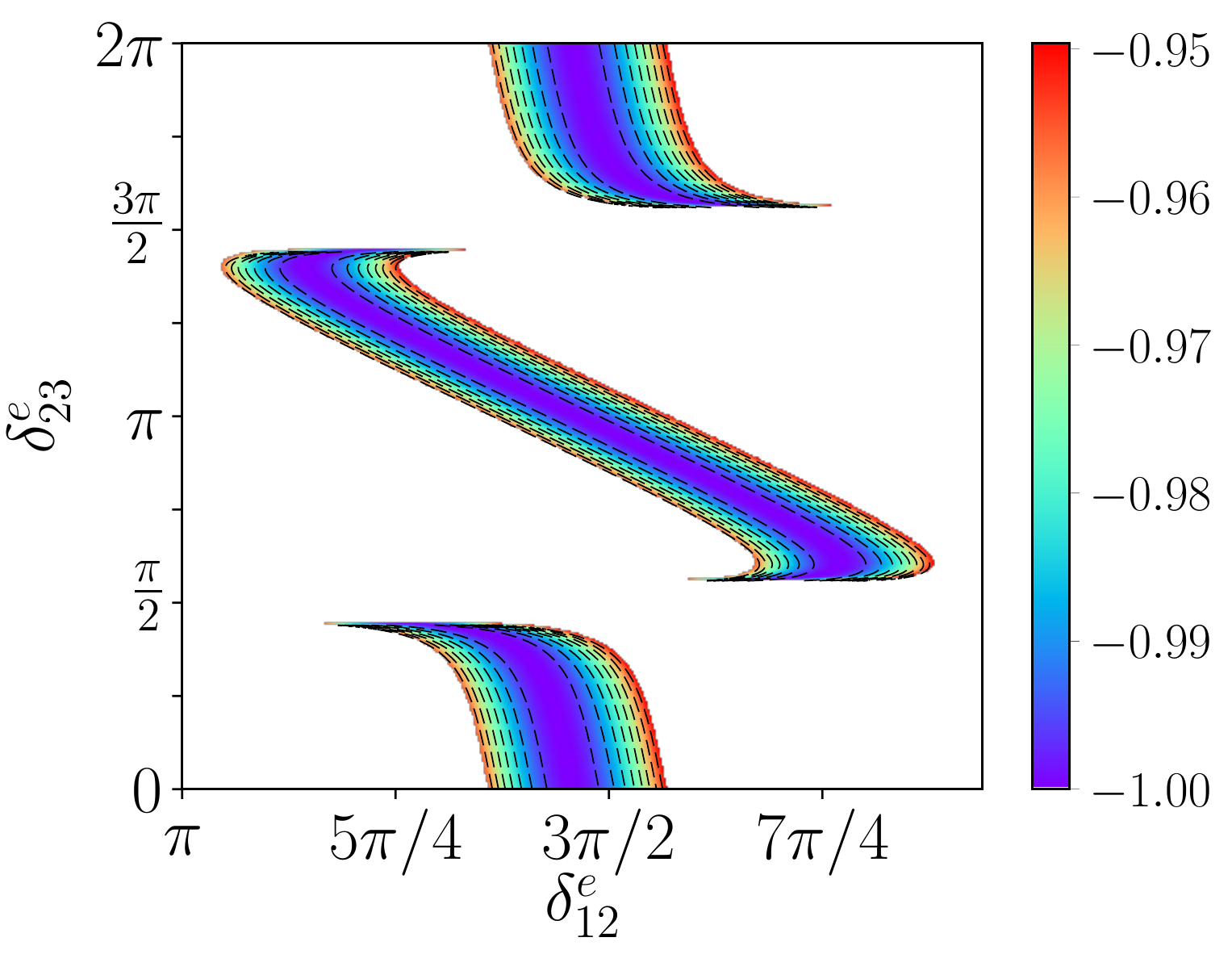}
  \end{subfigure}
  \caption{Distributions of $\cos \delta$ and $\sin \delta$ as a
  function of $\delta_{12}^{e}$ and $\delta_{23}^{e}$ for TBM mixing, for the case that $U_{e}=U^{e}_{23}(\theta_{23}^{e},\delta_{23}^{e})
U^{e}_{12}(\theta_{12}^{e},\delta_{12}^{e})$.}
\label{fig:1223tbm}
\end{figure}

\begin{figure}[H]
  \begin{subfigure}[b]{0.475\textwidth}
  \caption{$\cos \delta$($\delta_{12}^e{},\delta_{23}^{e}$) }
    \includegraphics[width=\textwidth]{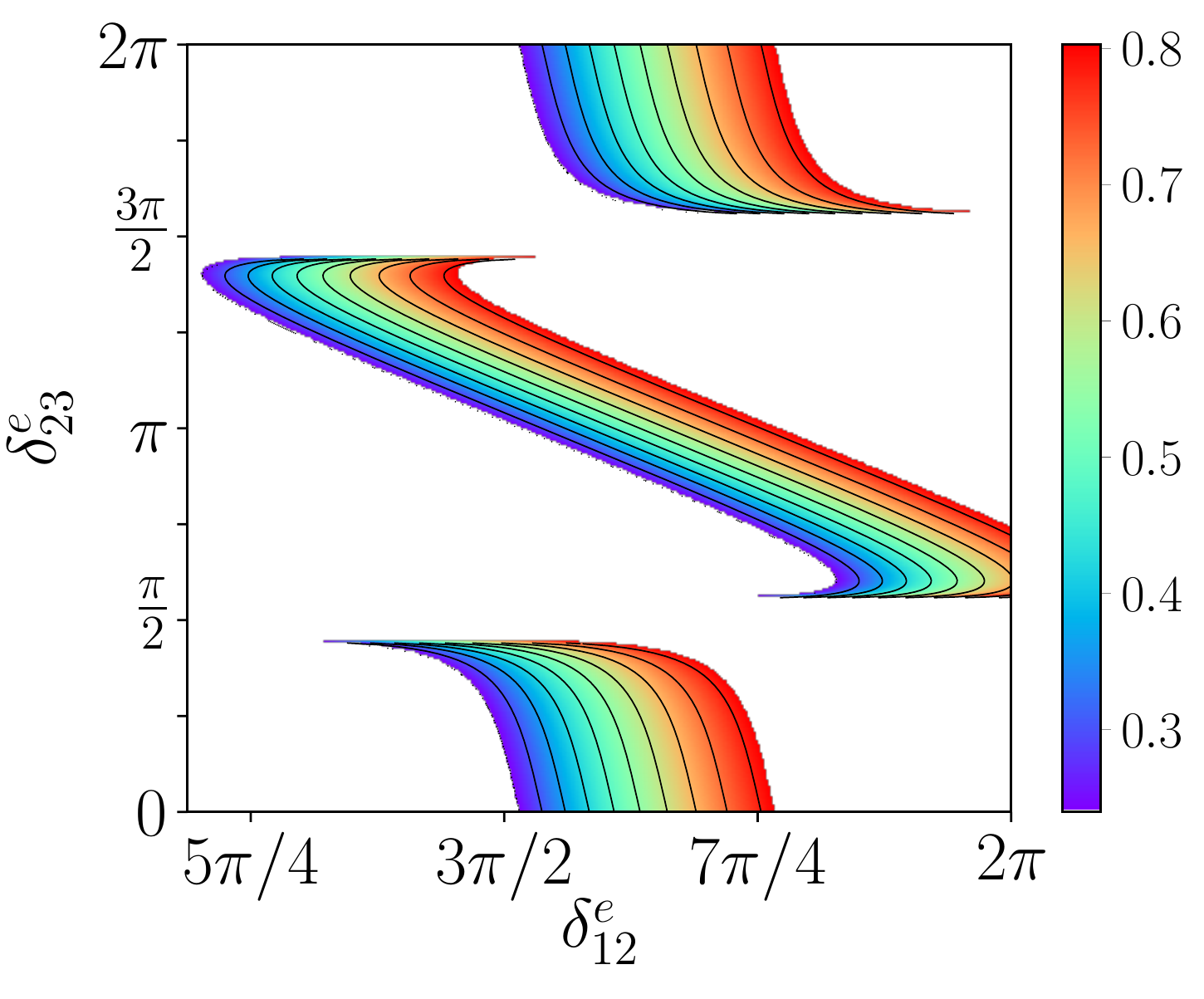}
  \end{subfigure}
  \hfill
  \begin{subfigure}[b]{0.475\textwidth}
  \caption{$\sin \delta$($\delta_{12}^e{},\delta_{23}^{e}$) }
    \includegraphics[width=\textwidth]{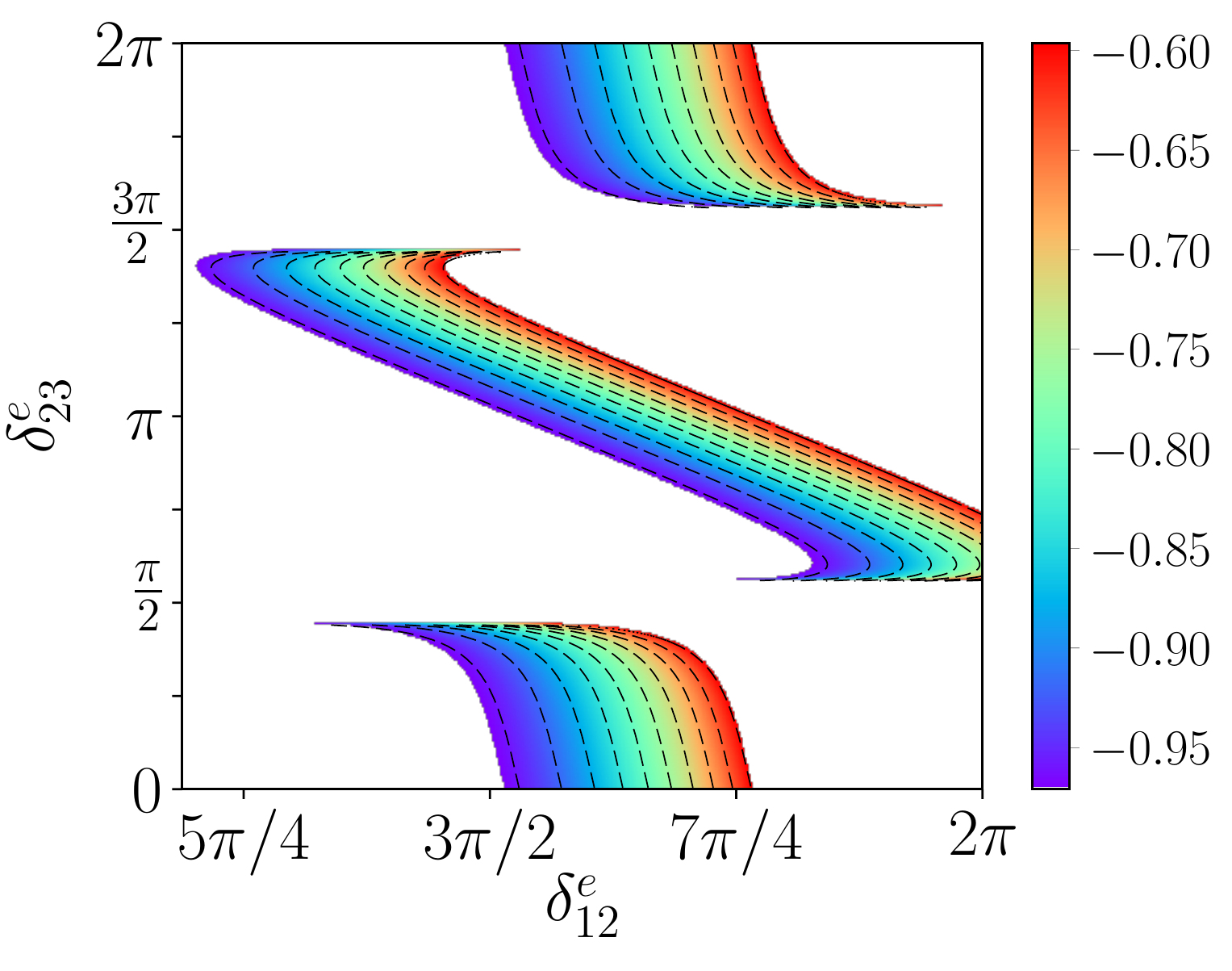}
  \end{subfigure}
  \caption{Distributions of $\cos \delta$ and $\sin \delta$ as a
  function of $\delta_{12}^{e}$ and $\delta_{23}^{e}$ for HEX mixing, for the case that $U_{e}=U^{e}_{23}(\theta_{23}^{e},\delta_{23}^{e})
U^{e}_{12}(\theta_{12}^{e},\delta_{12}^{e})$.}
  \label{fig:1223hex}
\end{figure}

\begin{figure}[H]
  \begin{subfigure}[b]{0.475\textwidth}
  \caption{$\cos \delta$($\delta_{12}^e{},\delta_{23}^{e}$) }
    \includegraphics[width=\textwidth]{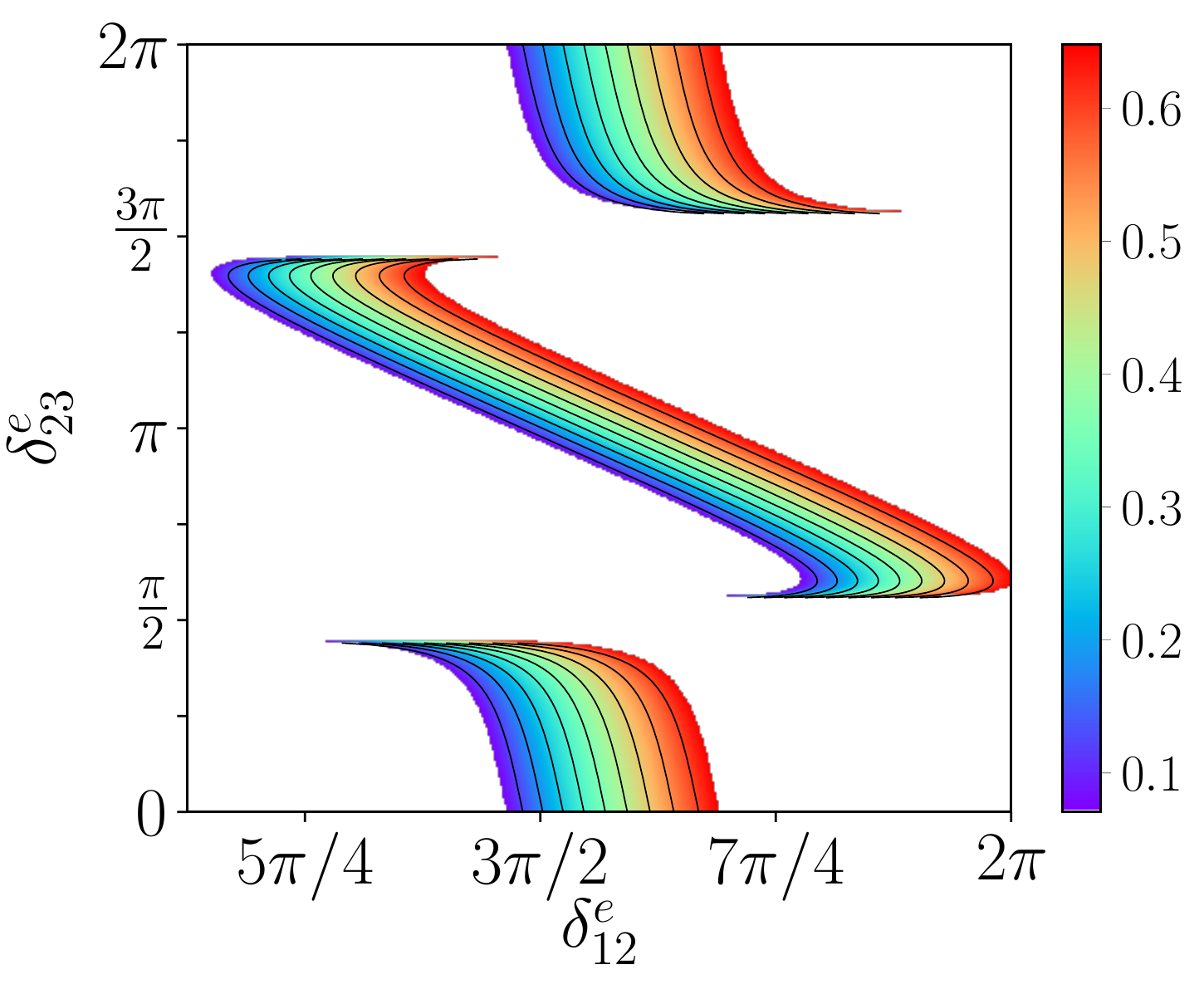}
  \end{subfigure}
  \hfill
  \begin{subfigure}[b]{0.475\textwidth}
  \caption{$\sin \delta$($\delta_{12}^e{},\delta_{23}^{e}$) }
    \includegraphics[width=\textwidth]{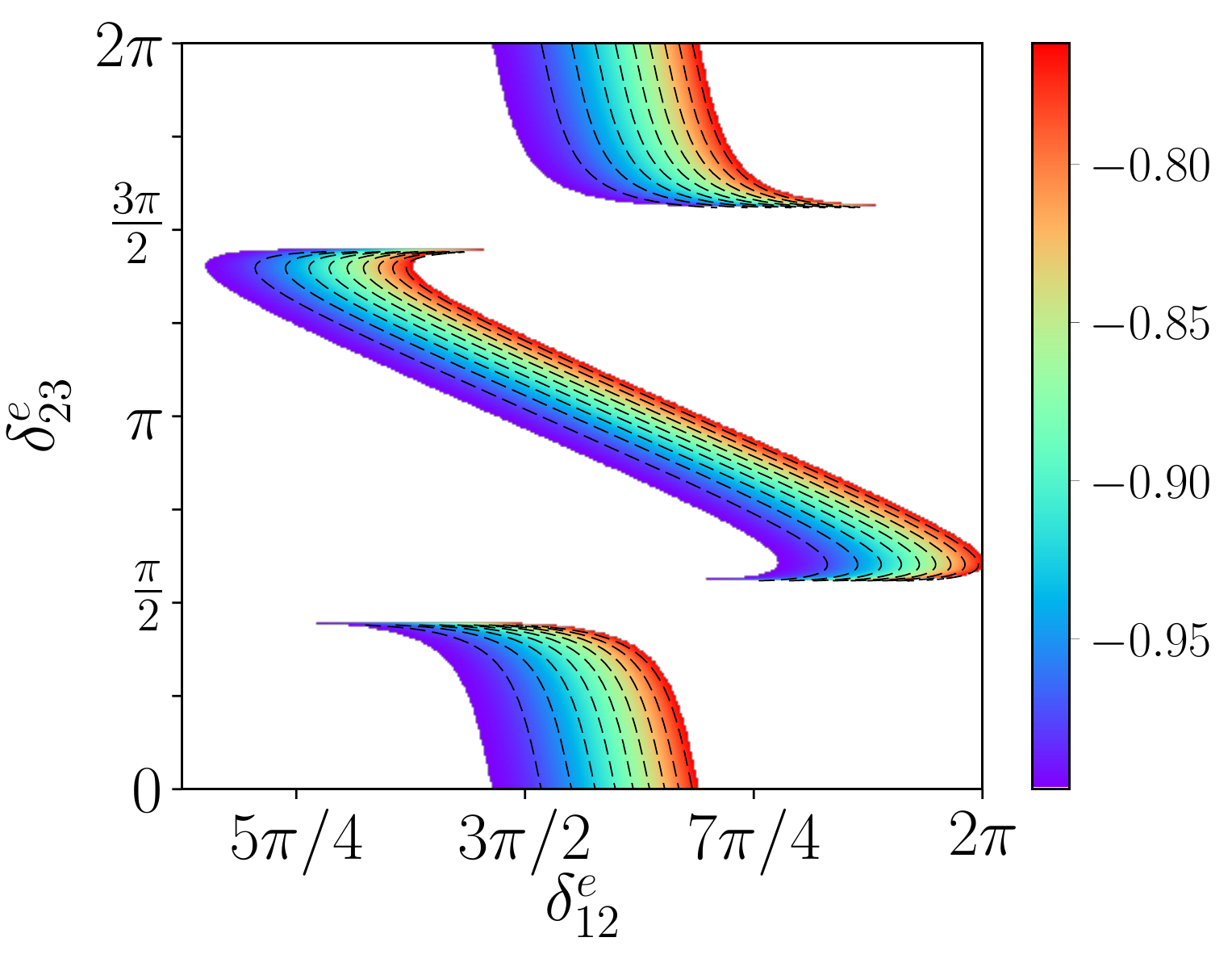}
  \end{subfigure}
  \caption{The predictions for $\cos \delta$ and $\sin \delta$ as a
  function of $\delta_{12}^{e}$ and $\delta_{23}^{e}$ for GR1 mixing, for the case that $U_{e}=U^{e}_{23}(\theta_{23}^{e},\delta_{23}^{e})
U^{e}_{12}(\theta_{12}^{e},\delta_{12}^{e})$.}
  \label{fig:1223gr1}
\end{figure}

\begin{figure}[H]
  \begin{subfigure}[b]{0.475\textwidth}
  \caption{$\cos \delta$($\delta_{12}^e{},\delta_{23}^{e}$) }
    \includegraphics[width=\textwidth]{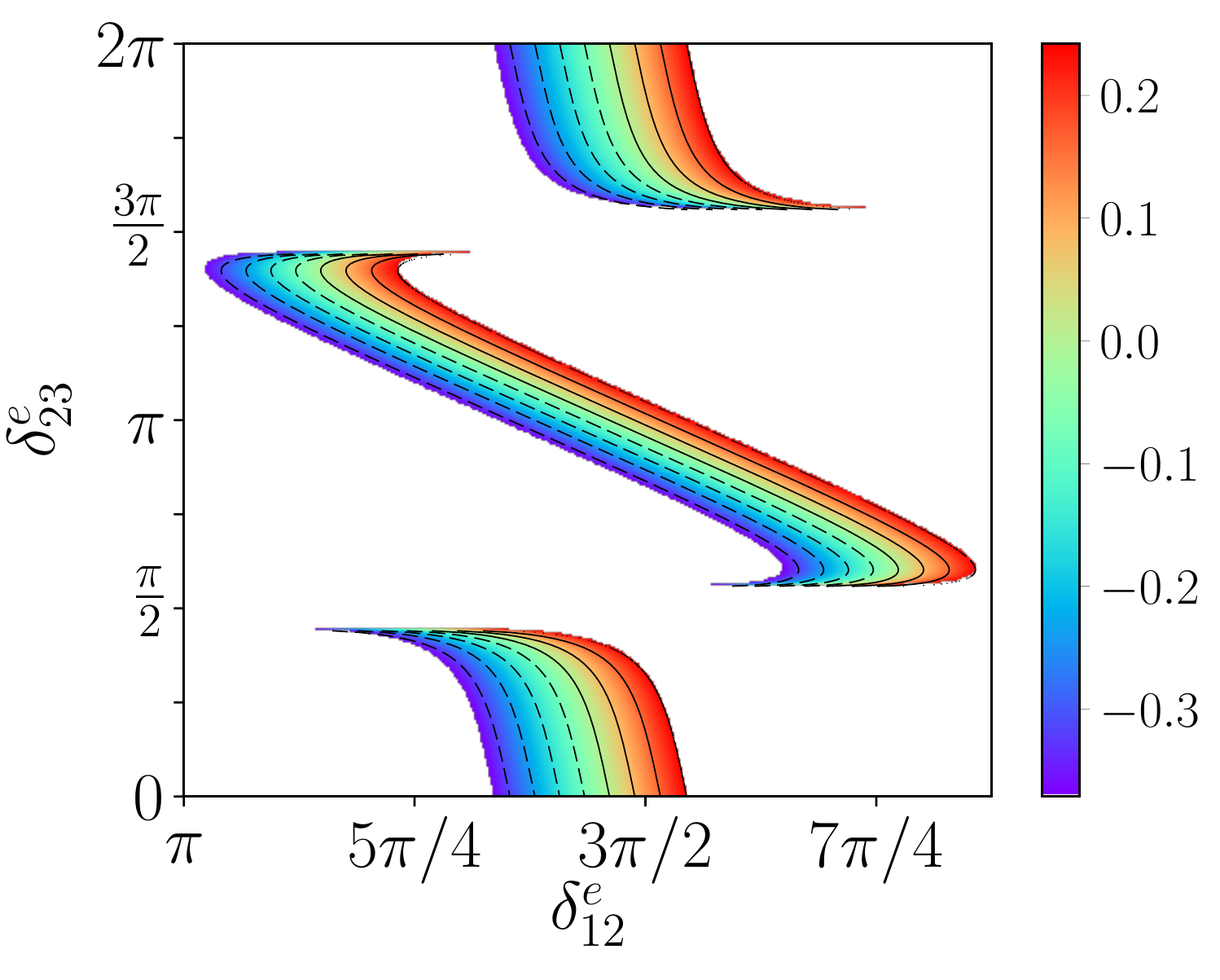}
  \end{subfigure}
  \hfill
  \begin{subfigure}[b]{0.475\textwidth}
  \caption{$\sin \delta$($\delta_{12}^e{},\delta_{23}^{e}$) }
    \includegraphics[width=\textwidth]{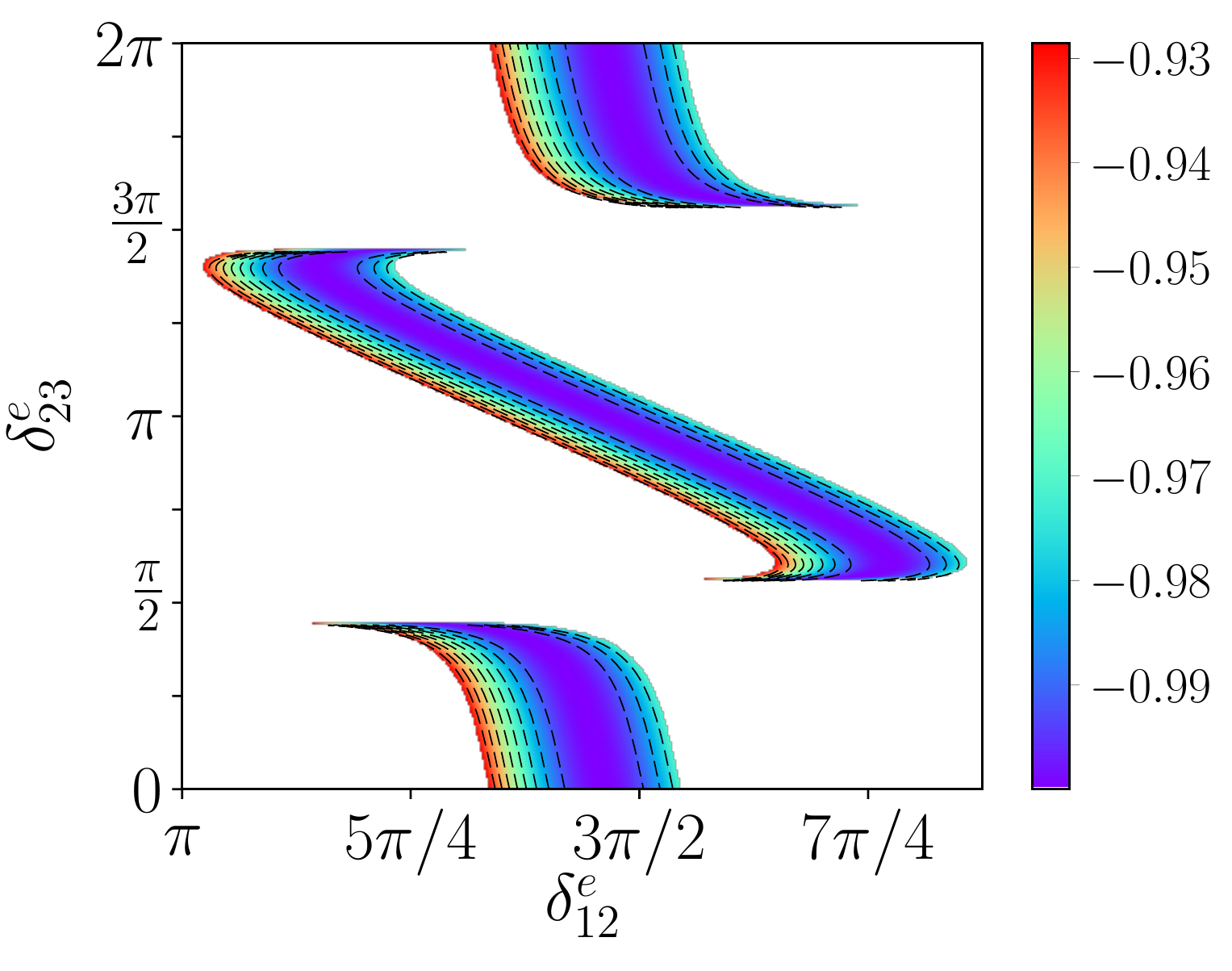}
  \end{subfigure}
  \caption{The predictions for $\cos \delta$ and $\sin \delta$ as a
  function of $\delta_{12}^{e}$ and $\delta_{23}^{e}$ for GR2 mixing, for the case that $U_{e}=U^{e}_{23}(\theta_{23}^{e},\delta_{23}^{e})
U^{e}_{12}(\theta_{12}^{e},\delta_{12}^{e})$.}
  \label{fig:1223gr2}
\end{figure}

The remaining set scenarios share many similar features with the BM mixing case, with the main difference being as usual that the solar mixing angle constraint is generally easier to satisfy due to the smaller values of $y$, which in turn affects the range of predictions for $\cos\delta$ and $\sin\delta$.  We show these results in Figures~\ref{fig:1223tbm}--\ref{fig:1223gr2}. Here we point out in particular the very tight range of predicted values for $\sin\delta$ in the TBM and GR2 mixing scenarios, whereas the HEX and GR1 cases allow for a wider $\sin\delta$ range.


\subsubsection*{Case 4: Two rotations in the 1--3 and 2--3 sectors ($U_{e}=U^{e}_{23}(\theta_{23}^{e},\delta_{23}^{e})U^{e}_{13}(\theta_{13}^{e},\delta_{13}^{e})$)} 
For the situation in which $U_{e}=U^{e}_{23}(\theta_{23}^{e},\delta_{23}^{e})U^{e}_{13}(\theta_{13}^{e},\delta_{13}^{e})$, we find many similarities with the just-discussed case in which $U_{e}=U^{e}_{23}(\theta_{23}^{e},\delta_{23}^{e})U^{e}_{12}(\theta_{12}^{e},\delta_{12}^{e})$, just as we found many similarities between single rotations in the $1-2$ and $1-3$ sectors.  More precisely, we have again that
\begin{equation}
\sin^2(\tilde{\theta}_{23}) = \frac{1}{2} \left (1-\cos(\delta^e_{23})s^{\prime e}_{23} \right )= \frac{1}{2} (1-z),
\end{equation}
where $z$ is defined as before to be $z=\cos(\delta^e_{23})s^{\prime e}_{23}$.  From Eq.~(\ref{eq:sin131323}), $s^2_{13}$ is thus given by
\begin{equation}
s^2_{13} = \frac{(s^e_{13})^2}{2}(1+z).
\end{equation}
In this case, we will fix $\theta^e_{13}=2.91$, and thus together with fixing $s^2_{13}$ to its central value, we obtain that  $z=-0.1890$ (i.e., it has the same magnitude and opposite sign as in the previous set of perturbations), and thus  $\sin^2(\tilde{\theta}_{23})=0.5910$.  As before, we use $z$ to eliminate $\theta^e_{23}$ for $\delta^e_{23}$, and we will again see the disallowed regions of $\delta^e_{23}$ centered at $\pi/2$ (mod $\pi$) for which it is impossible to satisfy the $z$ constraint.  From Eq.~(\ref{eq:sin231323}), $s^2_{23}$ now takes the form
\begin{equation}
s^2_{23} = \frac{1-z}{2(1-s^2_{13})}.
\end{equation}
As before, $s^2_{23}$ thus takes on a fixed value in our numerical analysis; this value is $s^2_{23}=0.6040$.  As in the single rotation case, for the $1-3$ perturbations $s^2_{23}>1/2$ for this set of parameters, whereas $s^2_{23}<1/2$ for the analogous parameters for the $1-2$ perturbations.

Once again in analogy with the previous subsection, it is the constraint on $s^2_{12}$ and $y$ and its effects on the two phase parameters that yield the allowed values of $\cos\delta$ for this particular choice of the $\theta^e_{jk}$.  The result for $s^2_{12}$ is given from
Eq.~(\ref{eq:sin121323}) to be
\begin{equation}\label{eq:solar1323}
s^2_{12} = \frac{(c^e_{13})^2y+\frac{1}{2}(1-y)(1-z)(s^e_{13})^2-s^{\prime e}_{13}\sqrt{\frac{y(1-y)}{2}}(s^e_{23}\cos(\delta^{\prime\prime})-c^e_{23}\cos(\delta^e_{13}))}{1-(s^e_{13})^2(1+z)/2},
 \end{equation} 
in which $\delta^{\prime\prime}=\delta^e_{23}-\delta^e_{13}$, and again $\theta^e_{23}$ is determined by the $z$ constraint. Therefore, as before, the solar mixing angle constraint fixes a preferred range for the phase angle $\delta^e_{12}$ as a function of the other model parameters.  From Eq.~(\ref{eq:sumruleorig13}) (or equivalently Eq.~(\ref{eq:cddouble1323form1}) or (\ref{eq:cddouble1323form2})),
\begin{equation}\label{eq:cosdel1323f}
\cos\delta = \frac{y(1-s^2_{13})(1+z)-(1+z)s^2_{12}-s^2_{13}(1-z)+s^2_{12}s^2_{13}(3-z)}{s^\prime_{12}s_{13}\sqrt{(1-z)(1+z-2s^2_{13})}},
\end{equation}
in which $s^2_{12}$ is given by Eq.~(\ref{eq:solar1323}). Hence, we see in complete analogy with Eq.~(\ref{eq:cosdel1223f}), we have an expression for $\cos\delta$ that depends on $s^2_{13}$, $s^2_{12}$, $y$, and $z$.  We note the sign flip in $z$ between Eq.~(\ref{eq:cosdel1223f}) and Eq.~(\ref{eq:cosdel1323f}), as well as in the solar angle expressions Eq.~(\ref{eq:solar1223}) and Eq.~(\ref{eq:solar1323}), but we also note that the numerical constraints force $z\rightarrow -z$ between the two cases.  


\begin{figure}[H]
  \begin{subfigure}[b]{0.475\textwidth}
  \caption{$\cos \delta$($\delta_{13}^{e}$, $\delta_{23}^{e}$) }
    \includegraphics[width=\textwidth]{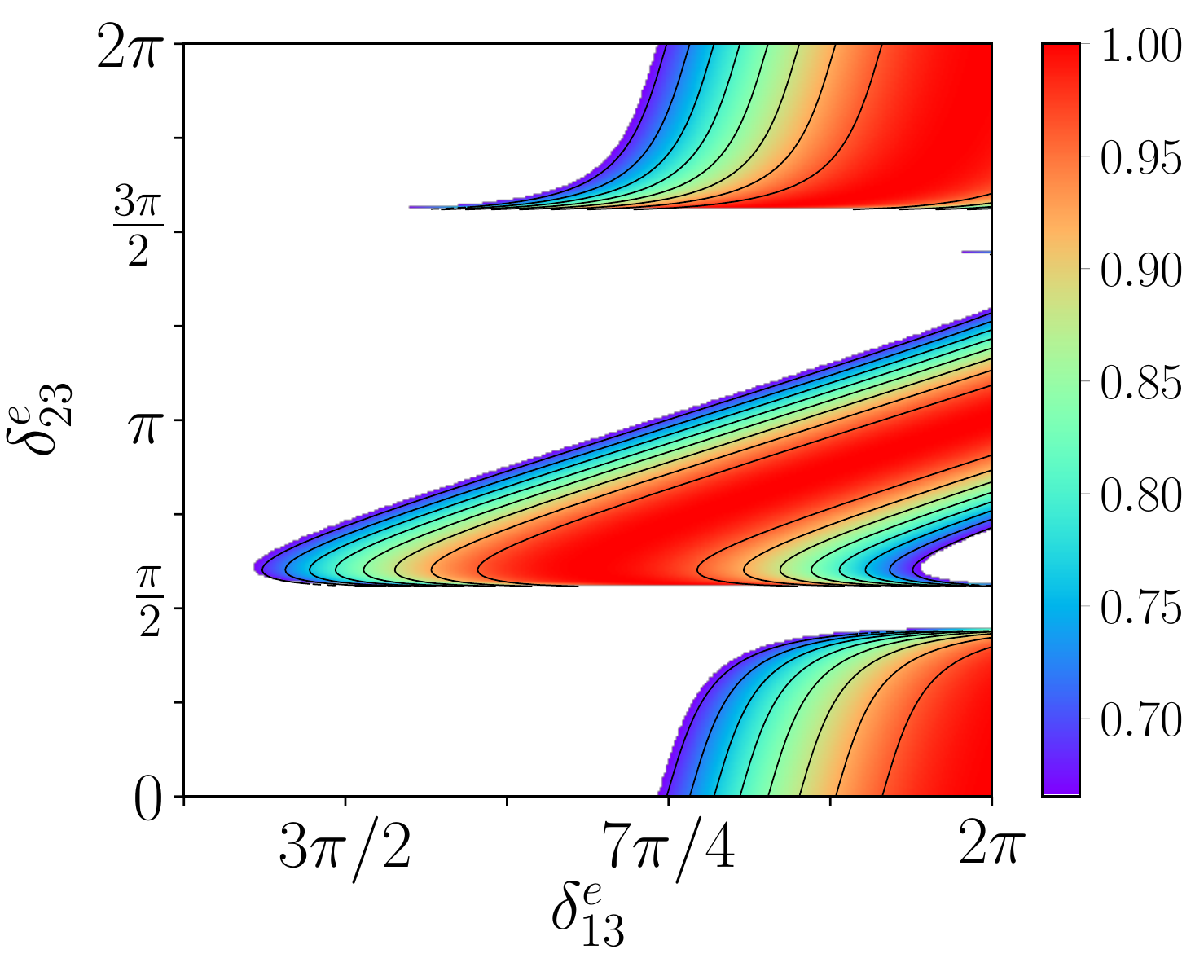}
  \end{subfigure}
  \hfill
  \begin{subfigure}[b]{0.475\textwidth}
  \caption{$\sin \delta$($\delta_{13}^{e}$, $\delta_{23}^{e}$) }
    \includegraphics[width=\textwidth]{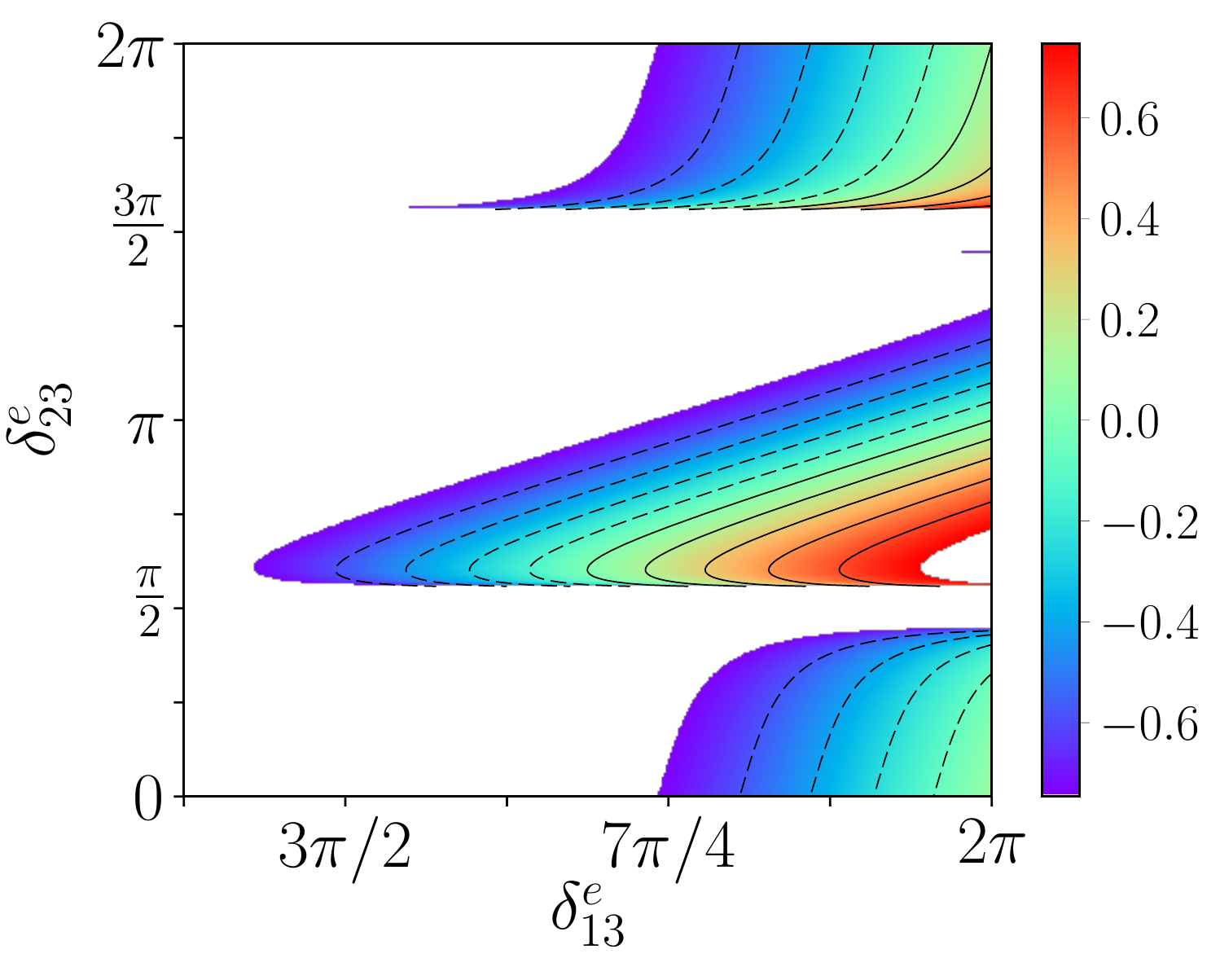}
  \end{subfigure}
  \caption{Predictions for $\cos \delta$ and $\sin \delta$ as a
  function of $\delta_{13}^{e}$ and $\delta_{23}^{e}$ for BM mixing,  for the case that $U_{e}=U^{e}_{23}(\theta_{23}^{e},\delta_{23}^{e})
U^{e}_{13}(\theta_{13}^{e},\delta_{13}^{e})$.}
  \label{fig:1323bm}
\end{figure}
\begin{figure}[H]
  \begin{subfigure}[b]{0.475\textwidth}
  \caption{$\cos \delta$($\delta_{13}^{e}$, $\delta_{23}^{e}$) }
    \includegraphics[width=\textwidth]{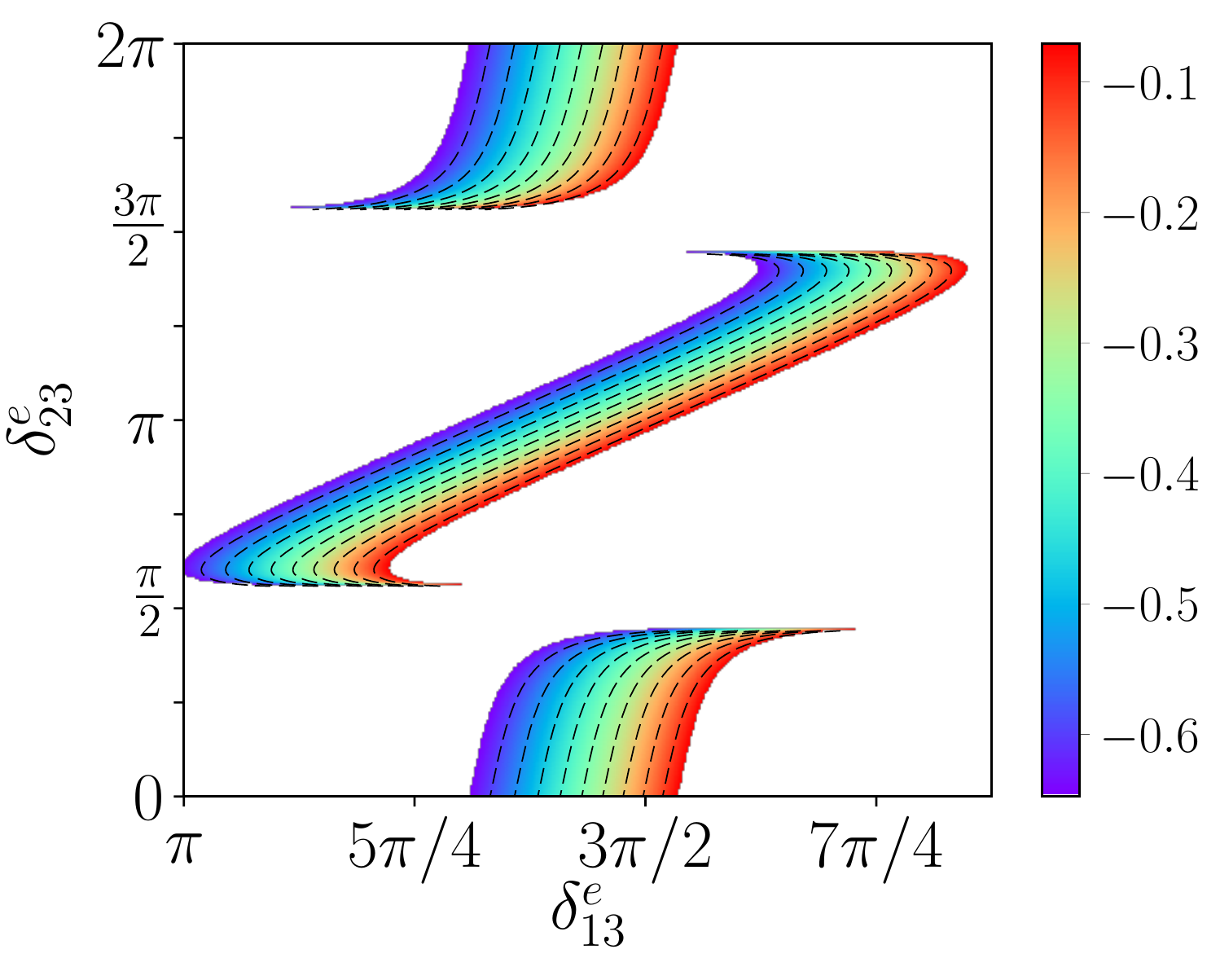}
  \end{subfigure}
  \hfill
  \begin{subfigure}[b]{0.475\textwidth}
  \caption{$\sin \delta$($\delta_{13}^{e}$, $\delta_{23}^{e}$) }
    \includegraphics[width=\textwidth]{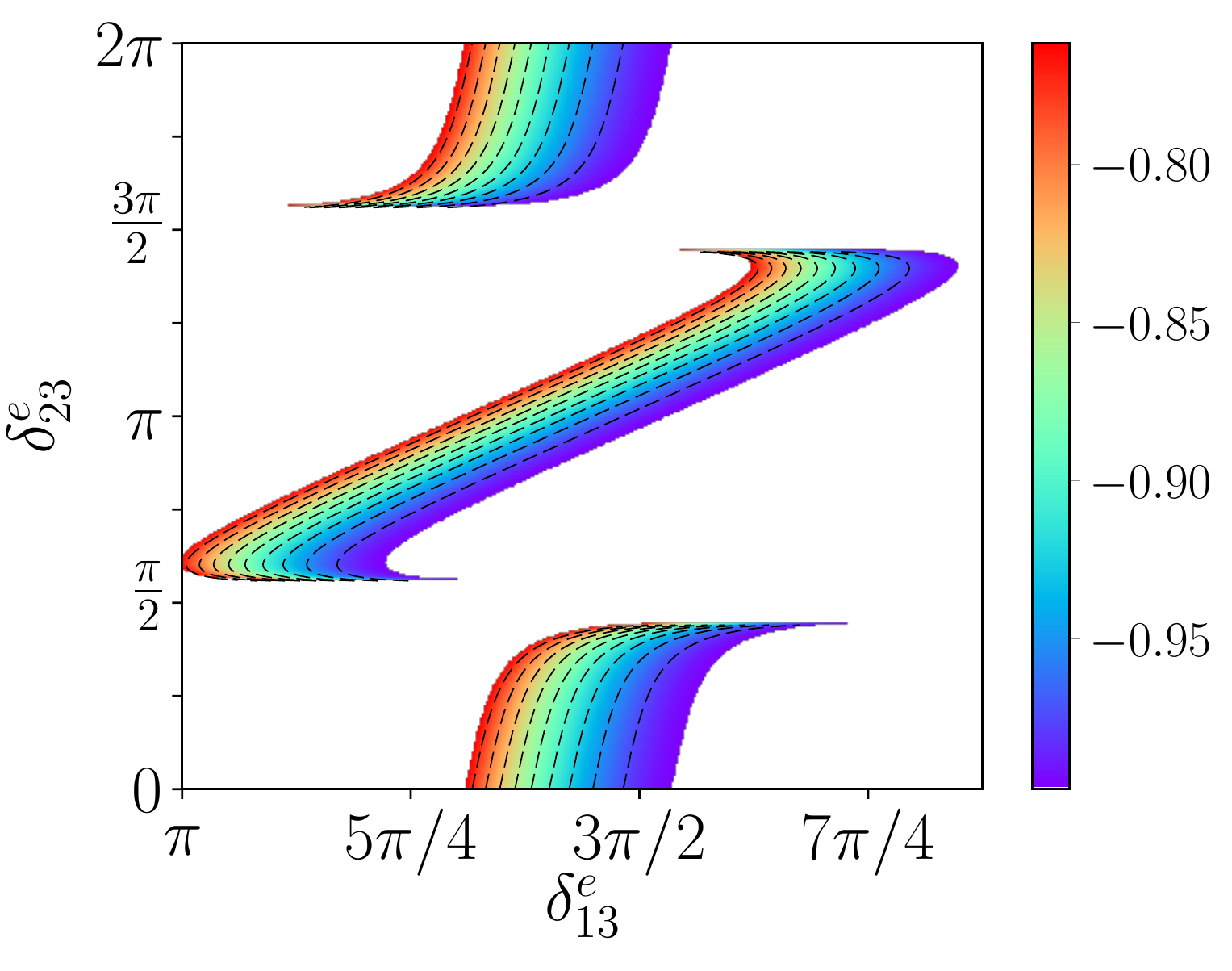}
  \end{subfigure}
  \caption{The predictions for $\cos \delta$ and $\sin \delta$ as a
  function of $\delta_{13}^{e}$ and $\delta_{23}^{e}$ for GR1 mixing, ,  for the case that $U_{e}=U^{e}_{23}(\theta_{23}^{e},\delta_{23}^{e})
U^{e}_{13}(\theta_{13}^{e},\delta_{13}^{e})$.}
  \label{fig:1323gr1}
\end{figure}
\begin{figure}[H]
  \begin{subfigure}[b]{0.475\textwidth}
  \caption{$\cos \delta$($\delta_{13}^{e}$, $\delta_{23}^{e}$) }
    \includegraphics[width=\textwidth]{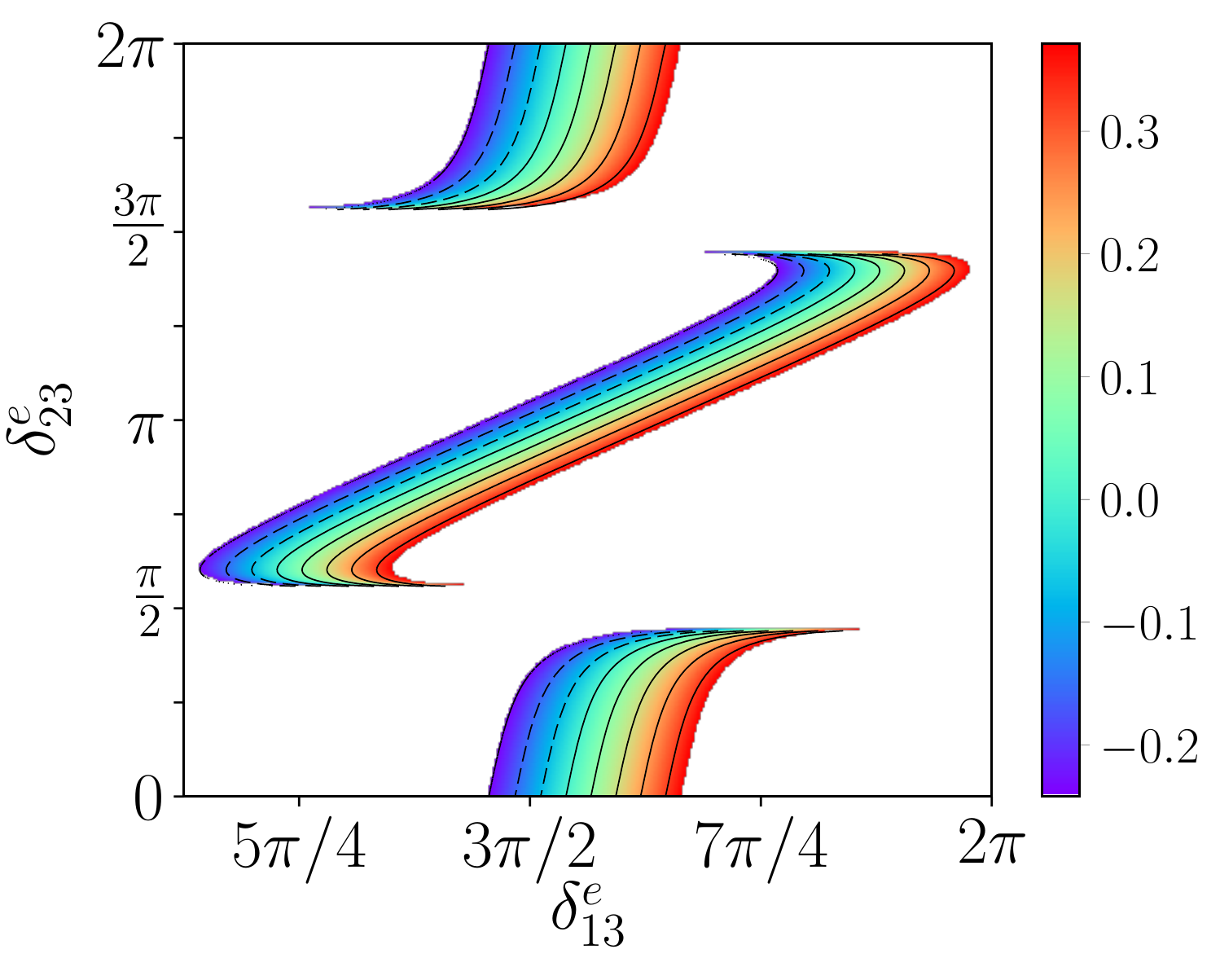}
  \end{subfigure}
  \hfill
  \begin{subfigure}[b]{0.475\textwidth}
  \caption{$\sin \delta$($\delta_{13}^{e}$, $\delta_{23}^{e}$) }
    \includegraphics[width=\textwidth]{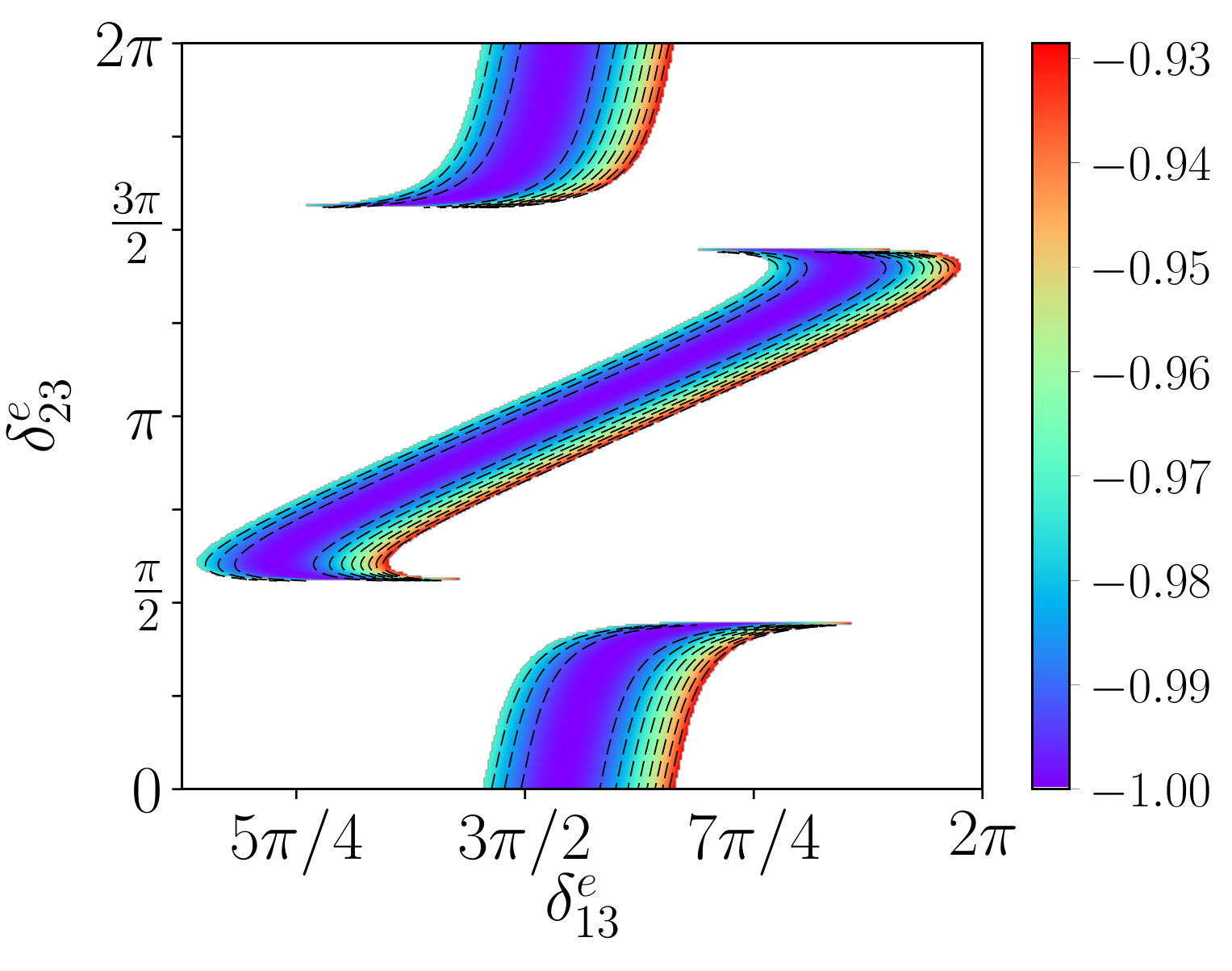}
  \end{subfigure}
  \caption{The predictions for $\cos \delta$ and $\sin \delta$ as a
  function of $\delta_{13}^{e}$ and $\delta_{23}^{e}$ for GR2 mixing,  for the case that $U_{e}=U^{e}_{23}(\theta_{23}^{e},\delta_{23}^{e})
U^{e}_{13}(\theta_{13}^{e},\delta_{13}^{e})$.}
  \label{fig:1323gr2}
\end{figure}

We begin as usual with the BM mixing case, as shown in Figure~\ref{fig:1323bm}.  As expected, there is a wide range for $\delta^e_{23}$ (other than the usual disallowed regions), and the allowed range of $\delta^e_{13}$ values is smaller due to the solar mixing angle constraint.  The characteristic allowed bands are similar to those of the previous subsection, again as expected.  Here the allowed $\delta^e_{13}$ region cuts off because of the artificial cutoff of our scan at $\delta^e_{13}=2\pi$, and the zoomed-in range of this parameter in the figure.   
\begin{figure}[H]
  \begin{subfigure}[b]{0.475\textwidth}
  \caption{$\cos \delta$($\delta_{13}^{e}$, $\delta_{23}^{e}$) }
    \includegraphics[width=\textwidth]{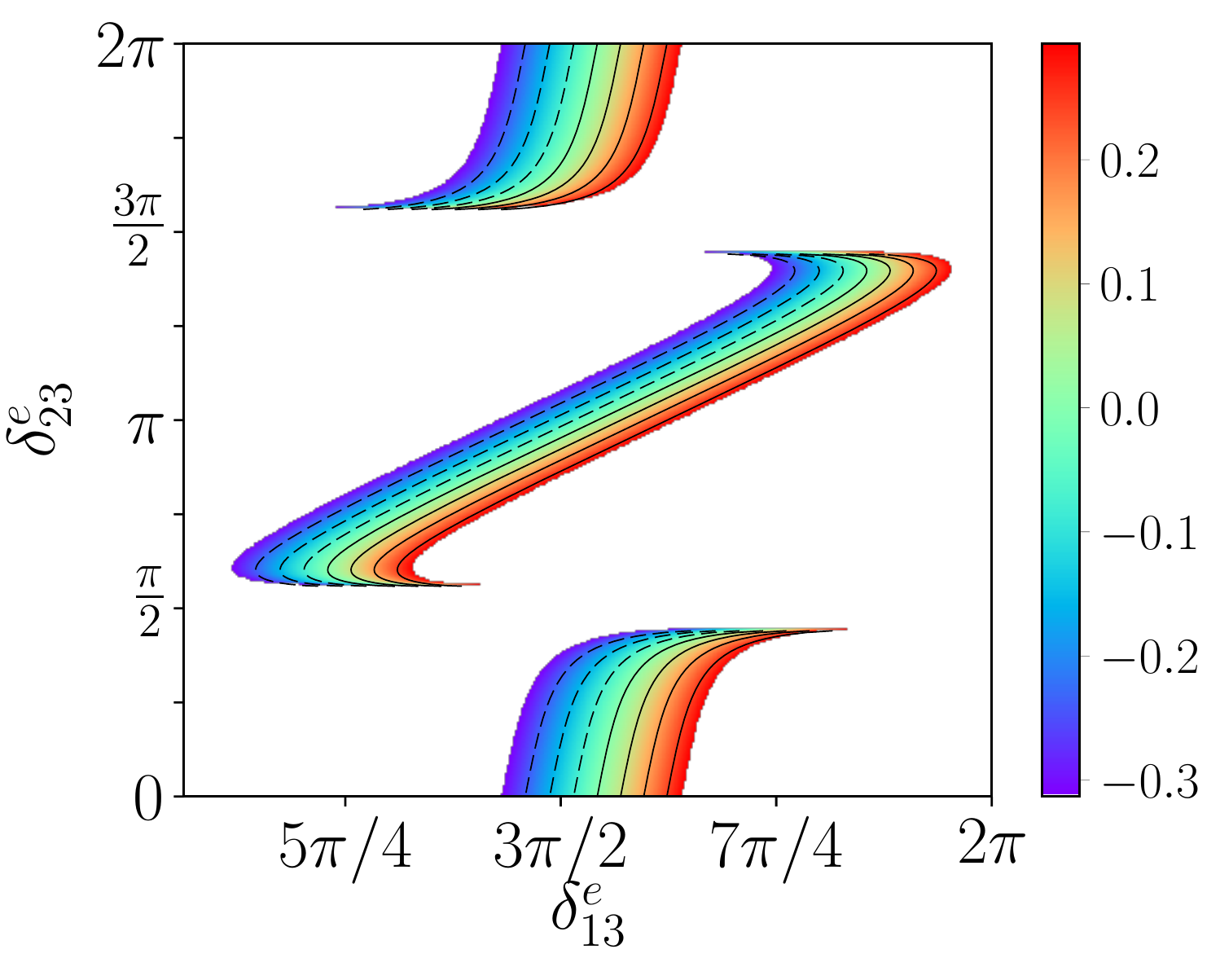}
  \end{subfigure}
  \hfill
  \begin{subfigure}[b]{0.475\textwidth}
  \caption{$\sin \delta$($\delta_{13}^{e}$, $\delta_{23}^{e}$) }
    \includegraphics[width=\textwidth]{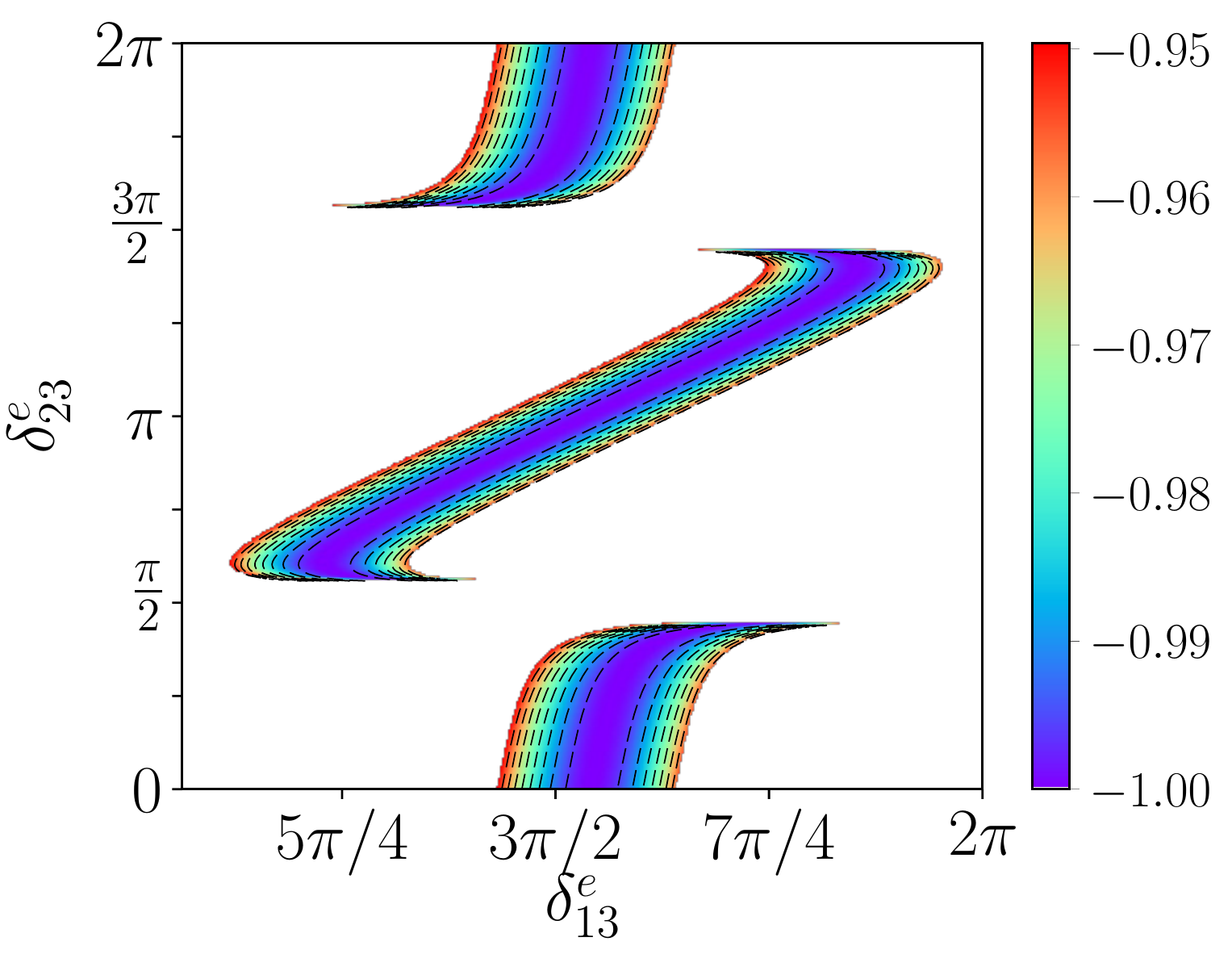}
  \end{subfigure}
  \caption{Predictions for $\cos \delta$ and $\sin \delta$ as a
  function of $\delta_{13}^{e}$ and $\delta_{23}^{e}$ for TBM mixing,  for the case that $U_{e}=U^{e}_{23}(\theta_{23}^{e},\delta_{23}^{e})
U^{e}_{13}(\theta_{13}^{e},\delta_{13}^{e})$.}
  \label{fig:1323tbm}
\end{figure}
\begin{figure}[H]
  \begin{subfigure}[b]{0.475\textwidth}
  \caption{$\cos \delta$($\delta_{13}^{e}$, $\delta_{23}^{e}$)}
    \includegraphics[width=\textwidth]{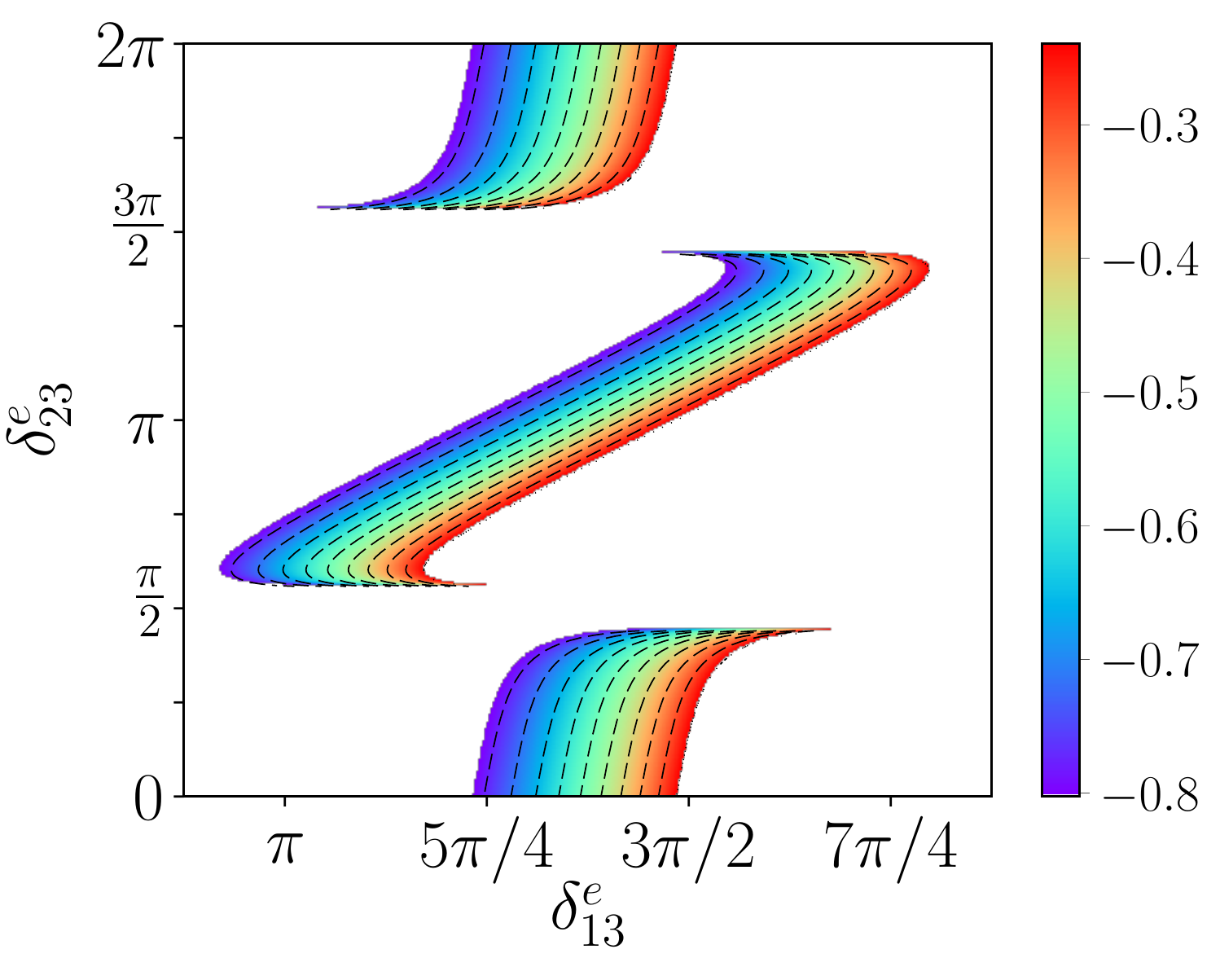}
  \end{subfigure}
  \hfill
  \begin{subfigure}[b]{0.475\textwidth}
  \caption{$\sin \delta$($\delta_{13}^{e}$, $\delta_{23}^{e}$) }
    \includegraphics[width=\textwidth]{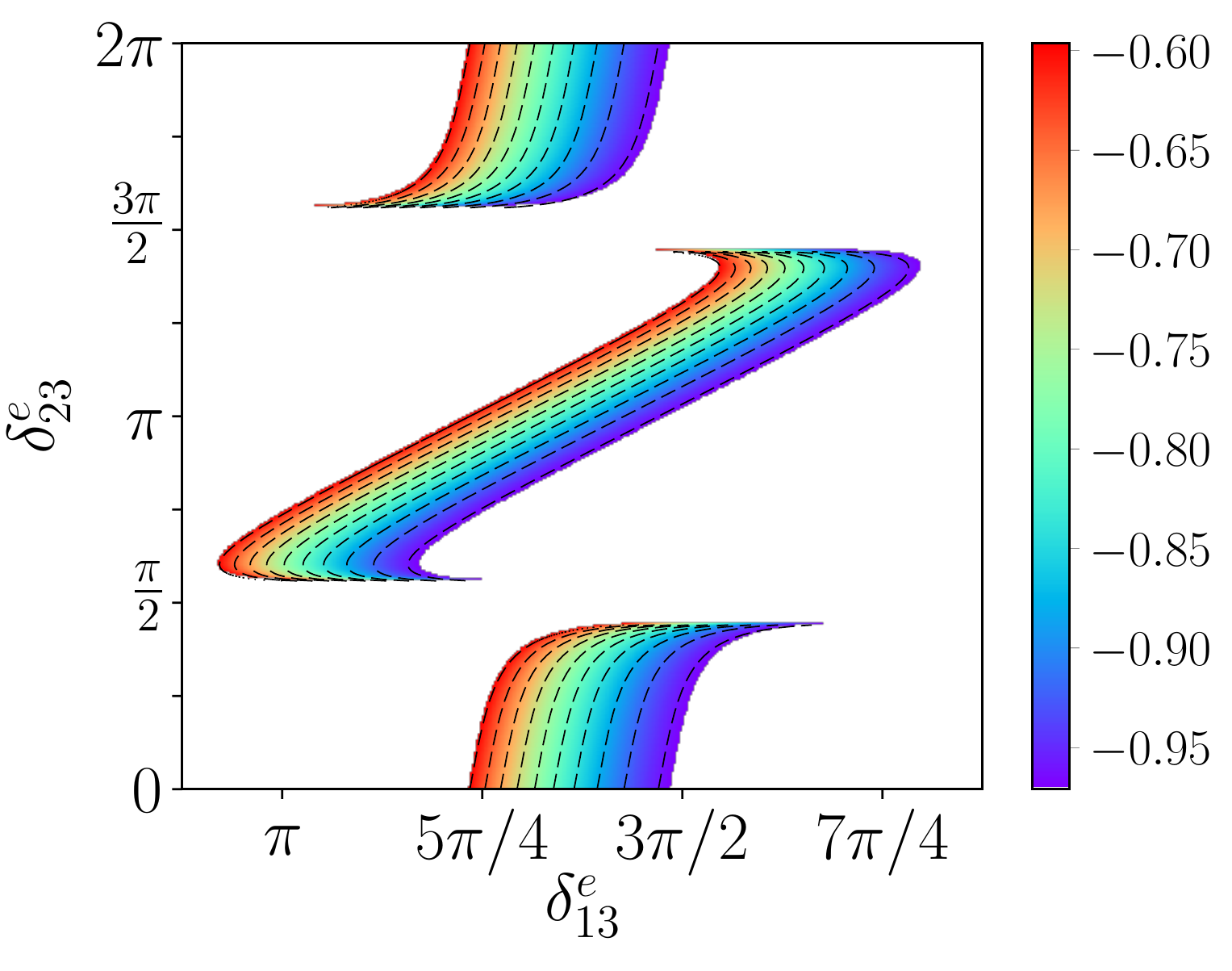}
  \end{subfigure}
  \caption{The predictions for $\cos \delta$ and $\sin \delta$ as a
  function of $\delta_{13}^{e}$ and $\delta_{23}^{e}$ for HEX mixing,  for the case that $U_{e}=U^{e}_{23}(\theta_{23}^{e},\delta_{23}^{e})
U^{e}_{13}(\theta_{13}^{e},\delta_{13}^{e})$.}
  \label{fig:1323hex}
\end{figure}
As expected, the remaining scenarios are not only similar in form to the BM case, but also to their counterparts in the previous subsection, just with shifts in the allowed values of $\delta^e_{13}$, and the corresponding values of $\cos\delta$ and $\sin\delta$. These results are displayed for completeness in Figures~\ref{fig:1323tbm}--\ref{fig:1323gr2}. Once again we see the very tight range of predicted values for $\sin\delta$ in the TBM and GR2 mixing scenarios, whereas the HEX and GR1 cases allow for a more broad range of values of $\sin\delta$.


\subsubsection*{Case 5: Two rotations in the 1--2 and 1--3 sectors ($U_{e}=U^{e}_{13}(\theta_{13}^{e},\delta_{13}^{e})U^{e}_{12}(\theta_{12}^{e},\delta_{12}^{e})$)} 
For the situation in which $U_{e}=U^{e}_{13}(\theta_{13}^{e},\delta_{13}^{e})U^{e}_{12}(\theta_{12}^{e},\delta_{12}^{e})$), we can see from the mixing angle predictions given in Eqs.~(\ref{eq:sin131213})--(\ref{eq:sin121213}) that this case is  slightly more involved than the double rotation cases discussed previously.  The reason is that now the relation between $\sin^2(\theta_{13})$ and $\sin^2(\theta_{23})$ is not as restrictive, such that fixing one of these two angles no longer fixes the other. There is an intricate set of constraints that must be satisfied, and hence the relation between the predictions for $\cos\delta$ (and thus $\sin\delta$) and the solar mixing angle constraint is not nearly as straightforward as it was in the previous subsections.  As we will see, this means that the allowed parameter space for the phases $\delta^e_{12}$ and $\delta^e_{13}$ no longer has the characteristic band structure of the previous cases, and instead there can be intriguing patterns of the allowed parameters.

To see this explicitly, let us begin with the form for $s^2_{13}$ as given in Eq.~(\ref{eq:sin131213}), for the scenarios at hand in which $\theta^\nu_{23}=\pi/4$:
\begin{equation}\label{eq:s13for1213}
s^2_{13} = \frac{1}{2} \left ((s^e_{12})^2+(c^e_{12})^2(s^e_{13})^2+s^{\prime e}_{12}s^e_{13}\cos(\delta^e_{12}-\delta^e_{23}) \right ).
\end{equation}
Hence, fixing $s^2_{13}$ to its central fit value of $0.02155$, and fixing $\theta^e_{12}=2.91$ for concreteness, constrains one combination of the two remaining model parameters, $\theta^e_{13}$ and $\delta^{\prime\prime} \equiv \delta^e_{12}-\delta^e_{13}$.  These quantities are further constrained by utilizing the relation for $s^2_{23}$ for $\theta^\nu_{23}=\pi/4$, which from Eq.~(\ref{eq:sin231213}) takes the form
\begin{equation}\label{eq:s23for1213}
s^2_{23} = \frac{1}{2(1-s^2_{13})} \left ((c^e_{12})^2+(s^e_{12})^2(s^e_{13})^2-s^{\prime e}_{12}s^e_{13}\cos(\delta^{\prime\prime}) \right ) = \frac{1+(s^e_{13})^2-2s^2_{13}}{2(1-s^2_{13})}.
 \end{equation}
 With these constraints, there are a set of values of $\sin(\theta^e_{13})$ and $\cos(\delta^{\prime\prime})$ that can yield the desired value of $s^2_{13}$ for the given input value of $\theta^e_{12}$, and simultaneously allow for a value of $s^2_{23}$ within the experimental range.  Here we note that for these given inputs, it is not possible to accommodate the full range of $3\sigma$ allowed values for $s^2_{23}$ from the global fit.  Instead, for these inputs the range of predictions for $\sin^2(\theta_{23})$ is given by $0.489<s^2_{23}<0.592$.  We further note that the reason for these boundaries ultimately is because both $\sin(\theta^e_{13})$ and $\cos\delta^{\prime\prime}$ are bounded functions, and thus there is an intricate relation between them that must be satisfied when applying the $s^2_{13}$ and $s^2_{23}$ constraints.
 
Turning now to the solar mixing angle, from Eq.~(\ref{eq:sin121213}) that for $\theta^\nu_{23}=\pi/4$, we have
\begin{equation}\label{eq:s12for1213}
s^2_{12} = \frac{(c^e_{12}c^e_{13})^2 y +\sqrt{\frac{y(1-y)}{2}}((c^e_{12})^2s^{\prime e}_{13}\cos(\delta^e_{13})-s^{\prime e}_{12}c^e_{13}\cos(\delta^e_{12}))+(1-y)(s^2_{13}-s^{\prime e}_{12}s^e_{13}\cos(\delta^{\prime\prime}))}{1-s^2_{13}}.
\end{equation}
 Hence, with $\theta^e_{12}$ and $s^2_{13}$ as inputs, and $\theta^e_{13}$ and $\delta^{\prime\prime} = \delta^e_{12}-\delta^e_{13}$ determined via the $s^2_{13}$ value and the range of allowed values for $s^2_{23}$, the solar mixing angle constraint provides a further bound on either of the individual phases $\delta^e_{12}$ and $\delta^e_{13}$ for a fixed value of $y$.   

The general formula for $\cos\delta$, as given in Eq.~(\ref{eq:cosdelformula1213}), takes the following form for $\theta^\nu_{23}=\pi/4$:
\begin{equation}\label{eq:cosdelfor1213}
\begin{aligned}
\cos\delta={}&
\frac{1}{2s^\prime_{12}s_{23}c_{23}s_{13}}\big \{ 2c_{12}^2c_{23}^2+c_{13}^2c_{23}^2+c_{13}^2c_{23}^2s_{12}^2-c_{12}^2-c_{23}^2\\
&+y(1-3c_{13}^2c_{23}^2)-2\vert c_{13}c_{23}\vert \sqrt{1-2c_{13}^2c_{23}^2}\sqrt{y(1-y)}\cos(\delta^e_{13})\big \},
\end{aligned}
\end{equation}
in which it is understood that the mixing angles are given by the relations of Eq.~(\ref{eq:s13for1213})--(\ref{eq:s12for1213}).  Therefore, together with the constraints on $\sin(\theta^e_{13})$ and $\delta^{\prime\prime}$ that result from the $s^2_{13}$ and $s^2_{23}$ constraints, there is an intricate interplay between the bounds on $s^2_{12}$ on the allowed model parameters and the resulting predictions for $\cos\delta$ and $\sin\delta$.
The numerical studies of the different model scenarios thus yield characteristically distinct results than in the other cases, where the $s^2_{13}$ and $s^2_{23}$ constraints were very tightly correlated, and $\cos\delta$ could be expressed simply in terms of the observed mixing angles and the parameter $y=\sin^2(\theta^\nu_{12})$.  

\begin{figure}[H]
  \begin{subfigure}[b]{0.475\textwidth}
  \caption{$\cos \delta$($\delta_{12}^{e}$, $\delta_{13}^{e}$) }
    \includegraphics[width=\textwidth]{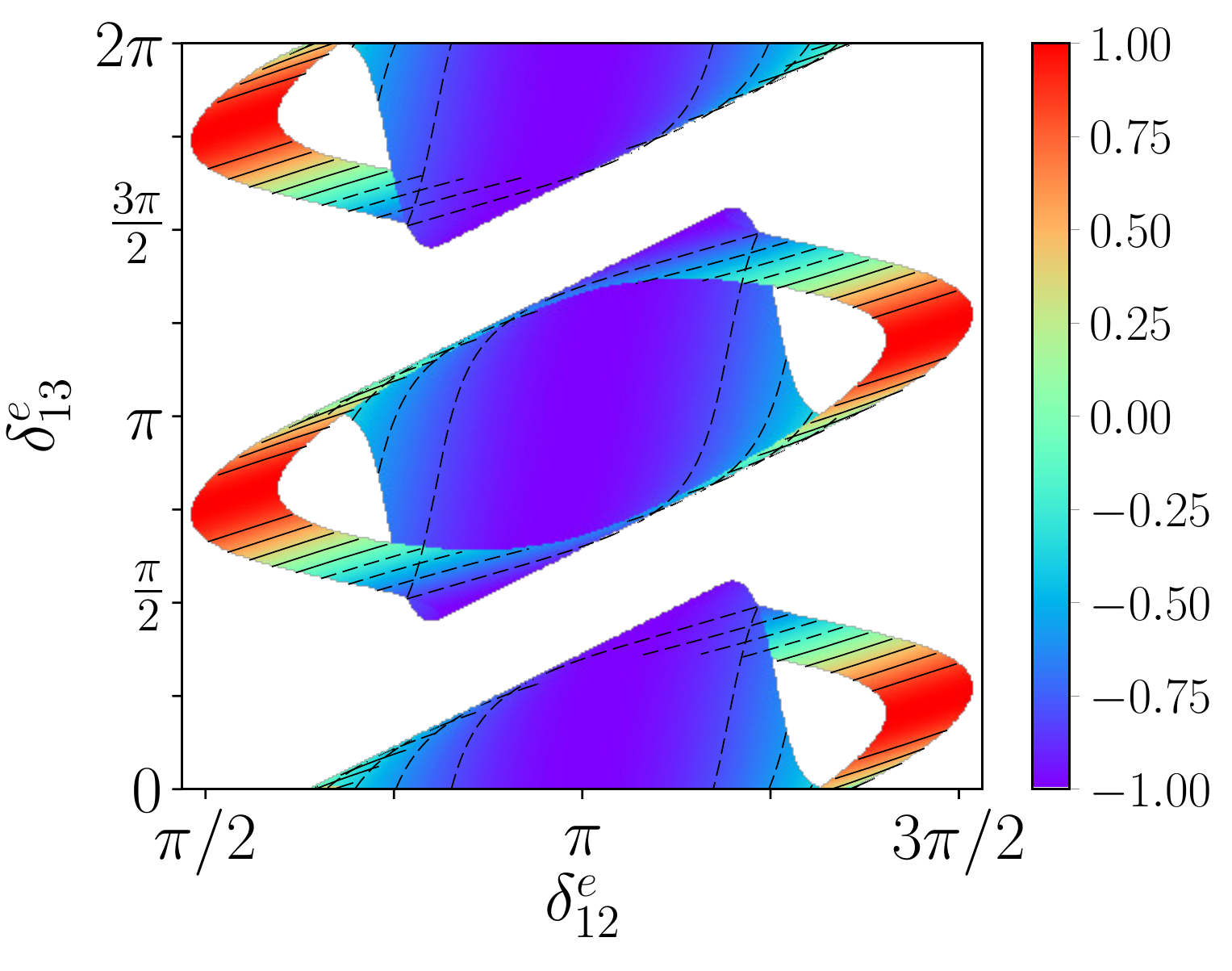}
  \end{subfigure}
  \hfill
  \begin{subfigure}[b]{0.475\textwidth}
  \caption{$\sin \delta$($\delta_{12}^{e}$, $\delta_{13}^{e}$) }
    \includegraphics[width=\textwidth]{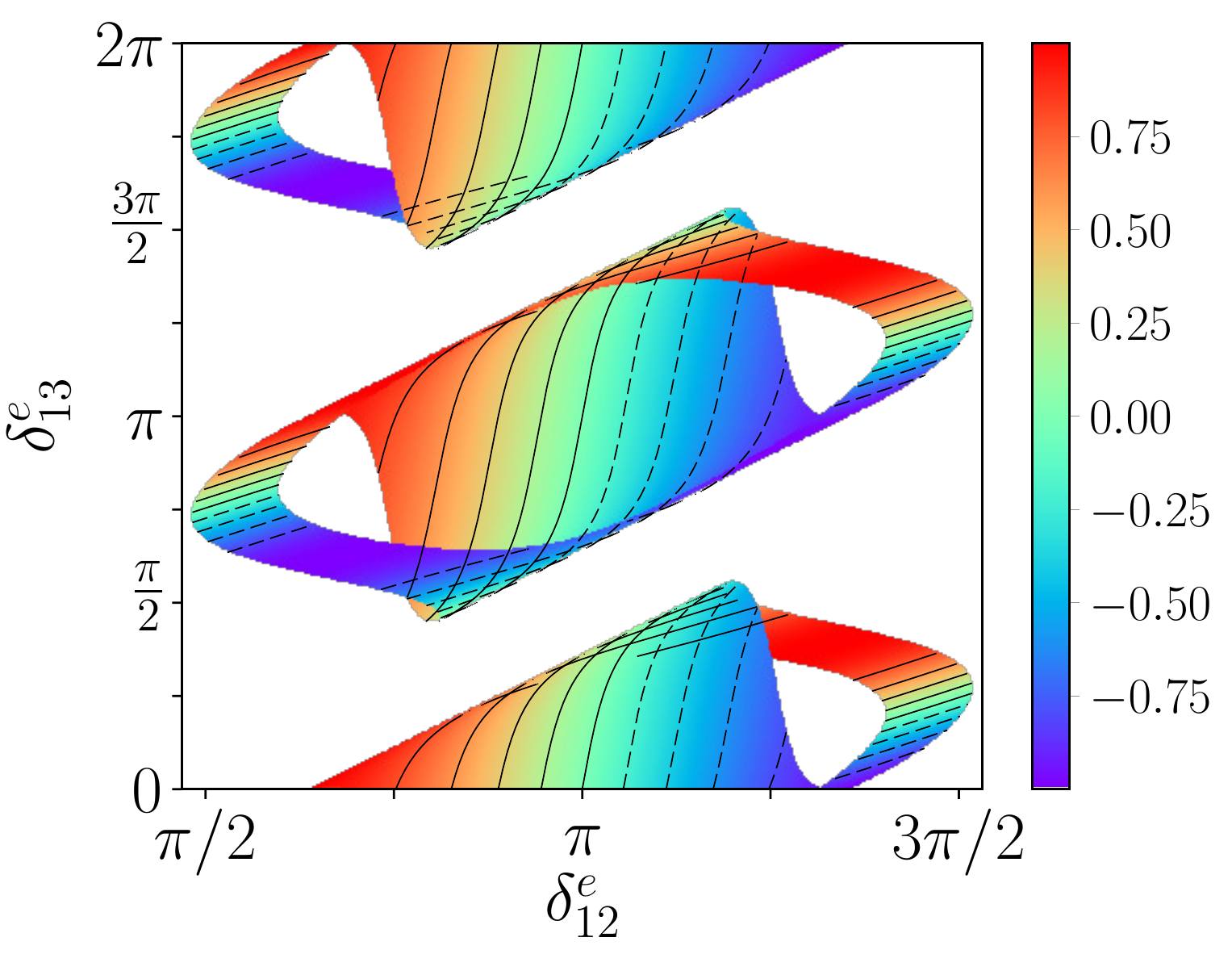}
  \end{subfigure}
  \caption{The predictions for $\cos \delta$ and $\sin \delta$  as a
  function of $\delta_{12}^{e}$ and $\delta_{13}^{e}$ for BM mixing,  for the case that $U_{e}=U^{e}_{13}(\theta_{13}^{e},\delta_{13}^{e})
U^{e}_{12}(\theta_{12}^{e},\delta_{12}^{e})$.}
  \label{fig:1213bm}
\end{figure}

These differences are easily seen in the numerical results for specific allowed regions of $\delta^e_{12}$ and $\delta^e_{13}$ for BM mixing, as shown in Figure~\ref{fig:1213bm}. The diagonal bands of disallowed combinations of $\delta^e_{12}$ and $\delta^e_{13}$ arise because not all values of $\delta^{\prime\prime}=\delta^e_{12}-\delta^e_{13}$ are consistent with the reactor and atmospheric angle relations, as discussed.  These distributions are also characterized by regions in which the same pair of phase angles can have different values for $\cos\delta$ and $\sin\delta$; ultimately this occurs because it is possible to have multiple solutions for $\theta^e_{13}$ that satisfy all constraints.  For BM mixing, there is a broad range of values for both phases that can satisfy all bounds, and the most probable value for $\cos\delta$ is clearly near $-1$.

\begin{figure}[H]
  \begin{subfigure}[b]{0.475\textwidth}
  \caption{$\cos \delta$($\delta_{12}^{e}$, $\delta_{13}^{e}$) }
    \includegraphics[width=\textwidth]{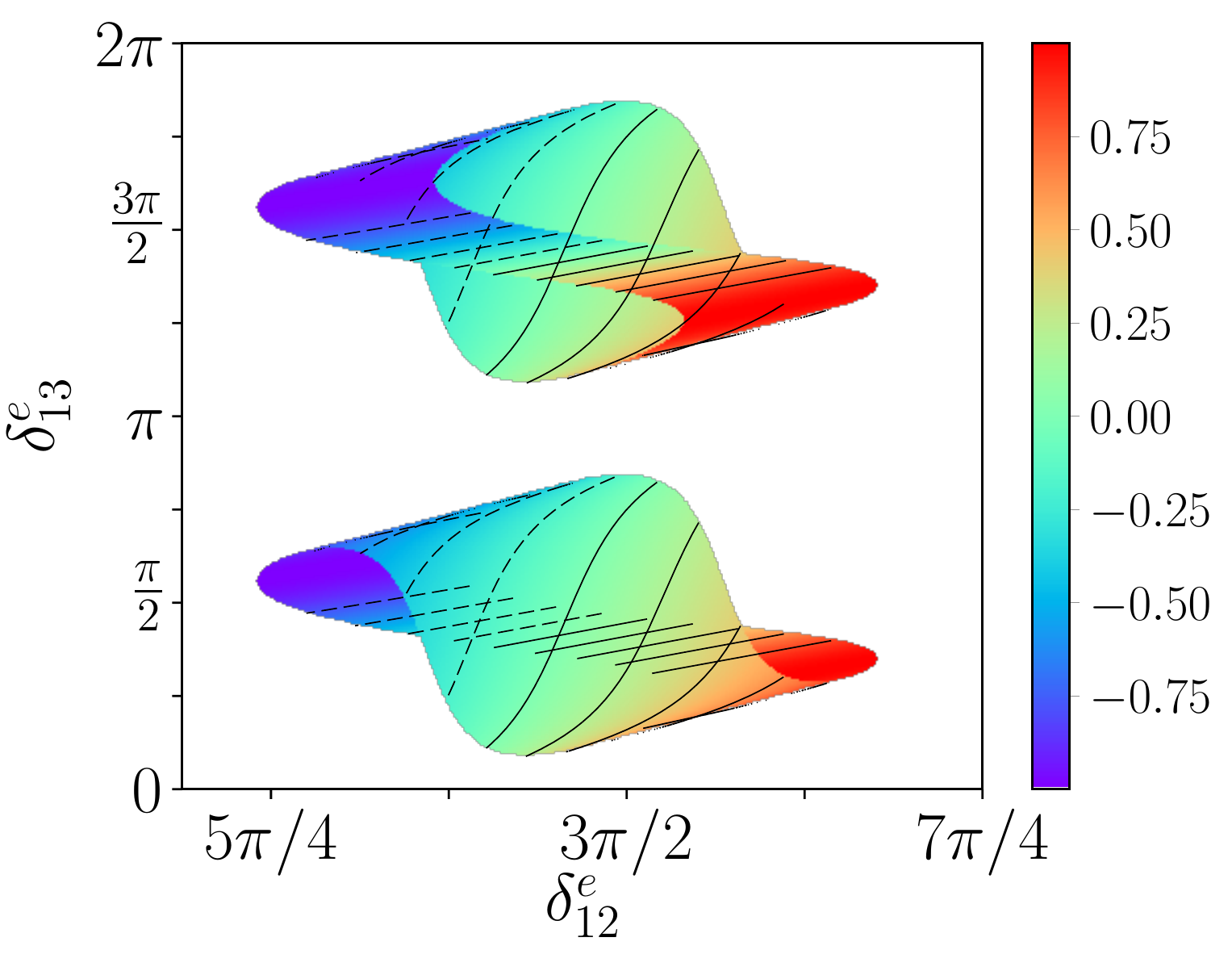}
  \end{subfigure}
  \hfill
  \begin{subfigure}[b]{0.475\textwidth}
  \caption{$\sin \delta$($\delta_{12}^{e}$, $\delta_{13}^{e}$) }
    \includegraphics[width=\textwidth]{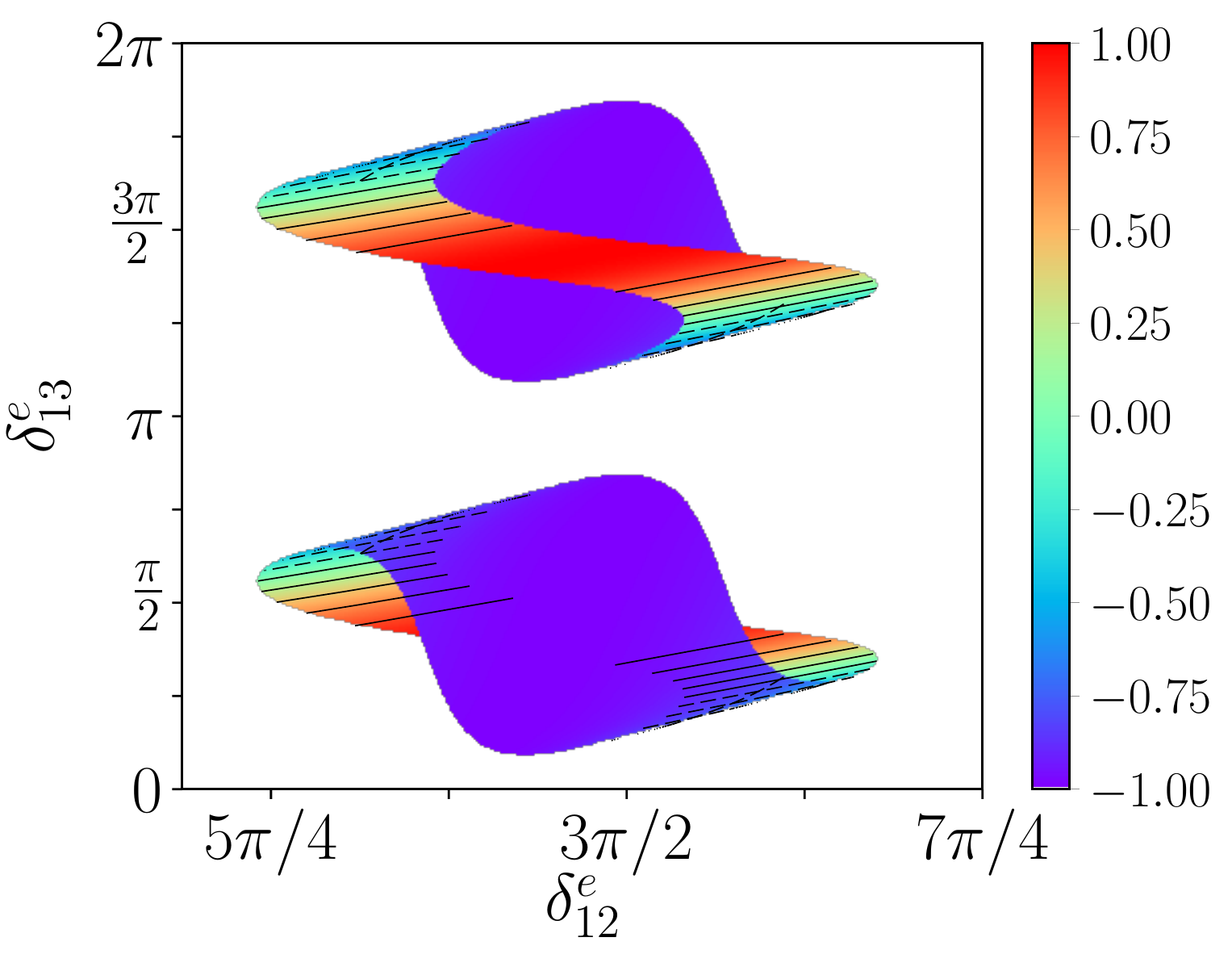}
  \end{subfigure}
  \caption{The predictions for $\cos \delta$ and $\sin \delta$  as a
  function of $\delta_{12}^{e}$ and $\delta_{13}^{e}$ for TBM mixing,  for the case that $U_{e}=U^{e}_{13}(\theta_{13}^{e},\delta_{13}^{e})
U^{e}_{12}(\theta_{12}^{e},\delta_{12}^{e})$.}
  \label{fig:1213tbm}
\end{figure}

\begin{figure}[H]
  \begin{subfigure}[b]{0.475\textwidth}
  \caption{$\cos \delta$($\delta_{12}^{e}$, $\delta_{13}^{e}$)}
    \includegraphics[width=\textwidth]{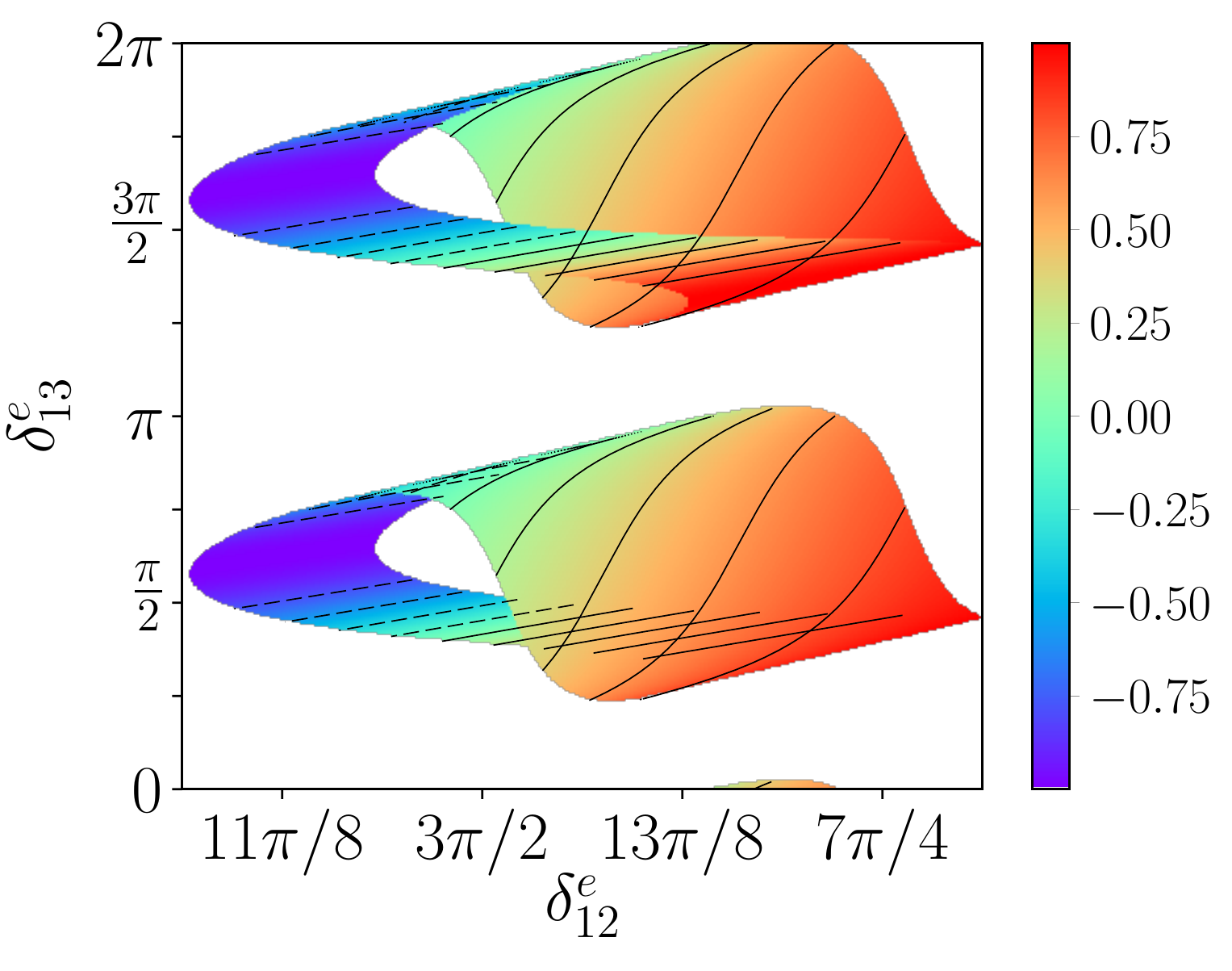}
  \end{subfigure}
  \hfill
  \begin{subfigure}[b]{0.475\textwidth}
  \caption{$\sin \delta$($\delta_{12}^{e}$, $\delta_{13}^{e}$) }
    \includegraphics[width=\textwidth]{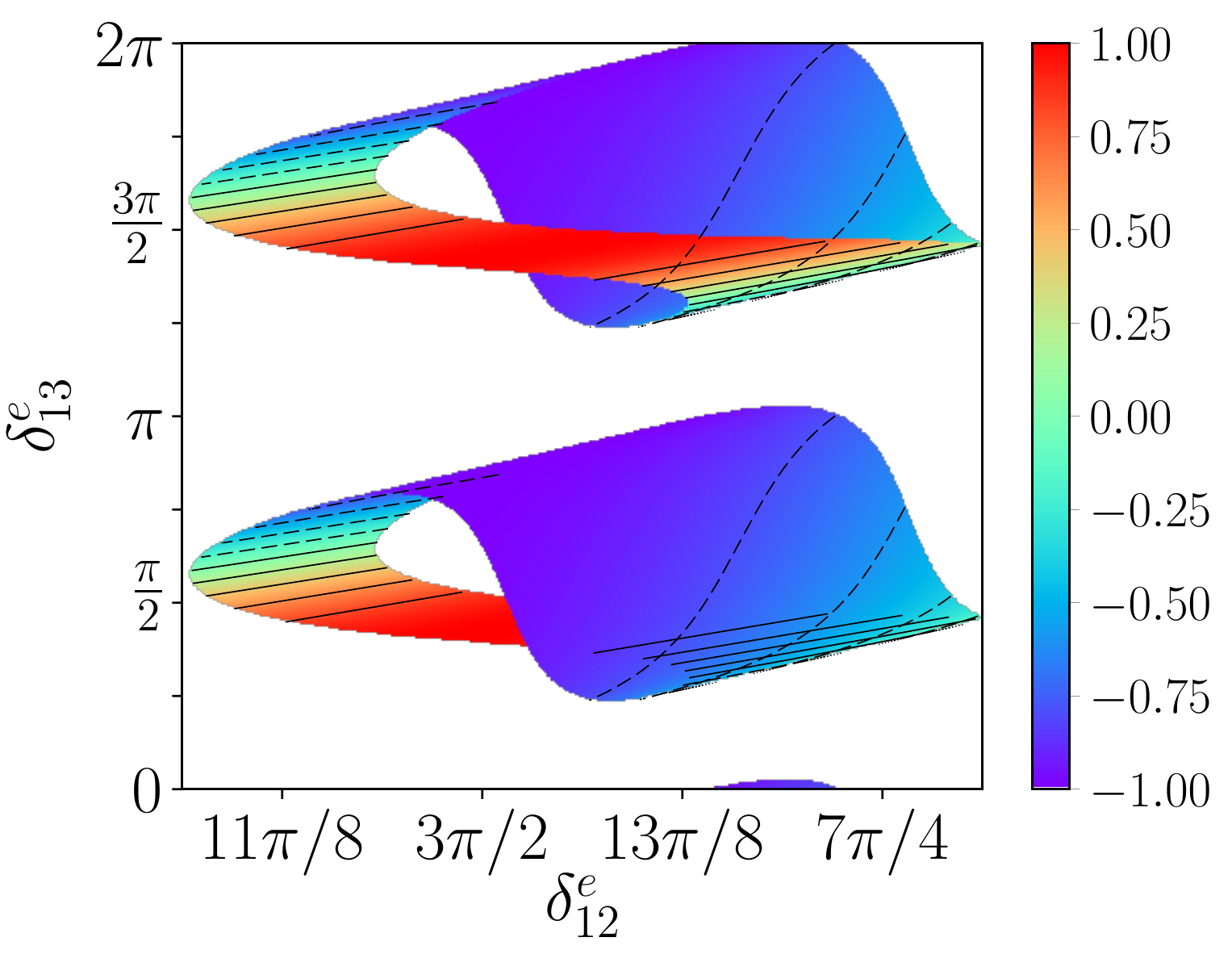}
  \end{subfigure}
  \caption{The predictions for $\cos \delta$ and $\sin \delta$ as a
  function of $\delta_{12}^{e}$ and $\delta_{13}^{e}$ for HEX mixing,  for the case that $U_{e}=U^{e}_{13}(\theta_{13}^{e},\delta_{13}^{e})
U^{e}_{12}(\theta_{12}^{e},\delta_{12}^{e})$.}
  \label{fig:1213hex}
\end{figure}

\begin{figure}[H]
  \begin{subfigure}[b]{0.475\textwidth}
  \caption{$\cos \delta$($\delta_{12}^{e}$, $\delta_{13}^{e}$) }
    \includegraphics[width=\textwidth]{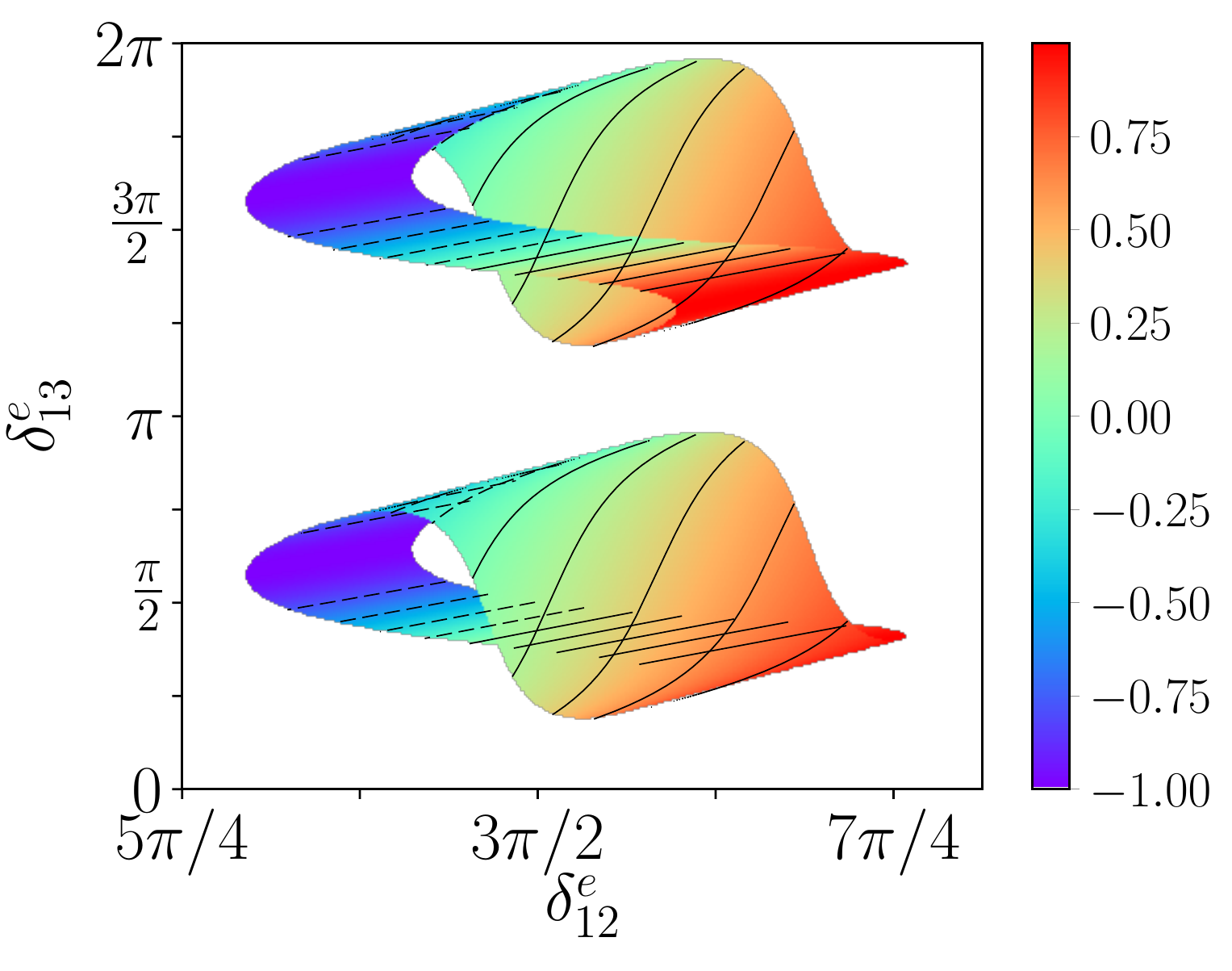}
  \end{subfigure}
  \hfill
  \begin{subfigure}[b]{0.475\textwidth}
  \caption{$\sin \delta$($\delta_{12}^{e}$, $\delta_{13}^{e}$) }
    \includegraphics[width=\textwidth]{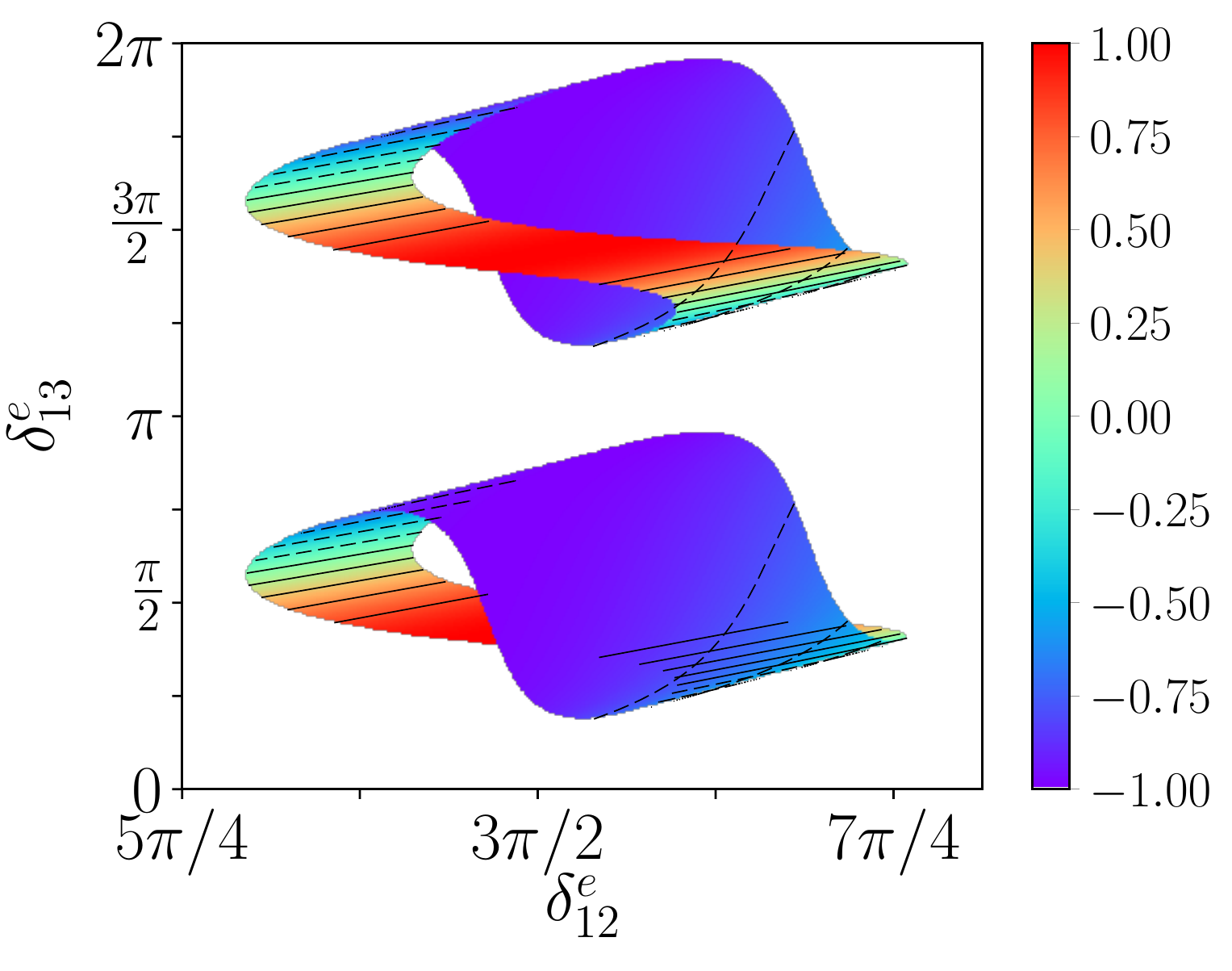}
  \end{subfigure}
  \caption{The predictions for $\cos \delta$ and $\sin \delta$  as a
  function of $\delta_{12}^{e}$ and $\delta_{13}^{e}$ for GR1 mixing,  for the case that $U_{e}=U^{e}_{13}(\theta_{13}^{e},\delta_{13}^{e})
U^{e}_{12}(\theta_{12}^{e},\delta_{12}^{e})$.}
  \label{fig:1213gr1}
\end{figure}

\begin{figure}[H]
  \begin{subfigure}[b]{0.475\textwidth}
  \caption{$\cos \delta$($\delta_{12}^{e}$, $\delta_{13}^{e}$) }
    \includegraphics[width=\textwidth]{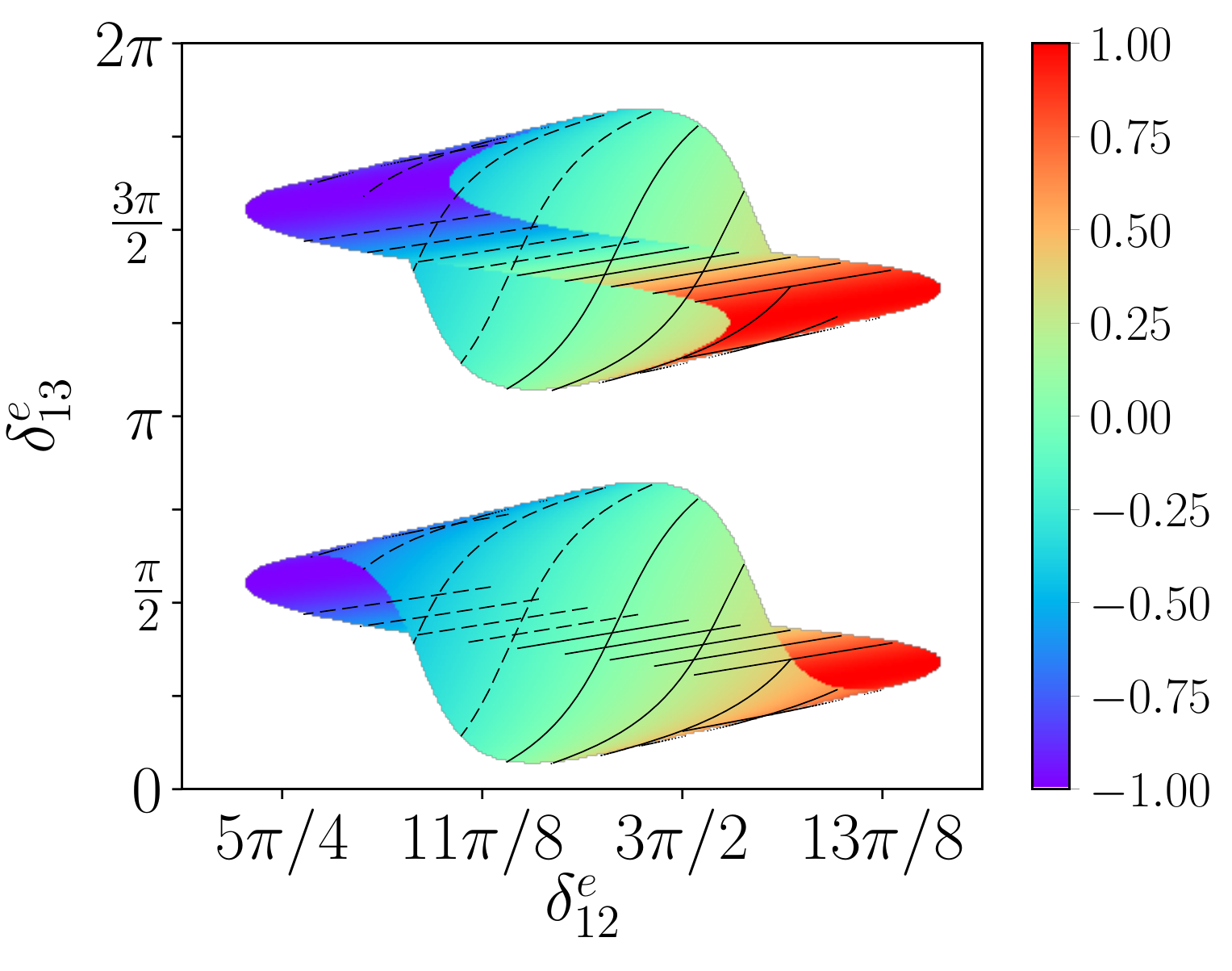}
  \end{subfigure}
  \hfill
  \begin{subfigure}[b]{0.475\textwidth}
  \caption{$\sin \delta$($\delta_{12}^{e}$, $\delta_{13}^{e}$) }
    \includegraphics[width=\textwidth]{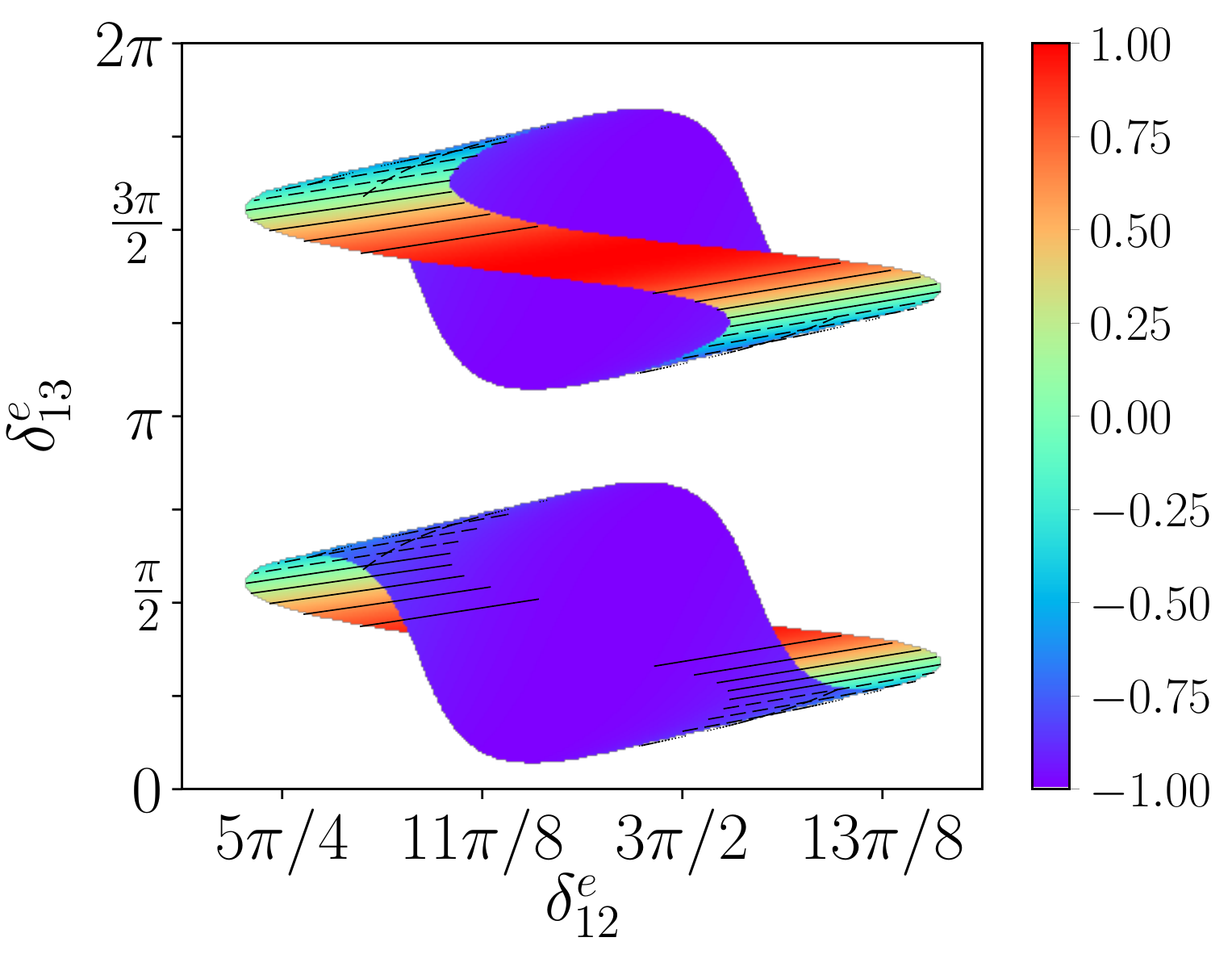}
  \end{subfigure}
  \caption{The predictions for $\cos \delta$ and $\sin \delta$ as a
  function of $\delta_{12}^{e}$ and $\delta_{13}^{e}$ for GR2 mixing,  for the case that $U_{e}=U^{e}_{13}(\theta_{13}^{e},\delta_{13}^{e})
U^{e}_{12}(\theta_{12}^{e},\delta_{12}^{e})$.}
  \label{fig:1213gr2}
\end{figure}

For the other mixing patterns, we have zoomed in on specific parameter regions for $\delta^e_{12}$, as in the previous subsections. We show the results for $\cos\delta$ and $\sin\delta$ for TBM mixing in Figure~\ref{fig:1213tbm}, for HEX mixing in Figure~\ref{fig:1213hex}, and for GR1 and GR2 mixing in Figures~\ref{fig:1213gr1} and \ref{fig:1213gr2}, respectively.  In each of these cases, we see a pattern that there is a relatively compact region of allowed parameters that is ``folded'' upon itself due to the possibility of double-valued solutions with different values of $\theta^e_{13}$.  
As in the previous subsections, we see similarities in the results for scenarios with similar values of $y$.  More precisely, the TBM and GR2 cases have many qualitative features in common, with a relatively compact allowed parameter range in $\delta^e_{12}$, and similar shapes to the distributions, with a most probable value of $\cos\delta$ near zero.  In contrast, while the HEX and GR1 scenarios also show a similar range of allowed values of $\delta^e_{12}$, their distributions in $\cos\delta$ and $\sin\delta$ are similar (and distinct from the TBM and GR2 cases), with a most probable value of $\cos\delta$ near 1.  We also note that in all cases, a broad range of values for $\cos\delta$ and $\sin\delta$ can be obtained, as opposed to the case for other types of charged lepton perturbations. 
In summary, we see that for the double rotations, our procedure for fixing $\sin^2(\theta_{13})$ to its central value and setting one the $\theta^e_{jk}$ to a specific input value, while fixing the remaining $\theta^e_{jk}$ from the remaining experimental bounds on the mixing angles, leads to characteristic results for the three classes of double rotations studied here.  For the perturbations that include a rotation in the $2-3$ plane, the value of $s^2_{23}$ is fixed by just one parameter once $s^2_{13}$ is specified, yielding specific bands of allowed values for the two phase parameters of the charged lepton sector.  In contrast, when the two rotations of the charged lepton sector are in the $1-2$ and $1-3$ planes, this tight correlation between the reactor and atmospheric mixing angles is modified, allowing for distinctive distributions for the viable parameter space and their resulting predictions for the MNSP phase $\delta$.

\section{Conclusions}\label{sec:con}


In this paper, we have analyzed sum rules for the lepton sector Dirac phase $\delta$ of $U_\textsf{MNSP}$, using the most recent global fits from the literature.  Our approach was to derive a comprehensive set of sum rules that when simultaneously imposed as required by unitarity of the lepton mixing matrix, result in a single sum rule for $\cos\delta$ as well as specific predictions for the lepton mixing angles that result in given top-down model scenarios. We then applied these constraints to a well-known class of theoretical models, and studied the implications for the allowed regions of model parameters and the resulting predictions for $\cos\delta$ and $\sin\delta$.  

To be more precise, we have investigated the phenomenologically allowed regions of the parameter space of $\theta_{ij}^{e}$ and $\delta_{ij}^{e}$ in the charged lepton mixing matrix $U_e$, assuming a set of well-known theoretical starting points for the neutrino mixing $U_{\nu}$ for which $\theta^\nu_{23}=\pi/4$ and $\theta^\nu_{13}=0$. These scenarios are thus characterized by their values of $\theta^\nu_{12}$, as well as the model parameters of the charged lepton perturbations.  We have considered five different sets of charged lepton perturbations: two single rotations in the $1-2$ and $1-3$ planes, and three sets of two non-commuting rotations in the $1-2/2-3$, $1-3/2-3$, and $1-2/1-3$  planes.  

The results of this analysis generally show that as a function of $\sin^2(\theta^\nu_{12})$ and the specific model parameters of the charged lepton sector, the very precise measurement of the reactor angle strongly constrains the possible values of the atmospheric mixing angle, such that improved determinations of $\sin^2(\theta_{23})$ can make definitive statements about the viability of these theoretical models.  For example, in the case of the single rotations with a fixed starting value of $\theta^\nu_{23}$, once the reactor angle is specified, the atmospheric angle is also fixed, independently of the value of $\sin^2(\theta^\nu_{12})$.  This means that in these scenarios, only a small portion of the experimentally allowed range for $\theta_{23}$ is allowed, with that small range governed by the allowed range of $\theta_{13}$. Further experimental bounds on the allowed range of $\theta_{23}$ will thus have the power to provide a definitive answer to the question of whether or not such simple models based on charged lepton corrections consisting of a single rotation are allowed or are ruled out.  This powerful statement does not depend on measurements of the mass ordering,  the Dirac CP-violating phase $\delta$, or the question of whether neutrinos are Dirac or Majorana fermions.   Furthermore, even if the atmospheric angle constraints are satisfied in a specific single rotation scenario, the results for $\theta_{12}$ and $\delta$ will further separate these scenarios as a function of  $\sin^2(\theta^\nu_{12})$.

In the case of double rotations, we have seen that there is more flexibility than is allowed by the single rotation cases (both with fixed $\theta^\nu_{23}$).  This is as expected given that the theoretical model involves additional degrees of freedom. Nonetheless, each scenario displays detailed correlations among the measured mixing parameters and the Dirac CP-violating phase, that can be probed in detail via the current and forthcoming neutrino experimental program, which include further improvements in the mixing angle ranges and constraints on CP violation. In this way, the combination of the types of rotations, and the value of $\sin^2(\theta^\nu_{12})$, can be jointly constrained by current and future data.

This work also showcases the predictive power within these simple theoretical scenarios for the preferred range for the Dirac CP-violating phase $\delta$.  Again, here we have seen that with the single rotations, which only have a single source of CP violation, there are strong correlations between the values of  $\sin^2(\theta^\nu_{12})$ and the value of $\delta$, while for double rotations, there is a more intricate set of correlations that result from the presence of two CP-violating sources.  Most importantly, studies such as this one highlight that with the anticipated improvements in the lepton mixing and CP phase measurements from neutrino oscillation experiments on the horizon, we may indeed be just at the cusp of making great progress in taking the next step toward a resolution to the flavor puzzle of the Standard Model.


\section*{Acknowledgments}


L.A.D. would like to thank A.~Aranda for useful discussions and acknowledge
support from  CONACYT project CB-2015-01/257655 (Mexico). The work of L.L.E.
is supported by the U.S.~Department of Energy under contract number
DE-SC0017647. A.S. would like to acknowledge support from PRODEP start-up
grant (511-67-612). R.R. is grateful for the hospitality of FdeC-CUICBAS
Universidad de Colima where part of this work was carried out. The
work of R.R. is supported in part by the Ministry of Science and Technology (MoST)
of Taiwan under grant numbers 106-2011-M001-113- and 104-2112-M-001-001-MY3.




 
\end{document}